\documentclass[]{pasj02} 
\usepackage[switch,mathlines]{lineno} % add line number to manuscript
\usepackage{color}
\newcommand{\red}[1]{#1}
\newcommand{\redtwo}[1]{#1}

\jyear{2026}
\Received{2026/05/17}%{yyyy/mm/dd}
\Accepted{2026/07/24}%{yyyy/mm/dd}
\doi{psag104}

\newcommand{\iac}{Instituto de Astrof\'\i sica de Canarias (IAC), 38205 La Laguna, Tenerife, Spain}
\newcommand{\komaba}{Department of Multi-Disciplinary Sciences, The University of Tokyo, 3-8-1 Komaba, Meguro, Tokyo 153-8902, Japan}
\newcommand{\komabains}{Komaba Institute for Science, The University of Tokyo, 3-8-1 Komaba, Meguro, Tokyo 153-8902, Japan}
\newcommand{\abc}{Astrobiology Center, 2-21-1 Osawa, Mitaka, Tokyo 181-8588, Japan}
\newcommand{\naoj}{National Astronomical Observatory of Japan, 2-21-1 Osawa, Mitaka, Tokyo 181-8588, Japan}
\newcommand{\astron}{Department of Astronomy, The University of Tokyo, 7-3-1 Hongo, Bunkyo, Tokyo 113-0033, Japan}
\newcommand{\sokendai}{Department of Astronomical Science, The Graduate University for Advanced Studies (SOKENDAI), 2-21-1 Osawa, Mitaka, Tokyo 181-8588, Japan}
\newcommand{\ep}{Department of Earth and Planetary Science, The University of Tokyo, 7-3-1 Hongo, Bunkyo, Tokyo 113-0033, Japan}

\newcommand{\ritsumei}{Department of Physical Sciences, Ritsumeikan University, 
1-1-1 Noji-higashi, Kusatsu, Shiga 525-8577, Japan}
\newcommand{\subaru}{Subaru Telescope, National Astronomical Observatory of Japan, 650 N. Aohoku Place, Hilo, HI 96720, USA}

\newcommand{\gron}{Kapteyn Astronomical Institute, University of Groningen, P.O. Box 800, 9700 AV Groningen, Netherlands}

\newcommand{\rikkyo}{Department of Physics, College of Science, Rikkyo University, 3-34-1 Nishi-Ikebukuro, Toshima-ku, Tokyo 171-8501, Japan}

\begin{document} 

%\title{A Search for helium in the atmospheres of four sub-Neptunes around M-dwarfs}
\title{A Search for helium in the atmospheres of three sub-Neptunes and a super-Earth around M-dwarfs}

%%% begin:list of authors
% Do NOT capitalize all letters in "textsc".
\author{Kiyoe \textsc{Kawauchi}\altaffilmark{1}\orcid{0000-0003-1205-5108}, 
% main support & analysis 
Norio \textsc{Narita}\altaffilmark{2,3,4}\orcid{0000-0001-8511-2981}, % intensive PI
Yuichi \textsc{Ito}\altaffilmark{5,4}\orcid{0000-0002-0598-3021}, % give some advice in the model part
Akifumi \textsc{Nakayama}\altaffilmark{6}\orcid{0000-0002-0998-0434}, % give some advice in the model part
Teruyuki \textsc{Hirano}\altaffilmark{4,5,7}\orcid{0000-0003-3618-7535},  %the reduction analysis on other days (without transit data) & observer on 2020/09/30
Masayuki \textsc{Kuzuhara}\altaffilmark{4,5}\orcid{0000-0002-4677-9182}, %gave me the reduction pipeline
Mayuko \textsc{Mori}\altaffilmark{4,5}\orcid{0000-0003-1368-6593}, %observer on 2021/01/30
%
% observation & intensive member (alphabet)
%
Izuru \textsc{Fukuda}\altaffilmark{8}\orcid{0000-0002-9436-2891},
Akihiko \textsc{Fukui}\altaffilmark{2,3}\orcid{0000-0002-4909-5763},
Yuya \textsc{Hayashi}\altaffilmark{8}\orcid{0000-0001-8877-0242},
Yasunori \textsc{Hori}\altaffilmark{4,9}\orcid{0000-0003-4676-0251},
Masahiro \textsc{Ikoma}\altaffilmark{5,10,4}\orcid{0000-0002-5658-5971},
Kai \textsc{Ikuta}\altaffilmark{11}\orcid{0000-0002-5978-057X}, 
%TOI-654の情報をいち早くくれていた
%
Taiki \textsc{Kagetani}\altaffilmark{5,8}\orcid{0000-0002-5331-6637},
Yugo \textsc{Kawai}\altaffilmark{8}\orcid{0000-0002-0488-6297},
Tadahiro \textsc{Kimura}\altaffilmark{12,13}\orcid{0000-0001-8477-2523},
Vigneshwaran \textsc{Krishnamurthy}\altaffilmark{14,4}\orcid{0000-0003-2310-9415},
%
%Kohei \textsc{Miyakawa},
%
Huan-Yu \textsc{Teng}\altaffilmark{15}\orcid{0000-0003-3860-6297},
Noriharu \textsc{Watanabe}\altaffilmark{8}\orcid{0000-0002-7522-8195},
Yujie \textsc{Zou}\altaffilmark{8}\orcid{0000-0002-5609-4427},
%
% instrument (alphabet)
%
Hiroki \textsc{Harakawa}\altaffilmark{16}\orcid{0000-0002-7972-0216},
Klaus \textsc{Hodapp}\altaffilmark{18}\orcid{0000-0003-0786-2140},
Tomoyuki \textsc{Kudo}\altaffilmark{17}\orcid{0000-0002-9294-1793},
Takashi \textsc{Kurokawa}\altaffilmark{
19},
Jun \textsc{Nishikawa}\altaffilmark{5,7,4}\orcid{0000-0001-9326-8134},
Masashi \textsc{Omiya}\altaffilmark{4,5}\orcid{0000-0002-5051-6027},
Motohide \textsc{Tamura}\altaffilmark{20,4,21}\orcid{0000-0002-6510-0681},
Akitoshi \textsc{Ueda}\altaffilmark{4,5,7},
S\'{e}bastien \textsc{Vievard}\altaffilmark{17,18}\orcid{0000-0003-4018-2569}
}
%\thanks{Example: Present Address is xxxxxxxxxx}}
\altaffiltext{1}{\ritsumei}
\altaffiltext{2}{\komabains}
\altaffiltext{3}{\iac}
\altaffiltext{4}{\abc}
\altaffiltext{5}{\naoj}
\altaffiltext{6}{\rikkyo}
\altaffiltext{7}{\sokendai}
\altaffiltext{8}{\komaba}
\altaffiltext{9}{Department of Earth Sciences, School of Science, Okayama University, 3-1-1 Tsushima-naka, Kita-ku, Okayama, 700-8530, Japan}
\altaffiltext{10}{\ep}
\altaffiltext{11}{Department of Social Data Science, Hitotsubashi University, 2-1 Naka, Kunitachi, Tokyo 186-8601, Japan}
\altaffiltext{12}{\gron}
\altaffiltext{13}{UTokyo Organization for Planetary Space Science (UTOPS), University of Tokyo, Hongo, Bunkyo-ku, Tokyo 113-0033, Japan}
\altaffiltext{14}{Trottier Space Institute at McGill, McGill University, 3550 University Street, Montreal, QC H3A 2A7, Canada}
\altaffiltext{15}{National Astronomical Observatories, Chinese Academy of Sciences, Beijing 100012, China}
\altaffiltext{16}{Department of Science, National Museum of Nature and Science, 4-1-1 Aamakubo, Tsukuba, Ibaraki, Japan, 3050005}
\altaffiltext{17}{\subaru}
\altaffiltext{18}{University of Hawaii, Institute for Astronomy, 640 N. Aohoku Place, Hilo, HI 96720, USA}
\altaffiltext{19}{Institute of Engineering, Tokyo University of Agriculture and Technology, 2-24-26 Nakacho, Koganei, Tokyo, 184-8588, Japan}
\altaffiltext{20}{\astron}
\altaffiltext{21}{Institute of Laser Engineering, The University of Osaka, 2-6 Yamadaoka, Suita, Osaka 565-0871, Japan}

% \author{
%  A-Firstname \textsc{A-Familyname},\altaffilmark{1}\altemailmark\orcid{0000-0000-0000-0000} \email{aaaaa@xxx.xxx.xx.xx} 
%  B-Firstname \textsc{B-Familyname},\altaffilmark{2}$^{,\dag}$\orcid{0000-0000-0000-0000}
%  C-Firstname \textsc{C-Familyname},\altaffilmark{3}\altemailmark \email{ccccc@xxx.xxx.xx.xx}
%  and 
%  D-Firstname \textsc{D-Familyname}\altaffilmark{2}\altemailmark\orcid{0000-0000-0000-0000} \email{ddddd@xxx.xxx.xx.xx}
% }
% \altaffiltext{1}{A-Address of Institute}
% \altaffiltext{2}{B-Address of Institute}
% \altaffiltext{3}{C-Address of Institute}

% \footnotetext[$\dag$]{Present address: ....}

%%% end:list of authors

%% !!! Select 3 to 5 words from PASJ's key words !!! 
%% List of Key Words: https://academic.oup.com/pasj/pages/Pasj_Keywords 
%% "\KeyWords{ }" always has to be placed before ``\maketitle'' 

\maketitle

\begin{abstract}
%Please read ``IMPORTANT NOTICE'' carefully before preparing a manuscript.  
Thousands of sub-Neptunes have been discovered mainly through space-based surveys such as Kepler and TESS. Their bulk compositions and internal structures are thought to reflect their formation and evolutionary pathways, and atmospheric observations provide constraints on these processes. The near-infrared helium triplet is a potential tracer of extended, escaping H/He atmospheres. Recent models that include geometric effects suggest that planets orbiting nearby late M dwarfs may offer favorable conditions for detecting this signal. Nevertheless, helium has been reported for only \red{three} planets around M dwarfs to date, in contrast to the larger number of detections around especially K dwarfs, motivating further surveys of M-dwarf systems.
We conducted high-resolution transmission spectroscopy of  
three sub-Neptunes (TOI-2136b, TOI-654b, and LP 791-18c) and a super-Earth (TOI-1634b) orbiting M dwarfs 
with the InfraRed Doppler (IRD) spectrograph on the Subaru Telescope. 
We find no statistically significant helium absorption in any target; accordingly, we derive 95\% confidence upper limits on the helium line depth of 1.36\%, 0.60\%, 2.07\%, and 3.00\%, and on the equivalent width of 7.3, 2.1, 7.4, and 9.1~m\mbox{\AA}, for TOI-2136b, TOI-1634b, TOI-654b, and LP~791-18c, respectively. 
We further explored constraints on the upper-atmospheric temperature and mass-loss rate by comparing these results with isothermal Parker-wind models. While we have compared with self-consistent \texttt{ATES} models of primordial H/He atmospheres spanning a range of assumed X-ray luminosities, 
changes in the assumed XUV flux do not appear to account for the non-detections. 
The results suggest that these planets have metal-enriched H/He primary atmospheres or non-primordial atmospheres, such as water-rich envelopes. 
Future observations of other absorption lines, such as Lyman-$\alpha$, H-$\alpha$, and H$_2$O, may provide further constraints on these atmospheres.

\end{abstract}

%\pagewiselinenumbers 

\noindent\small
%\textcolor{red}{
This is a pre-copyedited, author-produced version of an
article accepted for publication in ``Publications of the
Astronomical Society of Japan'' following peer review.
The version of record is available
online at [https://doi.org/10.1093/pasj/psag104].
%}
\normalsize

\section{Introduction}

Over the past three decades, ground-based and space telescopes, particularly the \redtwo{Kepler \red{ (\cite{Borucki+2010})} and Transiting Exoplanet Survey Satellite (TESS; \red{\cite{Ricker+2015}})}, have discovered more than 6,000 exoplanets. These discoveries have revealed that small close-in planets (1 R$_\oplus$ < R$_p$ < 4 R$_\oplus$), which do not exist in our solar system, are most common in the Milky Way (e.g., \cite{Howard+2012, Fulton+2017}). 
Recent advances in radial velocity (RV) measurements have increased the number of small transiting planets with known masses (e.g., \cite{Barkaoui+2025, Ilaria+2025}). By combining mass and radius measurements, we can estimate their bulk compositions. 
However, for sub-Neptune-sized planets (1.7--4 R$_\oplus$), mass and radius alone cannot uniquely constrain their compositions, leaving a range of possibilities, such as a rocky core surrounded by an H/He envelope or water-rich interiors (e.g., \cite{Zeng2019}). 
The latter includes hycean worlds, in which an H$_2$O ocean exists beneath an $\rm H_2$-rich atmosphere \citep{Madhusudhan+2021}.
\red{To unveil such compositional diversity, which allows us to better understand their formation and evolution pathways, one needs to obtain additional information by observing their atmosphere.}

In addition, NASA's Kepler space telescope uncovered a bimodal distribution in \red{the small, close-in planet population}, known as the "radius valley" \citep{Fulton+2017}. 
Although this feature may also reflect differences in formation location that lead to compositional diversity \red{(e.g. \cite{Luque+2022})}, it is commonly interpreted as the transition from rocky to non-rocky planets, potentially shaped by atmospheric escape driven by photoevaporation (e.g. \cite{Owen&Wu2013,Owen&Wu2017,Jin+2014,Lopez&Rice2018,Mordasini2020}) and/or core-powered mass-loss \citep{Ginzburg+2016,Ginzburg+2018,Gupta&Schlichting2019,Gupta&Schlichting2021}. To confirm this theory, it is important to observe the mass-loss rates of small planets.

The near-infrared helium triplet\red{, located around 10830 \mbox{\AA} in air wavelengths,} is an effective indicator for probing current mass-loss rates in exoplanetary atmospheres, as it can be observed from ground-based telescopes and is not significantly affected by interstellar medium absorption, unlike Lyman-$\alpha$. Moreover, it serves as a useful tracer of primordial atmospheres accreted from the protoplanetary nebula in hydrogen-rich exoplanets, since helium in any significant quantity is difficult to create through outgassing processes \citep{Elkins-Tanton2008}. 

The strength of the near-infrared helium triplet lines depends on the population of metastable helium atoms, which is primarily governed by the balance between stellar extreme-ultraviolet (XUV) and mid-ultraviolet (mid-UV) radiation \citep{Oklopcic+2018}. Based on this balance, previous theoretical studies suggested that planets orbiting K-type stars are the most favorable targets for helium detection \citep{Oklopcic2019}.

To date, helium has been searched for in the atmospheres of more than 70 exoplanets, with the majority of detections associated with planets orbiting K-type stars \citep{Orell-Miquel+2024, Krishnamurthy+2024}, which is consistent with the \red{theoretical expectations}. 
%However, recent models incorporating more detailed thermal structures of planetary outflows have predicted that planets orbiting M dwarfs show the strongest helium triplet signals 
\red{However, recent models incorporating more detailed thermal structures of planetary outflows predict stronger helium triplet signals for planets orbiting M dwarfs than for those orbiting earlier-type stars} \citep{Biassoni+2024, Ballabio+2025}. 

\begin {figure} [htbp]
% \begin{minipage}{0.45\hsize}
\begin{center}
 \includegraphics[width=0.48\textwidth] {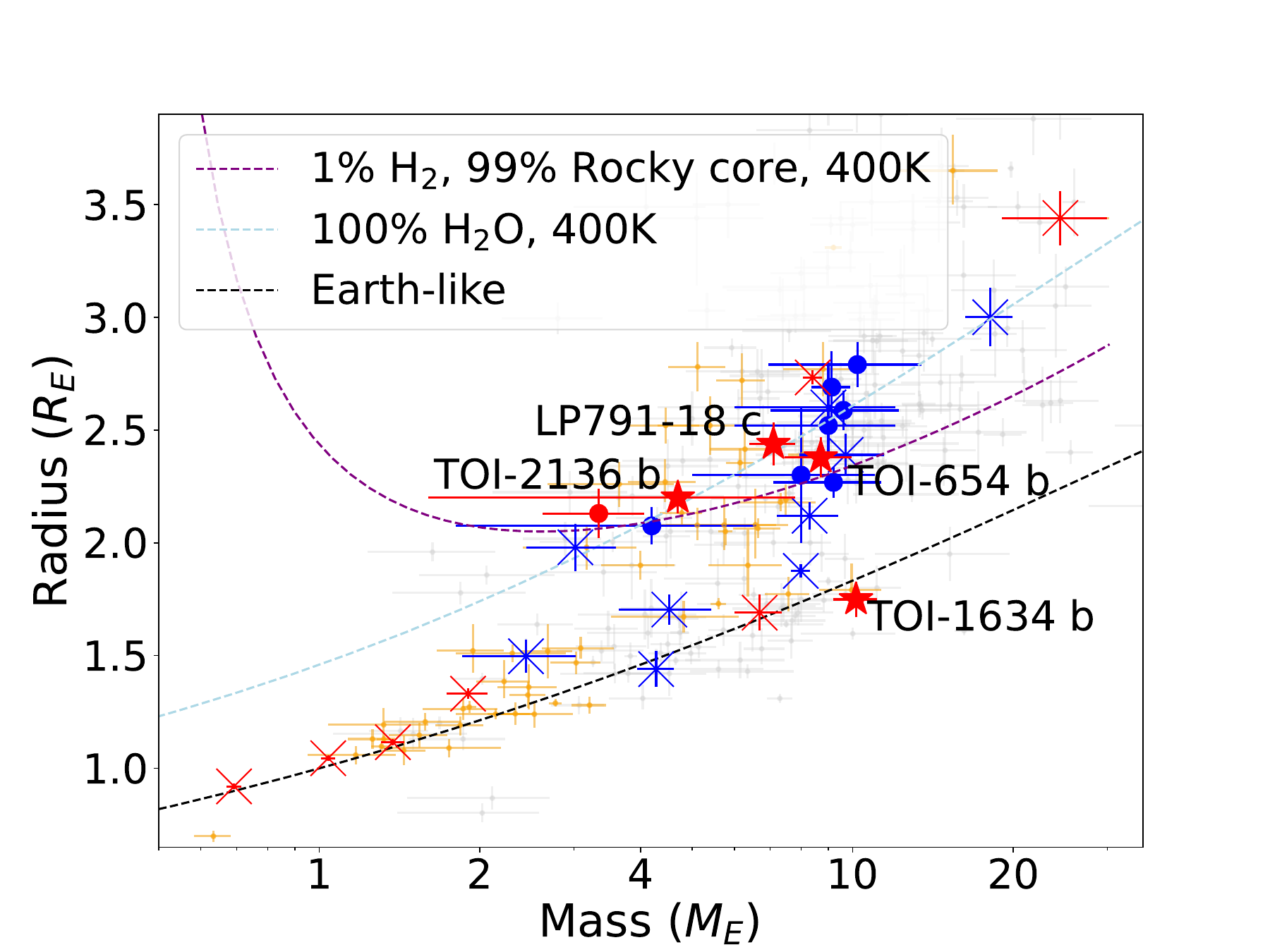}
\end{center}
\caption{Mass--radius diagram of known exoplanets. Data are from the NASA Exoplanet Archive. Red points mark planets orbiting M dwarfs that have been searched for atmospheric helium, including our targets (\redtwo{red stars labeled}), \red{where circles indicate helium detections and crosses indicate non-detections}. Blue points mark planets orbiting K dwarfs that have been searched for helium \red{(with the same symbol convention)}.
Orange points show other known planets around M dwarfs, and gray points show planets around other spectral types. The black, light-blue, and purple dashed lines represent bare rocky planets with Earth-like compositions, pure water worlds, and Earth-like rocky cores with 1\% H+He envelopes at a radiative-equilibrium temperature of 400~K, respectively, as used in figure 9 of \cite{Kawauchi+2022b}. {Alt text: Scatter plot of planet mass in Earth masses versus radius in Earth radii for our targets and other known exoplanets. The dashed lines indicate theoretical mass--radius relations for different compositions.}
}
\label{fig:MR}
\end{figure}

While previous works have attempted to detect helium in 13 systems \red{around M dwarfs} (e.g. \red{\cite{Krishnamurthy+2021}, \cite{Krishnamurthy+2023}, \cite{Palle+2023},} \cite{Orell-Miquel+2024}), most of these planets are super-Earths (< 1.7 R$_\oplus$), which are unlikely to retain substantial H-He envelopes. Only \red{three} sub-Neptunes orbiting M dwarfs, K2-25b \citep{Gaidos+2020}, GJ1214b (e.g., \cite{Orell-Miquel+2022,Spake+2022}), and \red{GJ3090b \citep{Ahrer+2025}} have been observed so far (figure~\ref{fig:MR}). 
Therefore, to better understand the low detection rate of helium around M dwarfs and to evaluate the new theoretical models, additional observations of sub-Neptunes in these systems are needed.

Here, we present a search for helium in the atmospheres of four small planets orbiting M dwarfs, based on high-resolution transmission spectroscopy. These planets were discovered by TESS, and their system parameters have been characterized through follow-up observations (\cite{Kawauchi+2022b, Hirano+2021, Cloutier+2021, Crossfield+2019, Peterson+2023, Ikuta+2025}). The system parameters are summarized in table~\ref{table:param}.
The helium observation of TOI-2136b was previously reported by \citet{Kawauchi+2022b}. In this study, we present updated results based on an improved telluric correction procedure, along with new constraints on upper atmospheric parameters derived from comparisons with theoretical models.
This paper is organized as follows. Sections~\ref{obs} and \ref{dr} describe the observations and data reduction procedures, respectively. In section \ref{res}, we show the results of transmission spectra. In Section~\ref{mod}, we use theoretical models to place limits on the temperature and mass-loss rate in the upper atmosphere, while assessing model dependence and the impact of uncertainties in the incident XUV flux.
We discuss the interpretation and implications of our results in section \ref{dis} and summarize this work in section \ref{sum}.

%%%%%%%%%%%%%%%%%%%%%%%%%%%%%%%%%%%%%%
%%%%%%%%%%%%%%%%%%%%%%%%%%%%%%%%%%%%%%%

%%%%%%%%%%%%%%%%%%%%%%%%%%%%%%%%%%%%%%
\begin{table*}[htbp]
\tbl{Planet and stellar parameters} {
\begin{tabular}{ccccc}
\hline \hline
 Parameter & TOI-2136b & TOI-1634b & TOI-654b & LP791-18c \\
%}
\hline
Star &  &  &  & \\
$V$ mag & $14.32 \pm 0.2 \redtwo{^{(1)}}$ & $13.24 \pm 0.04 \redtwo{^{(4)}}$ & $14.54 \pm 0.010 \redtwo{^{(7)}}$  & $16.9 \pm 0.2 \redtwo{^{(9)}}$\\
$J$ mag & $10.184 \pm 0.024 \redtwo{^{(2)}}$ & $9.484 \pm 0.021 \redtwo{^{(5)}}$ & $10.744 \pm 0.024 \redtwo{^{(7)}}$ & $11.559 \pm 0.024 \redtwo{^{(5)}}$\\
$K_s$ mag & $9.343 \pm 0.022 \redtwo{^{(2)}}$ & $8.600 \pm 0.014 \redtwo{^{(5)}}$ & $9.918 \pm 0.021 \redtwo{^{(7)}}$ & $10.644 \pm 0.023 \redtwo{^{(5)}}$\\
%Spectral type & M & M & M & M\\
$R_{*}$ (R$_\odot$) & $0.3440 \pm 0.0099 ^{\redtwo{(3)}}$ & $0.450 \pm 0.016 ^{\redtwo{(6)}}$ & $0.430 \pm 0.013 \redtwo{^{(8)}}$ & $0.182 \pm 0.007 \redtwo{^{(10)}}$\\
$M_{*}$ (M$_\odot$)& $0.3272 \pm 0.0082 ^{\redtwo{(3)}}$ & $0.451 \pm 0.015 ^{\redtwo{(6)}}$ & $0.419 \pm 0.009 \redtwo{^{(8)}}$ & $0.139 \pm 0.005 \redtwo{^{(11)}}$\\
$T_\mathrm{eff}$ (K) & $3373 \pm 108 ^{(3)}$ & $3472 \pm 70 ^{(6)}$ & $3521^{+33 \redtwo{(8)}}_{-48}$ & $2960 \pm 55 \redtwo{^{(11)}}$\\
$\left[\rm Fe/H\right]$ (dex) & $0.02 \pm 0.14 ^{\redtwo{(3)}}$ & $0.19 \pm 0.12 ^{\redtwo{(6)}}$ & $0.12 \pm 0.18 \redtwo{^{(8)}}$ & $-0.09 \pm 0.19 \redtwo{^{(11)}}$\\
$L_{(1)}$ (L$_\odot$) & $0.0137^{+0.0019 \redtwo{(3)}}_{-0.0017}$ & $0.0264^{+0.0030 \redtwo{(6)}}_{-0.0027}$ & $0.0246\pm 0.0005 \redtwo{^{(8)}}$ & $0.00230 \pm 0.00001 \redtwo{^{(10)}}$ \\
$P_{rot}$ (days) & $82.56 \pm 0.45 ^{\redtwo{(3)}}$ & $77^{+26 \redtwo{(4)}}_{-20}$ & $> 100 \ ({\rm estimated}) \redtwo{^{(12)}}$ & $> 100 \ ({\rm estimated}) \redtwo{^{(12)}}$ \\
distance (pc) & $33.361 \pm 0.019 ^{\redtwo{(3)}}$ & $35.072 \pm 0.023 ^{\redtwo{(6)}}$ & $57.851 ^{+0.055 \redtwo{(8)}}_{-0.056}$ & $26.65 \pm 0.03 \redtwo{^{(10)}}$ \\
Transit &  &  &  & \\
$T_{c}$  (BJD) & $2459214.00322^{+0.00045 \redtwo{(3)}}_{-0.00042}$ & $2458791.51495 \pm 0.00053 ^{\redtwo{(6)}}$ & $ 2459243.0584 \pm 0.0003 \redtwo{^{(8)}}$ & $2458771.055182^{+0.000091 \redtwo{^{(10)}}}_{-0.000097} $ \\
$P$ (days) & $7.851925 \pm 0.000016 ^{\redtwo{(3)}}$ & $0.9893436 \pm 0.0000020 ^{\redtwo{(6)}}$ & $ 1.527561\pm0.000001 \redtwo{^{(8)}}$ & $4.9899093^{+0.0000074 \redtwo{^{(10)}}}_{-0.0000072}$ \\
\red{$T_{14}$ (hour)} & \red{$\sim 1.6 ^{\redtwo{(3)}}$} & \red{$1.027 \pm 0.028 ^{\redtwo{(4)}}$} & \red{$1.200 \pm 0.013 ^{(8)}$}  &  \red{$1.1672 \pm 0.0085$}\\
%Duration &  &  & 720 & \\
$b$ & $0.462^{+0.067 \redtwo{(3)}}_{-0.078}$ & $0.375 \pm 0.049 ^{(6)}$ & $ 0.290 ^{+ 0.104 \redtwo{(8)}}_{- 0.112}$ & $0.134 \pm 0.082 \redtwo{^{(10)}}$ \\
$e$ & $0.07^{+0.09 \redtwo{(3)}}_{-0.05}$ & $0 ^{\redtwo{(6)}}$ & $0 \redtwo{^{(8)}}$ & $0.00008 \pm 0.00004 \redtwo{^{(10)}}$ \\
$a$ (au) & $0.0533 \pm 0.0015 ^{\redtwo{(1)}}$ & $0.01490 \pm 0.00017 ^{\redtwo{(6)}}$ & $0.01944 \pm 0.00015\redtwo{^{(8)}}$& $0.02961^{+0.00035 \redtwo{^{(10)}}}_{-0.00036}$ \\
Planet &  & &  & \\
$R_{p}$ ($\rm R_{\oplus}$) & $2.20 \pm 0.07 ^{\redtwo{(3)}}$ & $1.749 \pm 0.079 ^{\redtwo{(6)}}$ & $ 2.378 \pm  0.089 \redtwo{^{(8)}}$ & $2.438 \pm 0.096 \redtwo{^{(10)}}$ \\
$M_{p}$ ($\rm M_{\oplus}$) & $4.7^{+3.1 \redtwo{(3)}}_{-2.6}$ & $10.14 \pm 0.95 ^{\redtwo{(6)}}$ & $8.71 \pm 1.25 \redtwo{^{(8)}}$ & $7.1 \pm 0.7 \redtwo{^{(10)}}$ \\
$T_{eq,A=0.3}$ (K) & $378 \pm 13 ^{\redtwo{(3)}}$ & $842 \pm 23 ^{\redtwo{(6)}}$& $727 \pm 14 \redtwo{^{(8)}}$ & $324.4^{+2.0 \redtwo{^{(10)}}}_{-1.9}$ \\
\hline
\end{tabular}}\label{table:param}
\begin{flushleft}
Notes. \redtwo{$(1)$} \citet{Lepine2005}, \ \redtwo{$(2)$} \citet{Cutri+2003}, \redtwo{$(3)$} \citet{Kawauchi+2022b}, \redtwo{${(4)}$} \citet{Cloutier+2021}, \ \redtwo{$(5)$} \citet{Skrutskie+2006}, \ \red{$(6)$} \citet{Hirano+2021}, \ \redtwo{$(7)$} \citet{Stassun+2019}, \ \redtwo{$(8)$} \citet{Ikuta+2025}, \ \redtwo{$(9)$} \citet{Stassun+2018}, \ \redtwo{$(10)$} \citet{Peterson+2023}, \ \redtwo{$(11)$} \citet{Crossfield+2019}, \ \redtwo{$(12)$} \citet{Newton+2017}.
%*, ¥dagger, ¥ddagger, ¥S, ¥|, ¥#
\end{flushleft}
\end{table*}
%%%%%%%%%%%%%%%%%%%%%%%%%%%%%%%%%%%%%%

\section{Observations}\label{obs}

We observed a full transit of TOI-2136b, TOI-1634b, TOI-654b, and LP791-18c with the InfraRed Doppler (IRD) instrument on the Subaru 8.2 m telescope under the Subaru-IRD TESS intensive follow-up program (ID:
S20B-088I, S21B-118I). IRD is a fiber-fed near-infrared (NIR) spectrometer that covers 970 to 1750 nm with a spectral resolution of 70,000 \citep{Tamura+2012,Kotani+2018}. 

The data for TOI-2136b are the same as those presented in \red{\citet{Kawauchi+2022b}}. 
TOI-1634b was observed from 08:53 UT to 12:05 UT on 2020/09/30 to obtain the 11 consecutive spectra with 720 s exposure time, of which 9 frames are out-of-transit spectra and 4 frames are in-transit spectra. 
TOI-654b was observed from 11:59 UT to 14:57 UT on 2021/01/28 to obtain the 16 consecutive spectra with 600 s exposure time, of which 11 frames are out-of-transit spectra and 5 frames are in-transit spectra.
LP791-18c was observed from 12:18 UT to 15:27 UT on 2021/01/30 to obtain the 15 consecutive spectra with 720 s exposure time, of which 10 frames are out-of-transit spectra and 5 frames are in-transit spectra.
The signal-to-noise ratios (SNRs) of the extracted spectra of TOI-1634, TOI-654, and LP791-18 are approximately 89, 36, and 21 per pixel at around 1000 nm. The airmass varied from 1.13 to 2.13 for TOI-1634b, from 1.11 to 1.28 for TOI-654b, and from 1.25 to 1.59 for LP791-18c during the observations.
The observation log in table~\ref{table:log} summarizes all of our observations.

All raw IRD data were reduced with \texttt{IRAF} \citep{Tody1993} and our custom pipeline based on \texttt{PyRAF} and Python.
This pipeline performs reduction processes such as correcting the bias patterns of the detectors and performing wavelength calibration with a Th-Ar lamp and a laser-frequency comb, as reported in detail in \citet{Kuzuhara+2018,Hirano+2020,Aoki+2022}.

%%%%%%%%%%%%%%%%%%%%%%%%%%%%%%%%%%%%%%
\begin{table*}[htbp]
\tbl{Observation log}{
\begin{tabular}{ccccccc}
\hline \hline
 Object & Date & Obtained spectra & Exp. time & Airmass & SNR \\
%}
 & (UT) & in-transit + out-of-transit & (s) & & Ave.\\
\hline
TOI-2136b & 2020/09/27 & 6 + 5 & 900 & 1.05--1.33 & 63\\
TOI-1634b & 2020/09/30 & 5 + 8 & 720 & 1.13--2.13 & 89\\
TOI-654b & 2021/01/28 & 7 + 9 & 600 & \red{1.14--1.11--1.28} & 36\\
LP791-18c & 2021/01/30 & 6 + 9 & 720 & 1.25--1.59 & 21\\
\hline
\end{tabular}}\label{table:log}
\end{table*}
%%%%%%%%%%%%%%%%%%%%%%%%%%%%%%%%%%%%%%

\section{Data reduction}\label{dr}

~To extract the absorption lines from the atmosphere of planets from each raw spectrum, we need to remove the stellar absorption lines and telluric lines. The following data reduction steps were performed following \red{\citet{Kawauchi+2022b}. However, for the telluric correction, we employed theoretical telluric \redtwo{absorption} lines instead of a telluric standard star, as used in \citet{Kawauchi+2022b}.}

We removed telluric \red{absorption} lines using theoretical telluric spectra that were individually fitted to each frame. We also observed a telluric standard star (A0 or A1 type) on the same day or nearby dates for telluric correction, but the resulting data did not have sufficiently high SNRs. As a result, telluric correction using the standard star introduced noise amplification in the corrected spectra. Therefore, we chose to use the theoretical telluric spectra for the correction, different from \cite{Kawauchi+2022b}.
Theoretical telluric spectra were generated following the method described in \citet{Kawauchi+2018}, using line parameters from the HITRAN database and the MIPAS model for the vertical structure of the Earth's atmosphere. 
%Figure~\ref{fig:telluric} indicates that the telluric absorption lines are eliminated from the spectra down to the noise level. 

\begin {figure} [htbp]
% \begin{minipage}{0.45\hsize}
\begin{center}
 \includegraphics[width=0.48\textwidth] {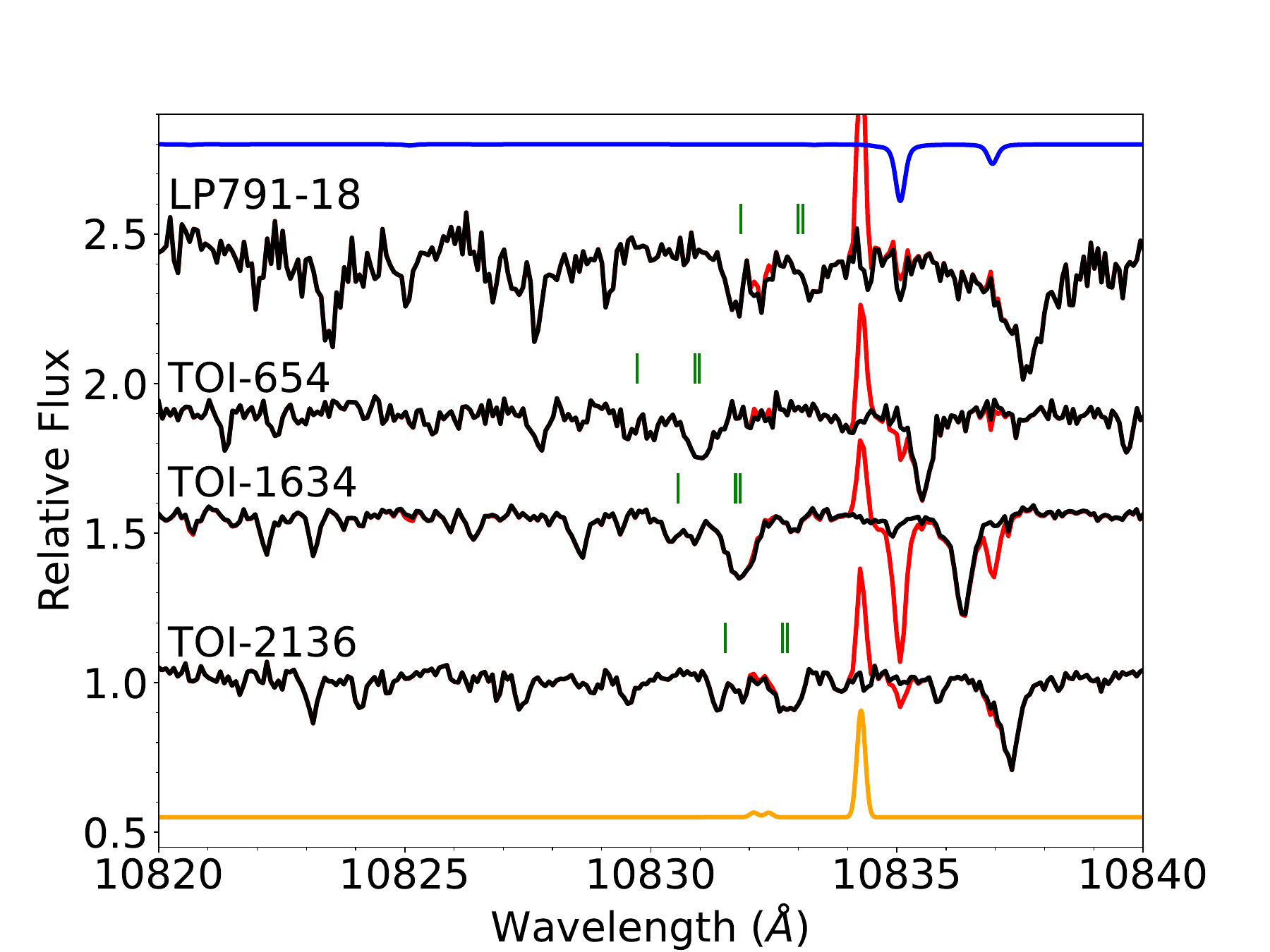}
\end{center}
\caption{Example for a single IRD spectrum of each TOI-2136, TOI-1634, TOI-654, and LP791-18 around the near-infrared helium triplet before (red) and after (black) telluric correction. 
\red{The spectra are shown in the observer rest frame.}
Blue and orange lines indicate telluric absorption lines and $\rm OH^{-}$ emission lines, respectively. \red{Green vertical lines mark the central wavelengths of the predicted near-infrared helium triplet lines for each target in the observer frame.}
The figure demonstrates the positions of emission lines and the removal of telluric absorption lines down to the noise level. {Alt text: Comparison of single exposure spectra near the near-infrared helium triplet for our four targets before and after telluric correction, with the positions of telluric absorption lines and OH emission lines indicated.}}
\label{fig:telluric}
\end{figure}

We corrected the $\rm OH^{-}$ emission around the near-infrared helium triplet using an empirical template. 
Since IRD does not provide simultaneous sky spectra, 
\redtwo{we adopted an empirical template derived from a high SNRs average of CARMENES fibre B spectra, which record the sky background during science exposures (\cite{Orell-Miquel+2023}; J.~Orell-Miquel, private communication).}
%derived from an average of high SNRs of CARMENES fibre B, which record the sky background during science exposures
%we constructed the template from CARMENES fiber B observations, which record the sky background during science exposures (e.g., \cite{Orell-Miquel+2022}).}
For each exposure, we fitted the strongest $\rm OH^{-}$ emission line at 10834.2873 \AA\ with a Gaussian profile. The line center was fixed to the template wavelength, while the line amplitude, width, and continuum offset were treated as free parameters. The two weaker $\rm OH^{-}$ lines at 10832.1071 and 10832.4081 \AA\ were not included in the fit because they are often blended with stellar absorption features, making the local continuum level difficult to determine reliably. The fitting range was adjusted when necessary to ensure a robust fit to the strongest emission line.
The best-fit amplitude and width scaling factors were applied to a three-line $\rm OH^{-}$ template \redtwo{based on the same CARMENES fibre B spectra}. This approach is motivated by the correlated variability of neighboring $\rm OH^{-}$ emission lines reported by \citet{Czesla+2022}. Each spectrum was corrected by dividing by the resulting template, as this procedure produced smaller residuals than a direct subtraction.
Figure~\ref{fig:telluric} indicates that both the telluric absorption lines and the $\rm OH^{-}$ emission lines are reduced to the noise level after the correction.

%%%%%%%%%%%%%%%%%%%%%%%%%%%%%%%%%%%%%%%

\begin {figure} [htbp]
% \begin{minipage}{0.45\hsize}
\begin{center}
 \includegraphics[width=0.48\textwidth] {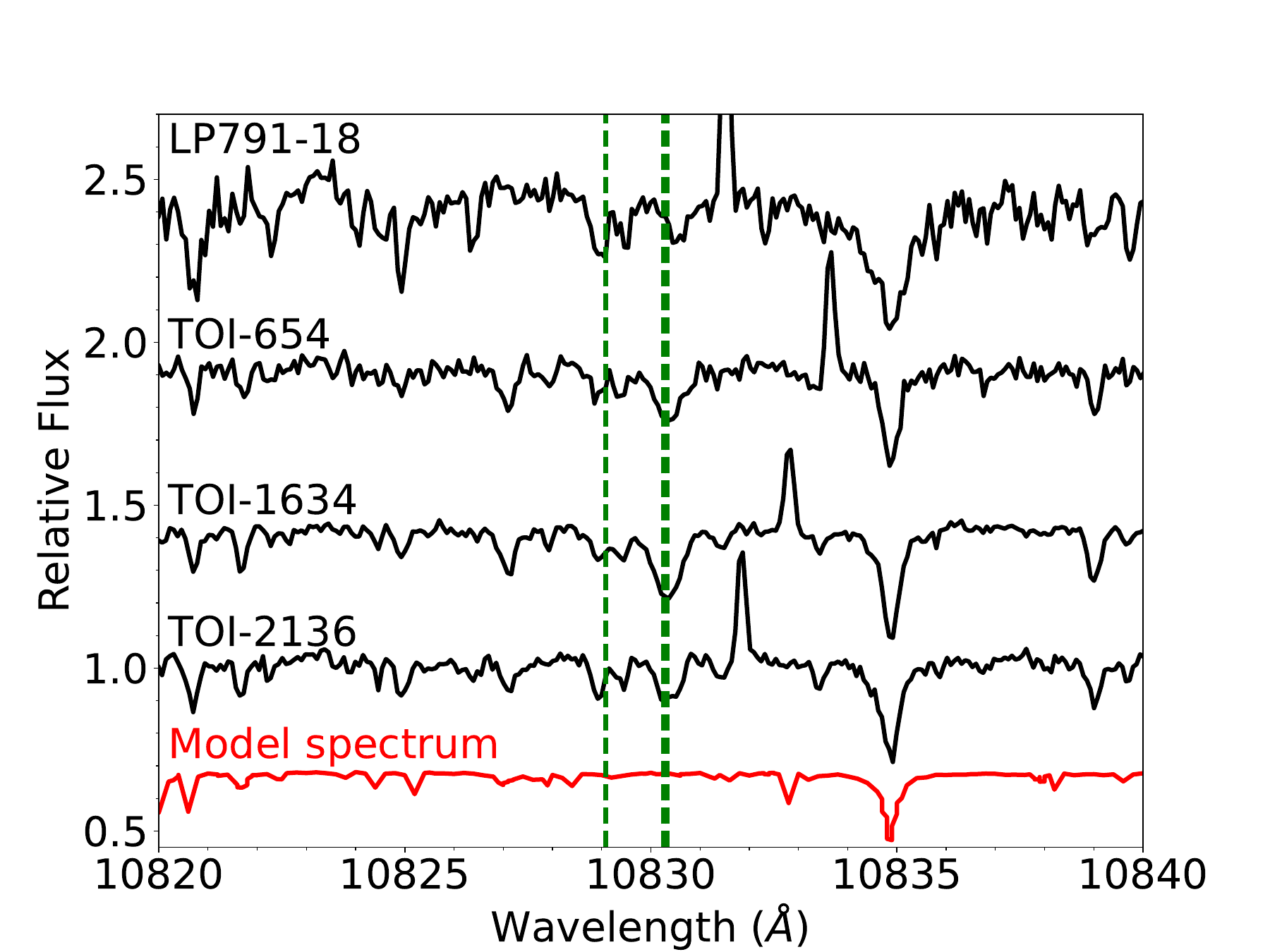}
\end{center}
\caption{Example for a single IRD spectrum of each TOI-2136, TOI-1634, TOI-654, and LP791-18 (black) after wavelength shift correction, compared with the BT-Settl model \citep{Allard+2014} spectrum in air (red). Vertical green dashed lines indicate the centers of the predicted near-infrared helium triplet lines in air. {Alt text: Single exposure IRD spectra for our four targets after wavelength shift correction, shown together with a stellar model spectrum. Vertical dashed lines mark the expected line-center wavelengths of the near-infrared helium triplet in air.} }
\label{fig:stellarRV}
\end{figure}

We corrected the wavelength shift of the stellar spectrum due to Earth’s rotation and revolution with \texttt{IRAF} task \texttt{rvcorrect}. However, the stellar spectrum still had the wavelength shift from the model stellar spectra at air and vacuum wavelengths. To correct the additional shift, we estimated it using the Least-squares deconvolution (LSD) method \citep{Donati+1997} with the observed spectra in 3 orders around the He lines (10681.55--10979.75 \mbox{\AA}), and the 426 absorption lines, assuming that the stellar temperature and $\log g$ are from VALD. The shift values obtained for TOI-2136, TOI-1634, TOI-654, and L791-18 are $-52.6$, $-64.3$, $-18.0$, and $-95.5$ km/s, respectively. \red{To confirm the accuracy of this shift correction, we compared our spectra with the BT-Settl model spectra \citep{Allard+2014}. As a result, the observed spectra are consistent with both the model spectra and those of other planets in air wavelengths (figure~\ref{fig:stellarRV}).}

%%%%%%%%%%%%%%%%%%%%%%%%%%%%%%%%%%%%%%
%%%%%%%%%%%%%%%%%%%%%%%%%%%%%%%%%%%%%%%

We created a template stellar spectrum to combine all out-of-transit spectra after the above correction and divide each frame by it. We shifted each frame by the calculated radial velocity of the planets. The radial velocity changes from +1.7 km\,s$^{-1}$ to $-1.5$ km\,s$^{-1}$ for TOI-2136b, from +20.3 km\,s$^{-1}$ to $-15.8$ km\,s$^{-1}$ for TOI-1634b, from +12.6 km\,s$^{-1}$ to $-8.8$ km\,s$^{-1}$ for TOI-654, and from +1.5 km\,s$^{-1}$ to $-1.5$ km\,s$^{-1}$ for LP791-18c during transit. We combined all frames during transit to create the final transmission spectrum.

\section{Results}\label{res}

\begin {figure*} [htbp]
\begin{center}
\includegraphics[width=\textwidth] {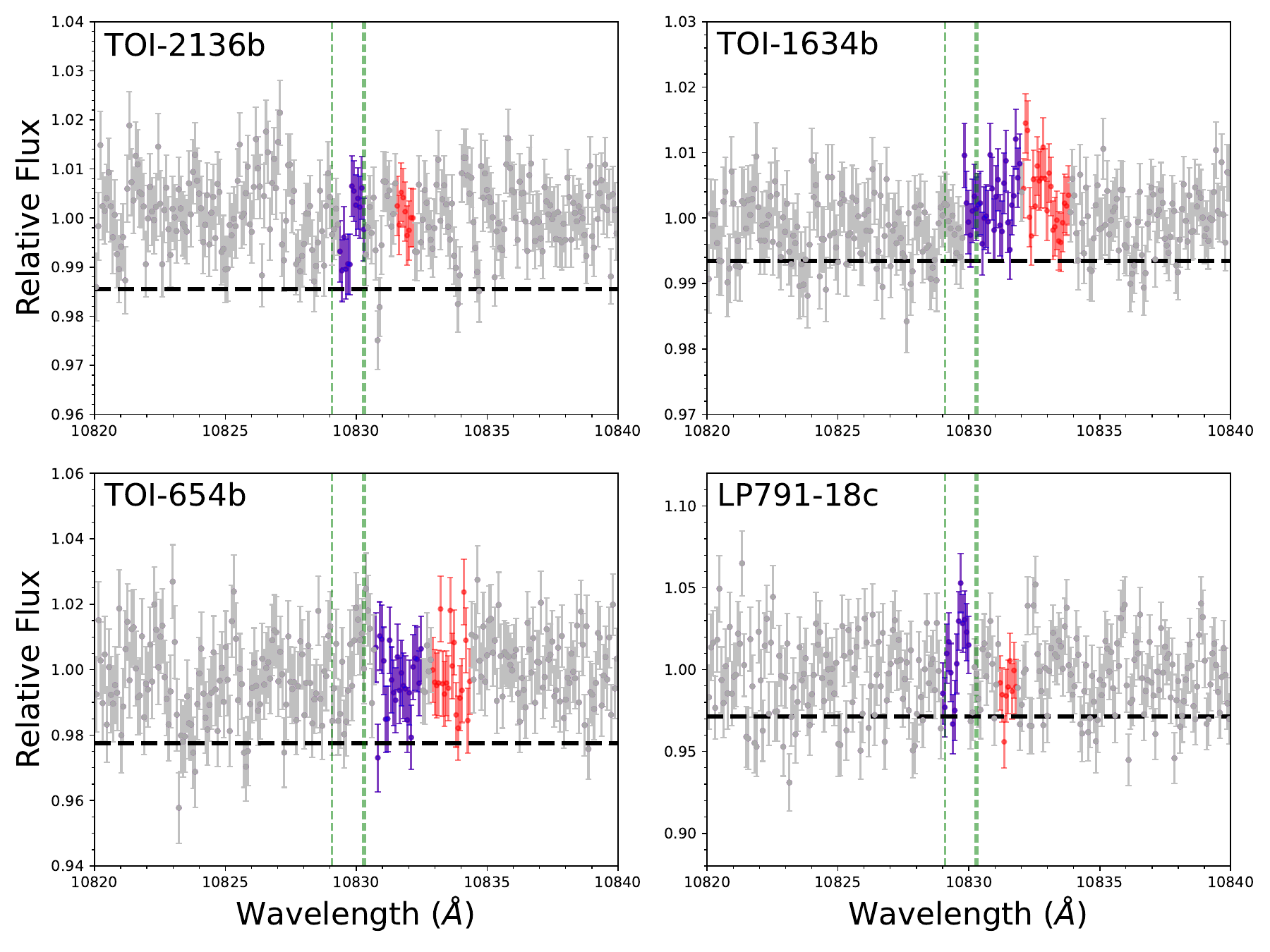}
\end{center}
\caption{The transmission spectra of TOI-2136b, TOI-1634b, TOI-654b, and LP791-18c around the near-infrared helium triplet. 
\red{Red and purple points indicate spectral regions contaminated by the strongest $\rm OH^{-}$ emission line and the other two weaker $\rm OH^{-}$ emission lines, respectively.}
%The red and purple plot show contaminated by the strongest and other shallower $\rm OH^{-}$ emission lines.
%Blank region is where we mask the strongest $\rm OH^{-}$ emission. 
%Horizontal black dashed lines represent 1 - $\sigma$, where $\sigma$ is the standard deviation of the relative fluxes calculated between 10820 and 10840 \mbox{\AA}.
\red{Horizontal black dashed lines represent the 95\% confidence upper limits obtained from the Gaussian model fitting.}
%are standard deviations between 10820 and 10840 \mbox{\AA}. 
Vertical green dashed lines indicate the centers of the predicted near-infrared helium triplet lines in air. {Alt text: Transmission spectra near the near-infrared helium triplet are shown in a two-by-two panel layout, one target per panel. Horizontal dashed lines mark \red{95\% confidence upper limits from Gaussian-model fits.} 
%A masked interval marks the strongest $\rm OH^{-}$ emission. 
\red{Red and purple markers highlight spectral regions contaminated by the strongest OH emission line and two weaker OH emission lines, respectively.}
Vertical dashed lines mark the expected line centers of the near-infrared helium triplet in air.}}
\label{fig:transitratio}
\end{figure*}

Figure~\ref{fig:transitratio} presents the transmission spectra of TOI-2136b, TOI-1634b, TOI-654b, and LP791-18c around the He triplet spectral region. 
\red{For TOI-2136b, the transmission spectrum differs slightly from that presented by \citet{Kawauchi+2022b} because we adopted theoretical telluric spectra instead of telluric standard-star observations to correct telluric absorption and additionally corrected $\rm OH^{-}$ emission lines. Although the overall scatter of the transmission spectrum outside the strongest $\rm OH^{-}$ emission line is similar to that reported by \citet{Kawauchi+2022b}, the adopted correction allows us to retain spectral regions that were previously affected by the strongest $\rm OH^{-}$ emission line.}
%This change reduces the scatter of the transmission spectrum.}
%The standard division of TOI-2136b is smaller than \cite{Kawauchi+2022b} since we used the model spectra on a telluric correction instead of the standard star.
%The result for TOI-2136b is 
%is based on a different \ref{telluric} correction method, as reported in \cite{Kawauchi+2022b}. 
%In \cite{Kawauchi+2022b}, we used the standard star to correct telluric, and without the correction of OH emission, but we used the telluric model 
%The result for TOI-2136b is identical to that reported in \cite{Kawauchi+2022b}. 
No significant He absorption signatures were detected in any of the targets. 
However, we still find an excess absorption signal at $\sim10830.8$~\mbox{\AA} in the transmission spectrum of TOI-2136b, as reported by \citet{Kawauchi+2022b}. The width of this feature is smaller than that of the instrumental profile inferred from the IRD spectral resolution, and this feature is present in multiple in-transit exposures, suggesting that it may reflect a stellar pseudo-signal (see \cite{Allart+2023}). 
\red{To investigate whether this feature is related to stellar activity, we examined the Paschen-$\beta$ line, which can be used as a stellar-activity indicator. Unfortunately, the Paschen-$\beta$ line was contaminated by telluric OH emission during the transit observations, preventing a reliable assessment of its origin.
}
\red{We further examined the residual maps for all four targets (appendix~1). Because the radial velocity of TOI-2136b during transit is small, corresponding to a wavelength shift of only about 0.06~\mbox{\AA} during transit, the residual map does not allow us to clearly identify the origin of the feature at $\sim$10830.8~\mbox{\AA}.}
%We also examined the residual maps for all four targets (appendix~1). It is difficult to identify the origin of the feature at $\sim10830.8$~\mbox{\AA} because the radial velocity of TOI-2136b is small, and the wavelength shift is about 0.06~\mbox{\AA}. }

In addition, we detected the telluric emission around 10833.3~\mbox{\AA} (vacuum wavelength) in the spectra before applying the barycentric correction in the following helium orbital phase variation investigation. \red{The feature was detected only in exposures obtained within approximately 1.5 hours of sunrise or sunset and was found in a subset of the observations of TOI-2136, TOI-654, and LP~791-18, while no clear detection was found for TOI-1634. Further details are provided in appendix~2.}
%This emission is visible only at the beginning and end of the night (see appendix~1).
Although this $\rm OH^{-}$ feature is not significantly visible in the TOI-2136 data taken during the transit night, it may still contribute to the out-of-transit spectra obtained before transit because our observations started at the beginning of the night (19:01 HST). Since the wavelength of the extended signal in the transmission spectrum is consistent with that of this $\rm OH^{-}$ emission, it is also possible that the signal arises from temporal variability in the $\rm OH^{-}$ emission.

%%%%%%%%%%%%%%%%%%%%%%%%%%%%%%%%%%%%%%
%%%%%%%%%%%%%%%%%%%%%%%%%%%%%%%%%%%%%%%

We estimated the upper limits of the He absorption signal by fitting a Gaussian model with the central wavelength, depth, sigma, and baseline as free parameters using the MCMC algorithm \texttt{emcee} \red{(\cite{Foreman-Mackey+2013})}. We applied a normal prior for the central wavelength between 10829.211 and 10831.379 \mbox{\AA} from the center of the stronger He lines. A uniform prior was adopted for sigma between 0.065 and \red{0.5} \mbox{\AA}, corresponding to the instrumental resolution limit and \red{covering the range of line widths typically observed in previous helium detections (e.g., \cite{Zhangb+2022}).}

%the thermal broadening at 30,000 K, respectively.
Our analysis indicates that any possible helium absorption signal would be 
% $<$ 1.36\% for TOI-2136b, % modified in 2025/05/17
% $<$ 0.60\% for TOI-1634b, % modified in 2025/05/16
% $<$ 2.07\% for TOI-654b, % modified in 2025/11/05
% and $<$ 3.00\% for LP791-18c % modified in 2025/05/17
\red{$<$ 1.54\% for TOI-2136b, % modified in 2026/05/31
$<$ 0.62\% for TOI-1634b, % modified in 2026/05/31
$<$ 1.97\% for TOI-654b, % modified in 2026/05/31
and $<$ 2.53\% for LP791-18c} % modified in 2026/05/31
at 95\% confidence. The corresponding equivalent widths (EW) are 
% $<$ 7.3 m\mbox{\AA} for TOI-2136b, % modified in 2025/05/17
% $<$ 2.1 m\mbox{\AA} for TOI-1634b, % modified in 2025/05/16
% $<$ 7.4 m\mbox{\AA} for TOI-654b, % modified in 2025/11/05
% and $<$ 9.1 m\mbox{\AA} for LP791-18c % modified in 2025/05/17
\red{$<$ 7.4 m\mbox{\AA} for TOI-2136b, % modified in 2025/05/17
$<$ 2.7 m\mbox{\AA} for TOI-1634b, % modified in 2025/05/16
$<$ 7.3 m\mbox{\AA} for TOI-654b, % modified in 2025/11/05
and $<$ 8.1 m\mbox{\AA} for LP791-18c} % modified in 2026/05/31
at 95\% confidence.
\red{The slight differences between these upper limits and those reported for TOI-2136b by \citet{Kawauchi+2022b} likely reflect a combination of differences in the telluric-correction procedure, the treatment of $\rm OH^{-}$ emission lines, and the setup of Gaussian fitting, including the adopted prior ranges.}

\begin {figure} [htbp]
% \begin{minipage}{0.45\hsize}
\begin{center}
 \includegraphics[width=0.48\textwidth] {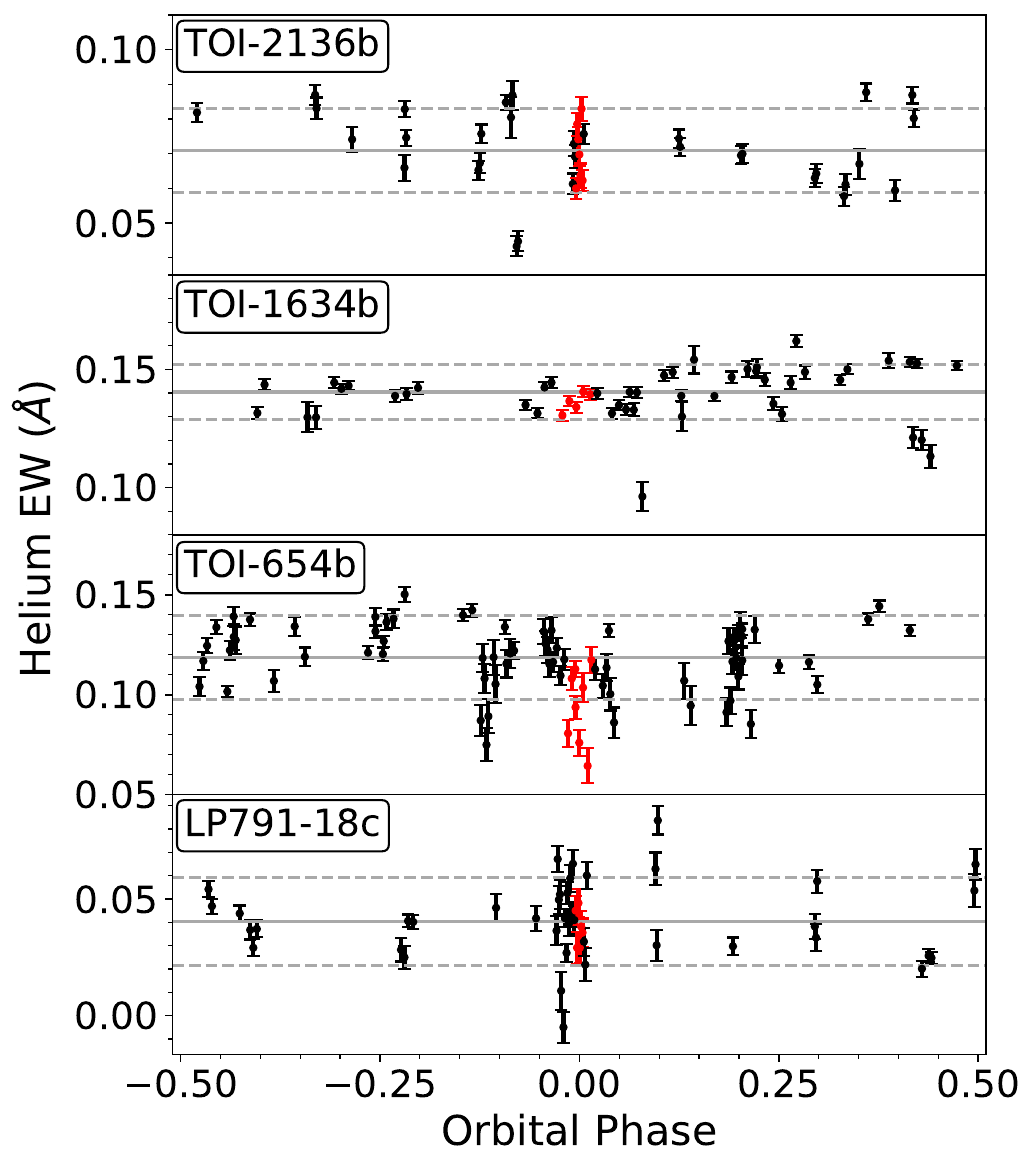}
\end{center}
\caption{Equivalent widths (EWs) of the near-infrared helium triplet for TOI-2136, TOI-1634, TOI-654, and LP 791-18, respectively. Red and black points indicate the data obtained during transit and out of transit. The gray horizontal solid and dashed lines indicate the median and standard deviation of all EW measurements for each planet. {Alt text: Equivalent-width measurements of the near-infrared helium triplet are shown in four vertically stacked panels, one target per panel. Separate points represent in-transit and out-of-transit data. Horizontal lines indicate the median and the standard deviation of the measurements for each target.} }
\label{fig:EWtv}
\end{figure}

We also investigate the orbital phase variation of the \redtwo{EW} of the \red{near-infrared helium} triplet in these stellar spectra to search for a giant tidal tail of escaping helium, similar to that reported for HAT-P-32b \citep{Zhang+2023} \red{and HAT-P-67 \citep{Gully-Santiago+2023}}.
We obtained phase-resolved spectra of all four stars for RV measurements with IRD under our intensive follow-up program.
\red{We corrected the telluric lines, including $\rm OH^{-}$ emission around the helium absorption lines, using the same procedure adopted for the transmission spectrum analysis.}
%by fitting model spectra to each frame. 
\red{The EWs of the near-infrared helium triplet were measured in the stellar rest frame over a fixed wavelength range encompassing the helium feature. In the absence of planetary absorption, the measured EW reflects only the stellar helium absorption, whereas any additional planetary helium absorption within this wavelength range would contribute to the measured EW. The local continuum was estimated from wavelength regions on both sides of the line and linearly interpolated across the integration region.}
%and then measured the EWs of the near-infrared helium triplet in the stellar and/or planetary spectra.
% As a result, we did not detect any large variation in EW, suggesting that the presence of an extended large helium tail is unlikely (figure~\ref{fig:EWtv}).
\red{As a result, we did not find any significant variation in the EW with orbital phase for any of the four systems (\redtwo{figure~\ref{fig:EWtv}}). In particular, no coherent enhancement comparable to the extended helium tails reported for HAT-P-32b or HAT-P-67 was detected.}

%%%%%%%%%%%%%%%%%%%%%%%%%%%%%%%%%%%%%%
%%%%%%%%%%%%%%%%%%%%%%%%%%%%%%%%%%%%%%%

\section{Modeling}\label{mod}

In this section, we assess what can be inferred from our results about the upper atmospheres of the observed planets through comparisons with theoretical models. In subsection~5.1, we constrain the allowed range of atmospheric temperature and mass-loss rate by comparing predictions from an isothermal Parker-wind model with our observational limits. In subsection~5.2, we compare these results with self-consistent hydrodynamic models to examine the model dependence of the inferred constraints. In subsection~5.3, we investigate how uncertainties in the stellar high-energy irradiation, in particular the X-ray-based estimate of the XUV flux, affect our conclusions.

\subsection{Atmospheric Constraints from Isothermal Parker-wind Models}

To constrain the upper-atmospheric temperature and mass-loss rate, we computed theoretical near-infrared helium triplet absorption profiles over a range of temperatures and mass-loss rates using the Parker wind model implemented in the \texttt{p-winds} code \citep{DosSantos+2022}. The models assume an isothermal upper atmosphere, an H/He number ratio of 90/10, and an XUV flux scaled from that of GJ~876 
\red{($R_{*}$ $\sim$ 0.35 $\rm R_\odot$, $M_{*}$ $\sim$ 0.34 $\rm M_\odot$, $T_\mathrm{eff}$ $\sim$ 3271 K; \cite{Stassun+2019})},
whose stellar parameters most closely resemble those of our targets, as provided by the \texttt{MUSCLES} database \citep{France+2016, Youngblood+2016, Loyd+2016}. 

As our targets are close-in planets, the stellar tidal potential can significantly affect the outflow structure. We therefore included the tidal gravity term in the momentum equation and computed the location of the sonic point and the Parker-wind solution using the routines provided in \texttt{p-winds}. This effect is particularly important for TOI-1634b and TOI-654b, for which the Roche-lobe radius \red{($R_{RL}$)} is small and comparable to the sonic radius, so that tidal forces substantially modify the atmospheric structure (e.g. \cite{Vissapragada+2022}).

In computing the near-infrared helium triplet absorption, we also need to specify an effective upper boundary of the atmosphere, since metastable helium may persist beyond the region that is strictly gravitationally bound to the planet. Outside this region, the gas dynamics are controlled primarily by the stellar wind, and the outflow can form large tails, as observed for systems such as HAT-P-32b \citep{Zhang+2023}. Such three-dimensional structures lie beyond the scope of our 1D spherically symmetric model and are not well constrained 
\red{from the transmission spectra alone, which sample only a limited range of orbital phases around transit.}
%from the narrow observation windows around the transits.
%by the narrow window around transit. 
Following \citet{Biassoni+2024}, we therefore adopt the Hill radius as the effective upper boundary of the atmosphere for the helium transmission calculations.

For reference, the mass-loss rate due to atmospheric escape can be estimated using the energy-limited formula (e.g., \cite{Erkaev+2007}):
\begin{equation}
\dot{M}_{\rm env} = \eta \frac{\pi R_p^3 F_{\rm XUV}}{G K(\xi) M_p},
\end{equation}

where $\eta$ is the evaporation efficiency, $R_p$ and $M_p$ are the planetary radius and mass, $F_{\rm XUV}$ is the stellar flux at wavelengths shorter than 912 \AA, and $G$ is the gravitational constant.
Here, $\xi$ is the ratio of the Roche lobe radius to the planetary radius ($\xi = R_{\rm RL} / R_p$), and
%%%%%%%%%%%%%
\begin{equation}
K(\xi) = 1 - \frac{3}{2\xi} + \frac{1}{2\xi^3}
\end{equation}
%%%%%%%%%%%%%%%%
is the correction factor that accounts for the reduction in gravitational potential energy due to stellar tidal forces.
$R_{\rm RL}$ is calculated using equation~(2) of \citet{Eggleton1983}.
Using this framework, we converted the mass-loss rates into the evaporation efficiency $\eta$.

The transmission spectrum is obtained by \red{integrating the contributions from all in-transit phases, and the model accuracy depends on the phase sampling used for the transit integration.}
%summing the contributions from each in-transit phase, so its model accuracy depends on the number of phase bins used to sample the transit. 
%To balance accuracy and computational cost, we divided the transit into 100 phase bins, computed the transmission spectrum in each bin, and adopted their mean as the model transmission spectrum. A comparison with a high-resolution calculation using 10,000 phase bins shows that the relative error is negligible compared to the observational uncertainties. This test was performed for TOI-654b, which has the strongest phase dependence (see appendix~2).
\red{To balance accuracy and computational cost, we divided the transit into 100 phase bins, computed the transmission spectrum at each phase, and integrated them to obtain the final model transmission spectrum. We evaluated the effect of the number of phase bins for all four targets and confirmed that the resulting relative error is negligible compared to the observational uncertainties when using 100 phase bins}  (see appendix~3).

\begin{figure*} [htb]
\begin{center}
\includegraphics[width=\textwidth] {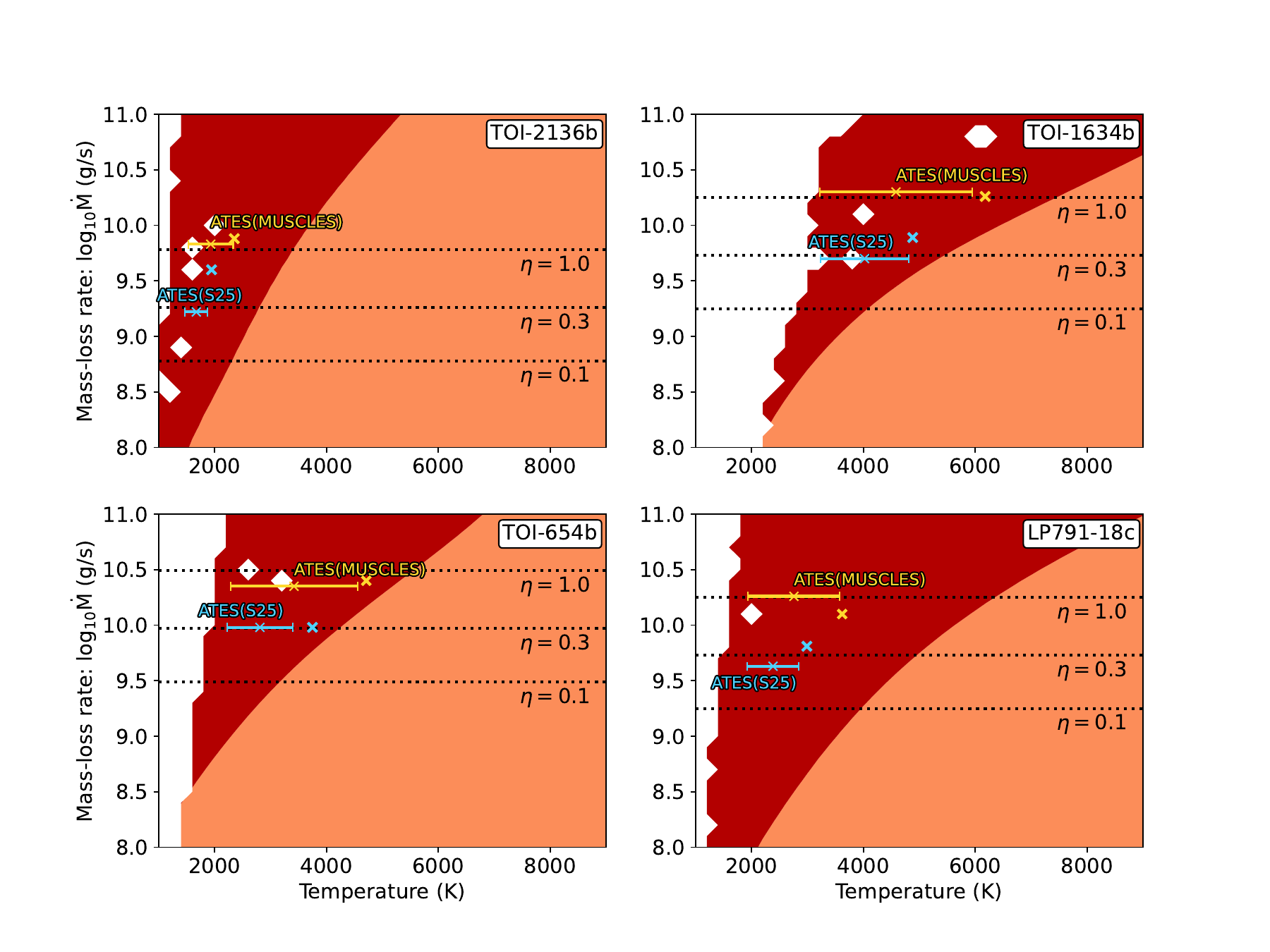}
\end{center}
\caption{\red{Map showing whether the model transmission spectra are consistent with the observational 95\% upper limits in the temperature--mass-loss-rate parameter space. Red regions are ruled out at the 95\% confidence level, while orange regions are not excluded by our observations. The white region on the left\red{, including isolated patches,} marks parameter combinations for which the \texttt{p-winds} calculations failed due to numerical issues. Crosses with error bars show the mass-loss rate and mean temperature obtained from the \texttt{ATES} code using the XUV flux from the \texttt{MUSCLES} database (yellow) and that estimated in S2025 (light blue). Crosses without error bars indicate the best-fitting mass-loss rate and temperature obtained by fitting the spectra generated with the \texttt{ATES} model with the isothermal model spectra computed with \texttt{p-winds} (yellow: \texttt{MUSCLES} XUV; light blue: S2025 XUV).
{Alt text: Four panels show temperature versus mass-loss rate for the four planets, one target per panel. Dark-red regions are excluded by the 95\% observational upper limits, and orange regions are not excluded. Crosses with horizontal error bars indicate the mean temperatures and mass-loss rates predicted by the ATES models under two XUV prescriptions, while crosses without error bars indicate the corresponding best-fitting isothermal p-winds models.}}}
\label{fig:comp_model}
\end{figure*}

We evaluated the statistical significance with which each theoretical model can be rejected by comparing the predicted absorption depths to the standard deviation (corresponding to 1$\sigma$) of our observed transmission spectra. Figure~\ref{fig:comp_model} shows the significance of model rejection for each planet in the mass-loss-rate--temperature parameter space. In the explored parameter space, a range of low temperature, high mass loss rate solutions (light-colored region in the figure) is ruled out by the non-detections.
We also overplotted the corresponding $\eta$ = 1.0, 0.3, and 0.1 in figure~\ref{fig:comp_model} as a guide to the range of mass-loss rates expected in this parameter space under energy-limited escape.

%%%%%%%%%%%%%%%%%%%%%%%%%%%%%%%%%%%%%%
%%%%%%%%%%%%%%%%%%%%%%%%%%%%%%%%%%%%%%%

\subsection{Comparison with Hydrodynamic Models}

We further explored the atmospheric structure, including advection, using the \texttt{ATES} code \citep{Caldiroli+2021,Caldiroli+2022,Biassoni+2024}.
 In \texttt{ATES}, the hydrodynamics of the atmospheric gas are computed by solving the Euler equations in a gravitational field under the assumption of spherical symmetry. The energy and ionization balance are evaluated assuming local photoionization equilibrium at each time step, and ion advection is included in a post-processing step once the hydrodynamic solution has converged.
 \red{For each planet, the \texttt{ATES} calculations were performed using the observed planetary and stellar parameters (e.g., planetary mass, radius, equilibrium temperature, and orbital separation) together with the adopted X-ray and EUV luminosities.}
The \texttt{ATES} models predict the logarithm of the mass-loss rate (in g\,s$^{-1}$) of 9.83, 10.30, 10.52, and 10.26 for TOI-2136b, TOI-1634b, TOI-654b, and LP 791-18c, respectively.
\red{These mass-loss rates lie close to the $\eta \simeq 1.0$ line shown in figure~\ref{fig:comp_model}, which was calculated using the planetary radius $R_p$ in the energy-limited prescription.}
%When these mass-loss rates are translated into an evaporation efficiency using the energy-limited formalism in the previous subsection, the implied values are close to $\eta \simeq 1.0$. This likely reflects the choice of radius adopted in the conversion.

Following \citet{Orell-Miquel+2024}, we estimated the XUV photosphere radius, $R_{\rm XUV}$, using their equations~(8) and (9)\red{. We then recomputed $\eta$ by replacing $R_p$ with $R_{\rm XUV}$ in the energy-limited expression while keeping the mass-loss rates fixed to the values predicted by the \texttt{ATES} models.}
We adopted the same values of $P$, $\mu$, and $\rho_{\rm atm,XUV}$ as in \citet{Orell-Miquel+2024}. The resulting efficiencies are $\eta \sim$0.24, $\sim$0.36, $\sim$0.13, and $\sim$0.43 for TOI-2136b, TOI-1634b, TOI-654b, and LP~791-18c, respectively. Although these estimates are approximate, they indicate that the inferred $\eta$ can depend sensitively on the adopted radius in the energy-limited conversion.

The \texttt{ATES} code also generates the model absorption line profiles for the near-infrared helium triplet, but by default, it only computes the transmission spectrum at the transit midpoint. We therefore generated the model absorption line profiles for the near-infrared helium triplet at each in-transit phase from the temperature and density profiles calculated with \texttt{ATES}, using the \texttt{radiative\_transfer\_2d} routine in the \texttt{p-winds} code, and then adopted the mean of these phase-resolved spectra as the final model.

The line depths of the \texttt{ATES} model spectra, after convolution with the instrumental profile, are 
11.7\%, %2025/11/22
1.09\%, %2025/11/22
2.58\%, %2025/11/22
and 24.0\% %2025/11/22
for TOI-2136b, TOI-1634b, TOI-654b, and LP 791-18c, respectively.
\red{All of these values exceed the 95\% upper limits derived from our observations.}
%Based on the standard deviation of our observed transmission spectra, these \texttt{ATES} model spectra are inconsistent with 
%16.2$\sigma$, %2025/11/22
%2.21$\sigma$, %2025/11/22
%2.22$\sigma$,  %2025/11/22
%and 10.2$\sigma$ %2025/11/22
%levels, respectively.

Recently, \citet{S2025} (hereafter S2025) recalculated the XUV flux using new coronal models. The X-ray and EUV luminosities of their template stars are substantially lower than those adopted in the \texttt{MUSCLES} database (table~\ref{table:XUV}).
We re-estimated the properties of the upper atmosphere and the model near-infrared helium triplet using the \texttt{ATES} code with the XUV flux from S2025.
\red{As a result, the predicted absorption depths are larger than those obtained using the XUV flux from the \texttt{MUSCLES} database. 
The corresponding depths are
13.5\%,
1.13\%,
2.76\%,
and 27.6\%}
%and these models are inconsistent with 
% 49.6$\sigma$,  
% 2.29$\sigma$, %2025/11/22
% 2.37$\sigma$, %2025/11/22
% and 11.8$\sigma$ %2025/11/22
for TOI-2136b, TOI-1634b, TOI-654b, and LP 791-18c, respectively.
\red{All of these values also exceed the 95\% upper limits derived from our observations.}
Therefore, while the XUV flux affects the near-infrared helium triplet, the theoretical models in both cases are in tension with our observations.

%%%%%%%%%%%%%%%%%%%%%%%%%%%%%%%%%%%%%%
\begin{table}[htbp]
\tbl{X-ray (5--100\,\AA) and EUV (100--920\,\AA) luminosities of the GJ~876 template}{
\begin{tabular}{lcc}
\hline \hline
 Object & $\log L_{\mathrm{X}}$ & $\log L_{\mathrm{EUV}}$ \\
        & (erg s$^{-1}$) & (erg s$^{-1}$) \\
\hline
GJ876 (\texttt{MUSCLES}) & 27.47 & 27.86\\
GJ876 (S2025) & 26.20 & 27.37\\
%GJ876 (from eq.~(4) of S2025) & 26.20 & 26.98\\
%GJ876 (from eq.~(2) of S2025) & 26.20 & 27.14\\
%GJ551 (MUSCLES) & 28.84 & 28.54 \\
%GJ551 (S2025) & 27.25 & 27.30\\
%GJ1214 (MUSCLES) & 26.91 & 27.87\\
%GJ1214 (S2025) & 25.88 & 26.27\\
\hline
\end{tabular}}\label{table:XUV}
\end{table}
%%%%%%%%%%%%%%%%%%%%%%%%%%%%%%%%%%%%%%

Since the temperature is not constant with altitude in the \texttt{ATES} models, we calculated the mean and standard deviation of the temperature \red{profile}, weighted by the metastable helium density, using equations~(4) and (5) of \citet{Linssen+2022} for comparison with the isothermal \texttt{p-winds} model. \redtwo{This procedure was applied to the \texttt{ATES} models computed using both the XUV flux from the \texttt{MUSCLES} database and that estimated in S2025 (figure~\ref{fig:comp_model}).}
%as shown by crosses with bars and labeled ATES (MUSCLES) and ATES (S25) in figure~\ref{fig:comp_model}.}
%(figure~\ref{fig:comp_model}). 
We also fitted the theoretical transmission spectra computed with \texttt{ATES} using the isothermal transmission spectra from \texttt{p-winds}. 
For all planets, the best-fit isothermal temperatures are higher than the corresponding weighted mean temperatures. 
This demonstrates that, even when averaging the temperature with a weight given by the metastable helium density, the mean temperature in a self-consistent model does not, in general, coincide with the temperature assumed in an isothermal model, implying that care is required when comparing self-consistent hydrodynamic models with the isothermal Parker-wind model. 
In this work, where our constraints are based on non-detections, we merely point out this difference and do not attempt to adjust the isothermal Parker-wind model to more closely match the self-consistent solution. 
However, when fitting detections of the near-infrared helium triplet with an isothermal model, it may be necessary to account explicitly for such differences by incorporating non-isothermal models, for example by using codes such as \texttt{sunbather} \citep{Linssen+2024}, in order to obtain more reliable atmospheric parameters.

Finally, we considered the EWs of the model lines. The EW of the \texttt{ATES} model spectra with the XUV flux from \texttt{MUSCLES} database are 
56.8 m\mbox{\AA}, %2025/12/01
6.68 m\mbox{\AA}, %2025/12/01
16.2 m\mbox{\AA}, %2025/12/01
and 138.4 m\mbox{\AA}  %2025/12/01
for TOI-2136b, TOI-1634b, TOI-654b, and LP 791-18c, respectively.
The corresponding EWs for the \texttt{ATES} spectra based on the XUV flux from S2025 are
%The EW of the ATES model spectra with the XUV flux from S2025 are 
55.4 m\mbox{\AA}, %2025/12/01
5.75 m\mbox{\AA}, %2025/12/01
14.4 m\mbox{\AA}, %2025/12/01
and 132.9 m\mbox{\AA},  %2025/12/08 improve Rp and Mp
in the same order.
In both cases, our results rule out the estimated EWs for all four planets at a confidence level exceeding 95\%.
\red{The predicted absorption depths and EWs of the \texttt{ATES} model spectra, together with the corresponding observational 95\% upper limits, are summarized in table~4.}

%%%%%%%%%%%%%%%%%%%%%%%%%%%%%%%%%%%%%%
\begin{table*}[htbp]
\tbl{Observed 95\% upper limits on the depth and EW of the near-infrared helium triplet, together with the corresponding model predictions based on the \texttt{MUSCLES} and S2025 XUV fluxes}{
\begin{tabular}{ccccccc}
\hline \hline
 Object & Depth UL (95\%) & EW UL (95\%) & Model Depth (MUSCLES) & Model EW (MUSCLES) & Model Depth (S25) & Model EW (S25)  \\
%}
 & (\%) & (m\AA) & (\%) & (m\AA) & (\%) & (m\AA)\\
\hline
TOI-2136b & < 1.54 & < 7.4 & 11.7 & 56.8 & 13.5 & 55.4 \\
TOI-1634b & < 0.62 & < 2.7 & 1.09 & 6.68 & 1.13 & 5.75 \\
TOI-654b & < 1.97 & < 7.3 & 2.58 & 16.2 & 2.76 & 14.4 \\
LP791-18c & < 2.53 & < 8.1 & 24.0 & 138.4 & 27.6 & 132.9 \\
\hline
\end{tabular}}\label{table:varX}
\end{table*}
%%%%%%%%%%%%%%%%%%%

\subsection{Impact of X-ray Flux Uncertainty}

%The uncertainty in the EUV flux estimated by S2025 is less than about 0.5 dex. However, for our targets, the X-ray emission has not been directly measured, and the inferred $L_{\mathrm{X}}$ is therefore uncertain. 
\red{The uncertainty associated with the conversion from X-ray to EUV luminosity is likely modest compared with the uncertainty in the X-ray luminosity itself. S2025 derived empirical relations between X-ray and EUV luminosities, with an RMS scatter of 0.28--0.36 dex around the fitted relations shown in their figure~3.}
However, for our targets, the X-ray emission has not been directly measured, and the inferred $L_{\mathrm{X}}$ \redtwo{($\rm erg \ s^{-1}$)} is therefore uncertain. 
\citet{Zhu+2025} report that, for early M dwarfs (M0--M4), the distribution of $\log(L_{\mathrm{X}}/L_{\mathrm{bol}})$ has mean values of about $-5.0$ to $-5.3$, with a minimum of roughly $-6.4$. \red{For M dwarfs later than M4}, the distribution peaks near $\log(L_{\mathrm{X}}/L_{\mathrm{bol}}) \sim -3$ and $\sim -5$, and the minimum is $\sim -5.5$. 
Furthermore, according to \citet{Caramazza+2023}, for early M dwarfs (M0--M4) within 10 pc, the surface X-ray flux ranges from $\log F_{\mathrm{X,surf}} \approx 7$ (corresponding to $\log(L_{\mathrm{X}}/L_{\mathrm{bol}}) \approx -3$) down to $\log F_{\mathrm{X,surf}} \approx 4$ \red{(corresponding to $\log(L_{\mathrm{X}}/L_{\mathrm{bol}}) \approx -6$)}. Even for the star with the faintest detected X-ray emission, which is comparable to the faintest solar coronal structures (coronal holes), the surface flux is $\log F_{\mathrm{X,surf}} \gtrsim 3.8$. 

Motivated by these ranges, we varied $\log(L_{\mathrm{X}}/L_{\mathrm{bol}})$ between $-3.0$ and $-6.5$ for the early M-dwarfs (TOI-2136, TOI-1634, TOI-654) and between $-3.0$ and $-6.0$ for the late M-dwarf (LP~791-18). For each value, we computed the depth and EW of the near-infrared helium triplet using the \texttt{ATES} model.
% \redtwo{Although the X-ray emission of our targets has not been directly measured, and X-ray luminosities lower than the adopted range cannot be completely excluded, the adopted ranges encompass the observed distributions of X-ray activity reported for M dwarfs reported by \citet{Zhu+2025}.}
\redtwo{We note that we cannot rule out the possibility that the X-ray luminosities of our targets are lower than the range previously observed for M dwarfs, although investigating this possibility is beyond the scope of the present study.}
The corresponding $L_{\mathrm{X}}$ values were obtained from the measured \red{bolometric luminosities of our targets}. We then constructed the XUV spectra by estimating the EUV luminosity in the 100--920\,\AA\ range using Eq.~(4) of S2025. 
We also repeated the calculation using Eq.~(2) of S2025 as an alternative EUV prescription. Although Eq.~(2) provides a poorer fit than Eq.~(4), the difference in the rms scatter is small (\(\Delta\mathrm{RMS}\approx 0.02\)). Comparing the results from both prescriptions allows us to check whether our conclusions depend on the adopted EUV estimate.

\begin {figure*} [tbp]
% \begin{minipage}{0.45\hsize}
\begin{center}
 \includegraphics[width=\textwidth] {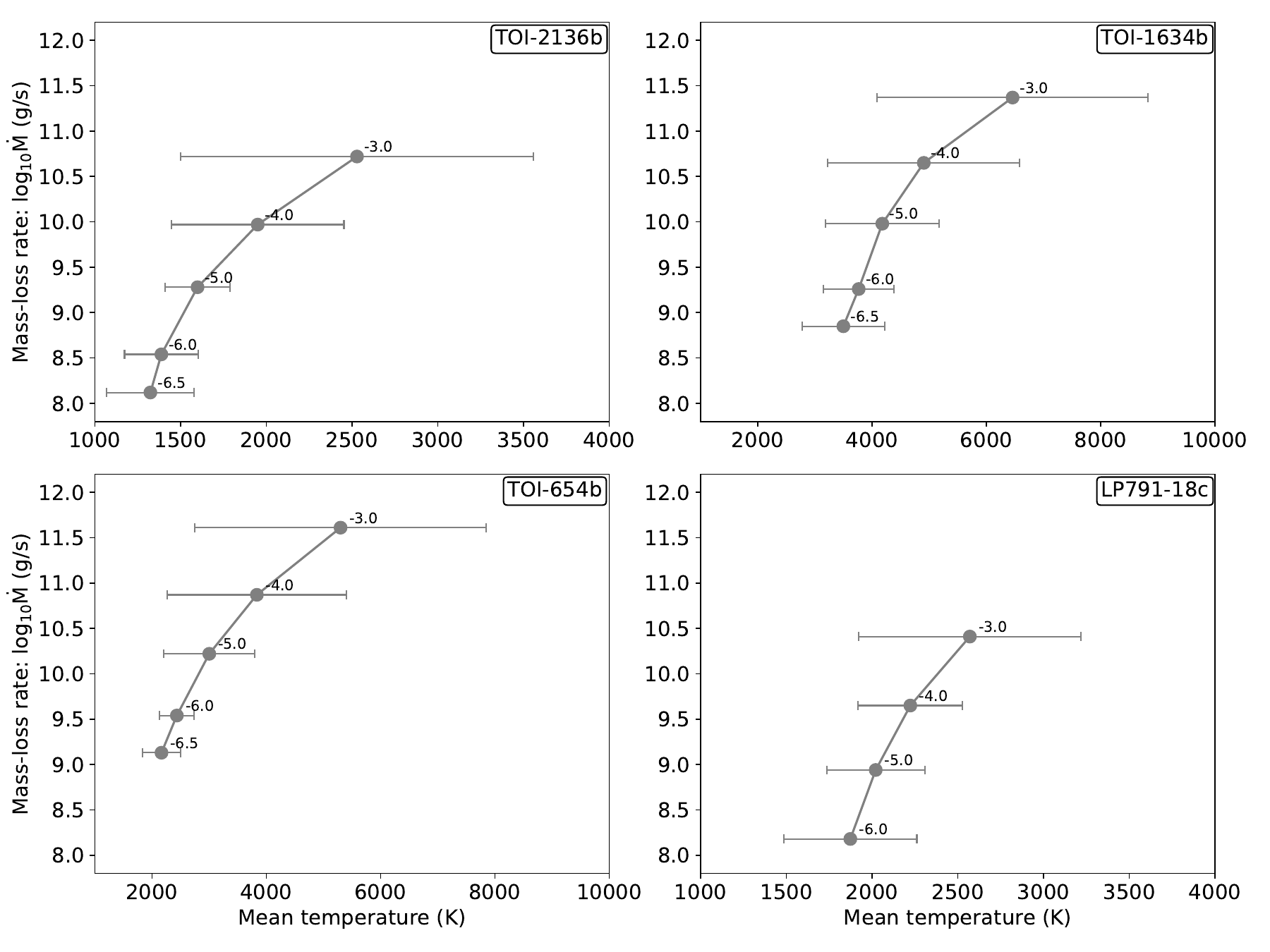}
\end{center}
\caption{Mass-loss rate and the mean temperature in the upper atmosphere for the \texttt{ATES} models of TOI-2136b, TOI-1634b, TOI-654b, and LP~791-18c, respectively. 
Each symbol corresponds to a model computed with a different X-ray luminosity, and the numbers next to the symbols indicate $\log (L_{\mathrm{X}}/L_{\mathrm{bol}})$.
\red{The EUV luminosity was estimated from the X-ray luminosity using equation~(4) of S2025. The horizontal error bars show the standard deviation of the temperature distribution weighted by the metastable helium density.}
%(erg s$^{-1}$). 
%Symbols are color-coded according to whether the model EW is below or above the 95\% confidence upper limit derived
%from our observations.
%The line indicates the values from the \texttt{ATES} models using the EUV flux estimated from X flux using the relation of equation~(4) in S2025. 
{Alt text: A two-by-two set of panels indicates mean upper-atmospheric temperature versus mass loss rate for four targets. \red{Each symbol represents an ATES model computed with a different assumed X-ray luminosity, labeled by the logarithmic X-ray to bolometric luminosity ratio. Symbols are connected by lines to illustrate the trend with X-ray luminosity, and horizontal error bars indicate the standard deviation of the temperature distribution weighted by the metastable helium density.}}
%Each symbol is one model run at a different X-ray luminosity, with the value of the logarithmic X-ray to bolometric luminosity ratio labeled.}
}
\label{fig:varX}
\end{figure*}

For each case, we computed the mean atmospheric temperature and mass-loss rate from the \texttt{ATES} atmospheric structures and assessed whether the predicted near-infrared He~I triplet exceeds the 95\% detection thresholds in both depth and EW for each observation. Figure~\ref{fig:varX} displays the \(\dot{M}\)--\(T\) map and the EW-based detectability obtained using the EUV luminosity estimated with equation~(4) of S2025. The corresponding results for the depth, as well as those obtained using equation~(2), are presented in the table in appendix~4. Across the tested range of XUV fluxes, we find no case in which the EW and the line depth fall below the 95\% detection thresholds.
\red{This suggests that uncertainties in the adopted XUV flux are unlikely to account for the absence of detectable helium absorption in our targets.}
%suggesting that XUV flux variability alone is not likely to explain the absence of a detectable helium absorption feature \red{in our targets}.

%%%%%%%%%%%%%%%%%%%%%%%%%%%%%%%%%%%%%%
%%%%%%%%%%%%%%%%%%%%%%%%%%%%%%%%%%%%%%%

\section{Discussion}\label{dis}

%\red{We changed the method to correct the telluric absorption lines with \citet{Kawauchi+2022b}, but this change did not have much effect on transmission spectra. This shows that we can correct the telluric absorption lines just by doing the best model fitting for each frame, not considering the airmass when we correct the specific region around he lines. In addition, the slight difference in the 95 \% confidence upper limits of infrared helium triplet on the line depth and EW with \citet{Kawauchi+2022b} is due to the difference in the range of priors of the fitting parameters when we fit a Gaussian model rather than due to the difference in transmission spectra.}

In previous sections, we derived upper limits on the line depth and EW of the near-infrared helium triplet and found that model spectra computed assuming primordial atmospheres are inconsistent with our results for all four planets. We also find that the non-detections are not readily explained by uncertainties in the incident X-ray flux alone.

These interpretations differ between the super-Earths and the sub-Neptunes. TOI-1634b is a super-Earth consistent with a rocky composition (figure~\ref{fig:MR}) and has a relatively high escape velocity (\(\sim 26.9\) km s\(^{-1}\)) and a short orbital separation. These properties suggest that \red{it is unlikely to retain a H/He-rich primordial atmosphere.}
Nevertheless, the presence of a secondary atmosphere cannot be excluded, and follow-up observations of other molecular species with JWST would be valuable.

The other three planets have low mean densities and are therefore more likely to retain atmospheres. \red{Our results may} be consistent with non-primordial atmospheres, such as water-rich envelopes.
%or secondary atmospheres. 
They may also be consistent with metal-enriched atmospheres that still contain H/He, since recent studies suggest that the near infrared helium triplet can be weaker at high atmospheric metallicity (e.g., \cite{Yan+2024,Zhang+2025}).

For LP~791-18c in particular, recent JWST transmission spectroscopy suggests an H$_2$-dominated atmosphere and infers a high atmospheric metallicity (\cite{Roy+2025}). That study further discusses a bulk composition with a higher fraction of H$_2$ than H$_2$O, which is interpreted as being more consistent with formation interior to the water-ice line.
Taken together, LP~791-18c may still host a primordial H/He atmosphere. In addition, recent studies have suggested mechanisms that could reduce the He~I signal, including strong stellar winds and vibrationally excited H$_2$ (\cite{Fossati+2023,Munoz+2025}). Further modeling that incorporates such processes will help clarify the origin of the weak or absent He~I absorption.

To further investigate the lack of helium absorption, it is useful to independently constrain atmospheric composition and metallicity. Recent JWST observations have begun to provide such constraints for sub-Neptune atmospheres, but they can remain highly uncertain when hazes or clouds are significant. Observations at longer wavelengths, where the influence of clouds and hazes is relatively reduced, together with measurements of additional molecular features such as H$_2$O, will be valuable for improving constraints on atmospheric composition and metallicity.

%Finally, the non-detection of helium may also be attributable to a deviation of the H/He ratio from the primordial value (e.g., \cite{Yan+2022}).
\red {Finally, the non-detection of helium may also be attributable to an atmospheric H/He ratio different from the value assumed in our models (H/He = 90/10; e.g., \cite{Yan+2022}).}
Complementary escape tracers such as Ly\(\alpha\) and H\(\alpha\) can provide an independent test of atmospheric escape and may help strengthen constraints on the origin of the non-detections.

\section{Summary} \label{sum}

We conducted high-resolution transmission spectroscopy of TOI-2136b, TOI-1634b, TOI-654b, and LP~791-18c with the \red{IRD} spectrograph on the Subaru Telescope. 
%Telluric \red{absorption} lines were removed by fitting theoretical telluric spectra to each frame, and wavelength shifts were corrected to align and subtract the stellar spectra.
\redtwo{Telluric absorption and emission lines were corrected by fitting theoretical telluric absorption spectra and empirical $\rm OH^{-}$ emission templates to each frame. Wavelength shifts were then corrected to align the stellar spectra before subtraction.}
After these corrections, we combined the in-transit spectra in the planetary rest frame, taking into account the radial velocities of the planets.

No significant helium absorption was detected for any of the planets. However, we placed 95\% confidence upper limits on the line depth of \red{$1.54$\% for TOI-2136b, $0.62$\% for TOI-1634b, $1.97$\% for TOI-654b, and $2.53$\% for LP~791-18c}. We also derived 95\% confidence upper limits on the equivalent width of \red{$7.4$~m\mbox{\AA} for TOI-2136b, $2.7$~m\mbox{\AA} for TOI-1634b, $7.3$~m\mbox{\AA} for TOI-654b, and $8.1$~m\mbox{\AA} for LP~791-18c}.

By comparing these upper limits with one-dimensional isothermal Parker-wind models computed with \texttt{p-winds}, we constrained the region of mass-loss-rate--temperature parameter space that is inconsistent with our observations. We further computed self-consistent models of primordial H/He atmospheres using \texttt{ATES} and indicated that the corresponding helium absorption signals predicted by these models are also inconsistent with our data.

Overall, our results suggest that these planets are unlikely to host a primordial H/He envelope with near-solar composition, as such atmospheres would be expected to produce detectable helium absorption in our models. At the same time, an H/He atmosphere enriched in heavy elements and/or with an increased H/He ratio, which may weaken the helium absorption, remains a plausible possibility. To distinguish among these scenarios, it will be important to obtain further constraints on atmospheric composition, including the H/He ratio and metallicity, using complementary diagnostics such as molecular features (e.g., H$_2$O) and atomic escape tracers (e.g., Ly\(\alpha\) and H\(\alpha\)).

\begin{ack}
%Acknowledgement should be placed at end of main text.
%(NOT after the Appendix.)

This research is based on data collected at Subaru Telescope, which is operated by the National Astronomical Observatory of Japan.
We are honored and grateful for the opportunity of observing the Universe from Maunakea, which has the cultural, historical and natural significance in Hawaii.
The part of our data analysis was carried out on common use data analysis computer system at the Astronomy Data Center, ADC, of the National Astronomical Observatory of Japan. 
Our data reductions benefited from PyRAF and PyFITS that are the products of the Space Telescope Science Institute, which is operated by AURA for NASA.

%% Kakenhi %%%
This study was partly supported by the JSPS KAKENHI Grant Numbers 
JP21K13955, 
JP24H00017, 
JP24K17083, JP24K00689, 
JP24H00017,
JP24K17082, JP24H00248,
JP19K14783, JP21H00035,
JP 18H05439
JSPS Bilateral Program Number JPJSBP120249910,
JSPS Grant-in-Aid for JSPS Fellows Grant Number JP24KJ0241, JP25KJ0091,
JP25KJ1036,
JP25KJ1040,
JST SPRING, Grant Number JPMJSP2108, JPMJSP2108,
MEXT/JSPS KAKENHI grant Nos. 18H05442, 15H02063, 22000005, and  24H00242.

\end{ack}

\appendix %%%%%%%%%%%%%%%%%%%%%%%%%%%%%%%%%%%%%%%%%%%%%%%%%%%%%%%%
% \section*{Case of single paragraph}
%  No section number is necessary. Add ``*'' after \verb/\section/.

% %%%% 
\red{\section{Residual maps}\label{sec:ap1}}

\begin {figure*} [htbp]
% \begin{minipage}{0.45\hsize}
\begin{center}
 \includegraphics[width=\textwidth] {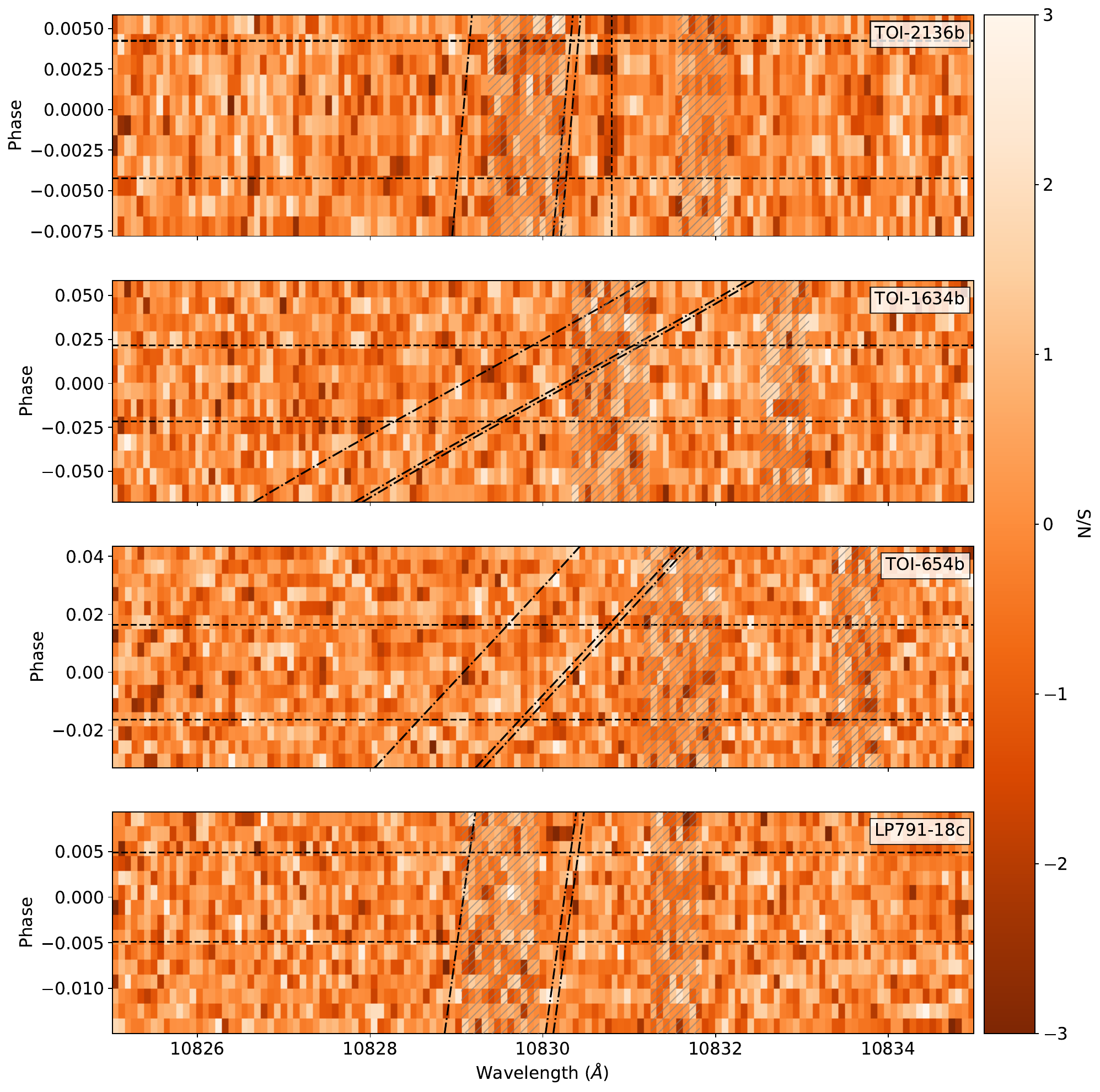}
\end{center}
\caption{Residual maps in units of S/N for TOI-2136b, TOI-1634b, TOI-654b, and LP~791-18c, computed for each exposure by subtracting the median residual flux and dividing by its standard deviation within the displayed wavelength range. The dash-dotted lines indicate the expected traces of the near-infrared helium triplet. Dashed horizontal lines show the beginning of ingress (T1) and the end of egress (T4). The hatched regions indicate wavelength ranges affected by $\rm OH^{-}$ emission lines. The dashed vertical line in the top panel marks the wavelength of the 10830.8~\mbox{\AA} feature.
{Alt text: Four residual maps around the near-infrared helium triplet. The horizontal axis shows wavelength, and the vertical axis shows orbital phase. Colors represent residual signal-to-noise ratio. Dash-dotted lines indicate the expected helium-triplet traces, hatched regions mark wavelength ranges affected by OH emission lines, and a dashed vertical line in the TOI-2136b panel indicates the 10830.8 angstrom feature.}}
\label{fig:residual_map}
\end{figure*}

\red{Figure~\ref{fig:residual_map} presents the residual maps around the near-infrared helium triplet region for all four targets. The hatched regions indicate wavelength ranges affected by $\rm OH^{-}$ emission lines after accounting for the stellar radial-velocity variation during the observations. The dashed vertical line in the panel of TOI-2136b marks the wavelength of the 10830.8~\mbox{\AA} feature discussed in the main text. No clear planetary absorption is visible in any of the targets.}

%%%%%%%%%%%%%%%%%%%%%%%%%%%%%%%%%%%%%%
%%%%%%%%%%%%%%%%%%%%%%%%%%%%%%%%%%%%%%%

\section{Additional telluric emission around the near-infrared helium triplet} \label{sec:ap2}

Telluric absorption and emission features can complicate the detection of planetary atmospheric absorption. In particular, numerous $\rm OH^{-}$ emission lines are present near the He I 10830~\mbox{\AA} triplet, and their wavelengths and potential impact on the observations merit careful characterization. At least three known $\rm OH^{-}$ emission lines are identified in this region: a strong line at $\sim$10834.4~\mbox{\AA} \red{(corresponding to an unresolved $\rm OH^{-}$ doublet at our spectral resolution)} and two weaker lines at $\sim$10832.1 and $\sim$10832.4~\mbox{\AA}. \red{All wavelengths are given in vacuum.} Although their absolute intensities vary with observing conditions, their relative intensity ratios are broadly similar across our observations.

During our analysis of the orbital-phase variation of the helium EW, we found anomalous EW values at several phases. Inspection of the corresponding spectra revealed an additional emission line near $\sim$10833.3~\mbox{\AA} in vacuum wavelength. In the observer frame before the barycentric correction, this line appears at approximately the same wavelength in data taken on different nights and at different times, suggesting a terrestrial origin rather than a stellar one.

\begin {figure} [htbp]
% \begin{minipage}{0.45\hsize}
\begin{center}
 \includegraphics[width=0.48\textwidth] {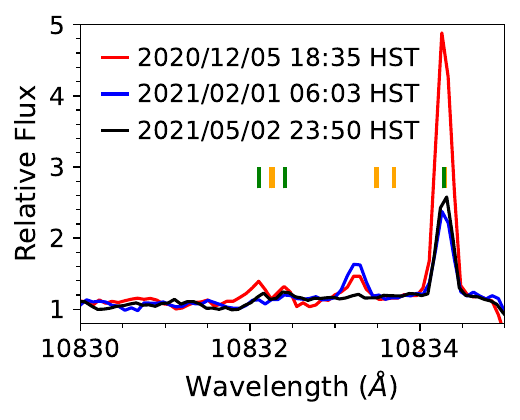}
\end{center}
\caption{IRD spectra of TOI-2136 between 10830 and 10835~\mbox{\AA} in vacuum wavelengths before telluric and barycentric corrections, observed at 2020-12-05 18:35 HST (red), 2021-02-01 06:03 HST (blue), and 2021-05-02 23:50 HST (black). The green vertical lines mark the central wavelengths of the well-known $\rm OH^{-}$ emission lines, and the yellow vertical lines indicate the $\rm OH^{-}$ line-list wavelengths from \citet{Rousselot+2000}. {Alt text: Three spectra of TOI-2136 are shown from 10830 to 10835 angstroms at vacuum wavelengths before telluric and barycentric corrections. Vertical markers indicate the expected wavelengths of known OH emission lines and the wavelengths from the OH line list.} }
\label{fig:telemission}
\end{figure}

This additional line shows no correlation with the intensities of the other $\rm OH^{-}$ lines. It was detected in spectra obtained within about 1.5 hours of local sunset or sunrise (figure~\ref{fig:telemission}). Here, the time from sunset or sunrise is computed from the difference between mid-exposure time and the sunset/sunrise time on each observing date. Among all exposures, the number of frames obtained within 1.5 hours of sunset or sunrise is 11 for TOI-2136, 5 for TOI-1634, 15 for TOI-654, and 5 for LP~791-18. Clear detections of the line by visual inspection are found in 4, 0, 5, and 1 frames for these targets, respectively, for a total of 10 frames. Nine of these 10 frames were taken before sunrise, which is consistent with the line being more likely to appear near dawn than near dusk.

For TOI-2136 and \red{TOI-1634}, some exposures obtained within 1.5 hours of sunset or sunrise do not clearly show the line. In \red{some of} these cases, the relevant wavelength region overlaps with stellar absorption lines, which may reduce line visibility. For \red{TOI-654} and LP~791-18, the spectra were not contaminated by stellar absorption in this region. However, even within the 1.5-hour window, many of these exposures were taken farther from twilight than those with clear detections, which may contribute to the non-detections. No correlation was found with azimuth. \red{The occurrence of the feature may reflect variations in atmospheric conditions and solar illumination from night to night.}

To assess whether this feature could be an $\rm OH^{-}$ line, we compared the spectra with an $\rm OH^{-}$ line list \citep{Rousselot+2000}. The line near $\sim$10833.3~\mbox{\AA} does not coincide with the listed $\rm OH^{-}$ wavelengths, and differences relative to the line list are also present for the two known weak $\rm OH^{-}$ lines (orange lines in figure~\ref{fig:telemission}). Therefore, a comparison with this line list alone does not allow us to identify the $\sim$10833.3~\mbox{\AA} feature as an $\rm OH^{-}$ emission line.

These results indicate that, \red{in a subset of the observations obtained} near sunset or sunrise, an additional telluric emission feature can appear around $\sim$10833.3~\mbox{\AA}.
\red{The observed behavior of the feature is consistent with telluric He emission from the terrestrial upper atmosphere \citep{Shefov1964, Kaifler+2022}, which has been reported to vary with solar illumination and to be enhanced near twilight (e.g., \cite{Suzuki1983}). Similar contamination near the He I 10830~\AA\ region has also been reported in previous helium observations (e.g., \cite{Orell-Miquel+2024}).}
%although its detailed identification remains uncertain. 
If this feature contaminates the out-of-transit reference spectra and varies with time, it can imprint an apparent absorption-like signal in the transmission spectrum. The possible presence and variability of this emission feature should therefore be considered when interpreting He absorption measurements obtained from data acquired near twilight.

%%%%%%%%%%%%%%%%%%%%%%%%%%%%%%%%%%%%%%
%%%%%%%%%%%%%%%%%%%%%%%%%%%%%%%%%%%%%%%

\section{Effect of phase grid size for transmission spectrum} \label{sec:ap3}

 In-transit spectra are obtained by co-adding the spectra over all phases during the transit. When constructing model spectra, we determine the start and end phases from the observation times and divide this interval into several bins over which the spectra are summed. 
When the phase bins are too small, the contributions from ingress and egress become significant, and the resulting transmission spectrum appears shallower than the intrinsic one. 
In contrast, using a very large number of bins greatly increases the computational time.
Therefore, an appropriate phase-bin number should be investigated.

\begin {figure} [htbp]
% \begin{minipage}{0.45\hsize}
\begin{center}
 \includegraphics[width=0.48\textwidth] {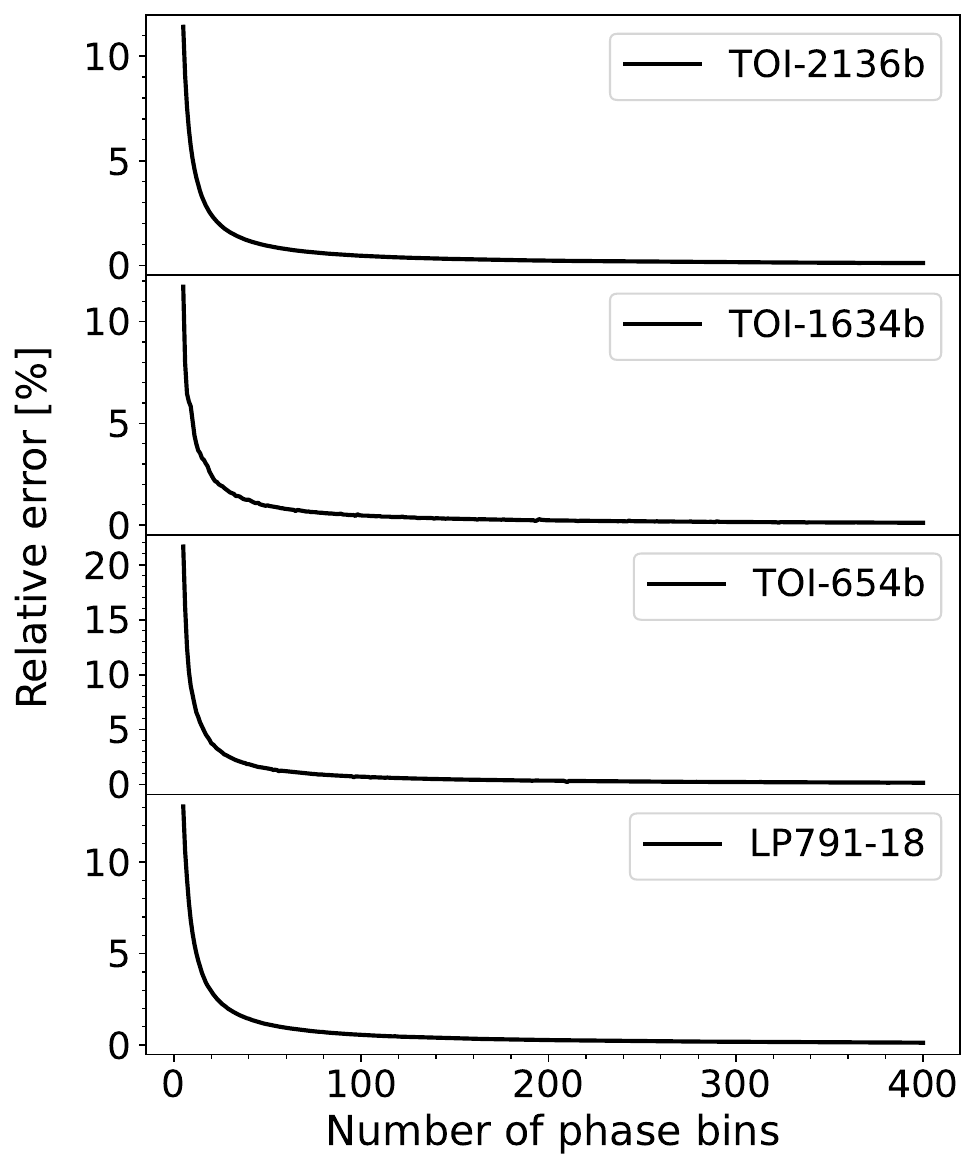}
\end{center}
\caption{Relative error in the inferred line depth as a function of the number of phase bins used to generate the planetary transmission spectra for TOI-2136b, TOI-1634b, TOI-654b, and LP~791-18c. The relative error is defined as $|D - D_{ref}|/D_{ref} \times 100 \ [\%]$. $D_{ref}$ is the line depth obtained with 10,000 phase bins. The line depth is measured from model spectra computed using the atmospheric composition derived with the \texttt{ATES} code, assuming the XUV flux from the \texttt{MUSCLES} spectrum. {Alt text: Four vertically stacked panels show the relative error in inferred line depth as a function of the number of phase bins used to generate transmission spectra, one target per panel. The error is computed relative to a reference line depth obtained using ten thousand phase bins.} }
\label{fig:numvar}
\end{figure}

In this study, we investigated the impact of phase binning by varying the number of phase bins between 5 and 400 within the interval between the start and end phases determined from the observation times and by examining the resulting changes in the depth of the near-infrared helium triplet. Figure~\ref{fig:numvar} shows the relative errors in the line depth as a function of the number of phase bins. For this calculation, we used atmospheric temperature structures computed with the \texttt{ATES} code, assuming the XUV flux derived from the \texttt{MUSCLES} spectrum, and we adopted the result obtained with 10,000 phase bins as the reference value.

When the number of phase bins is set to 5, the relative uncertainties in the line depth are approximately 11.4\%, 11.7\%, 21.7\%, and 13.0\% for TOI-2136b, TOI-1634b, TOI-654b, and LP~791-18c, respectively. These values indicate that, when generating the in-transit absorption lines with depths of approximately 8.8$\sigma$, 8.6$\sigma$, 4.6$\sigma$, and 7.7$\sigma$ for the four planets, the inferred line depths can deviate from the reference values by roughly 1$\sigma$ due to the coarse phase binning. To keep the relative errors sufficiently small ($\sim$0.5--0.7\%), we adopt 100 phase bins when constructing the model.

%%%%%%%%%%%%%%%%%%%%%%%%%%%%%%%%%%%%%%
%%%%%%%%%%%%%%%%%%%%%%%%%%%%%%%%%%%%%%%

\section{Summary table for the X-ray flux uncertainty analysis}

Table~\ref{table:varX} lists the expected depth and EW of the near-infrared helium triplet, the calculated mean temperature, and the mass-loss rate predicted by the \texttt{ATES} model for each $\log L_{X}$. The table also includes the EUV flux estimated from X flux using equation~(2) or equation~(4) of S2025.

%%%%%%%%%%%%%%%%%%%%%%%%%%%%%%%%%%%%%%
\begin{table*}[htbp]
\tbl{Expected depth and EW of the near-infrared helium triplet, mean temperature ($T$), and mass-loss rate ($\dot{M}$) predicted by \texttt{ATES} model for each $\log L_{X}$}{
\begin{tabular}{cccccccc}
\hline \hline
 Object & $\log (L_{X} / L_{bol})$  & $\log L_{X}$ & $\log L_{EUV}$ & $T$ & $\dot{M}$ & Depth & EW \\
%}
 & & ($\rm erg \ s^{-1}$) & ($\rm erg \ s^{-1}$) & (K) & ($\rm g \ s^{-1}$) & (\%) & (m\AA)\\
\hline
TOI-2136b & $-3$ & 28.72 &  28.49 (eq.4)& 2529 $\pm$ 1030 & 10.72 & 12.17 & 90.40\\
 & $-3$ & 28.72 &  29.21 (eq.2) & 3386 $\pm$ 1851 & 10.93 & 9.78 & 92.57\\
 & $-4$ & 27.72 &  27.99 (eq.4) & 1950 $\pm$ 503  & 9.97 & 11.85 & 65.09 \\
 & $-4$ & 27.72 &  28.39 (eq.2) & 2253 $\pm$ 793 & 10.21 & 9.08 & 55.03 \\
 & $-5$ & 26.72 &  27.36 (eq.4)&  1599 $\pm$ 190 & 9.28 & 13.16 & 54.52 \\
 & $-5$ & 26.72 &  27.57 (eq.2) & 1721 $\pm$ 270 & 9.45 & 13.19 & 59.67 \\
 & $-6$ & 25.72 &  26.61 (eq.4)& 1388 $\pm$ 215 & 8.54 & 5.20 & 15.92 \\
 & $-6$ & 25.72 &  26.75 (eq.2) & 1419 $\pm$ 202 & 8.67 & 6.42 & 20.32 \\
 & $-6.5$ & 25.22 &  26.19 (eq.4)& 1324 $\pm$ 254 & 8.12 & 2.59 & 7.44 \\
 & $-6.5$ & 25.22 &  26.34 (eq.2) & 1349 $\pm$ 243 & 8.27 & 3.32 & 9.71 \\
\\
TOI-1634b & $-3$ & 29.00 & 28.78 (eq.4)& 6453 $\pm$ 2373 & 11.37 & 1.80 & 21.03\\
 & $-3$ & 29.00 & 29.44 (eq.2) & 7888 $\pm$ 2825 & 11.50 & 2.01 & 30.46\\
 & $-4$ & 28.00 & 28.27 (eq.4) & 4904 $\pm$ 1678 & 10.65 & 1.21 & 9.23\\
 & $-4$ & 28.00 & 28.62 (eq.2) & 5317 $\pm$ 2251 & 10.82 & 1.28 & 11.56\\
 & $-5$ & 27.00 & 27.65 (eq.4) & 4178 $\pm$ 991 & 9.98 & 1.08 & 6.02\\
 & $-5$ & 27.00 & 27.80 (eq.2) & 4283 $\pm$ 1160 & 10.10 & 1.06 & 6.15\\
 & $-6$ & 26.00 & 26.90 (eq.4) & 3765 $\pm$ 614 & 9.26 & 1.37 & 6.41\\
 & $-6$ & 26.00 & 26.98 (eq.2) & 3818 $\pm$ 626 & 9.33 & 1.32 & 6.23\\
 & $-6.5$ & 25.50 & 26.48 (eq.4) & 3497 $\pm$ 723 & 8.85 & 1.45 & 6.12\\
 & $-6.5$ & 25.50 & 26.57 (eq.2) & 3560 $\pm$ 694 & 8.93 & 1.48 & 6.40\\
\\
TOI-654b & $-3$ & 28.97 & 28.75 (eq.4)& 5299 $\pm$ 2549 & 11.61 & 3.67 & 40.02\\
 & $-3$ & 28.97 & 29.42 (eq.2) & 7074 $\pm$ 3258 & 11.71 & 3.70 & 56.83\\
 & $-4$ & 27.97 & 28.24 (eq.4) & 3835 $\pm$ 1567 & 10.87 & 2.50 & 19.07\\
 & $-4$ & 27.97 & 28.60 (eq.2) & 4361 $\pm$ 2156 & 11.02 & 2.37 & 21.07\\
 & $-5$ & 26.97 & 27.62 (eq.4) & 3002 $\pm$ 792 & 10.22 & 2.47 & 13.99\\
 & $-5$ & 26.97 & 27.78 (eq.2) & 3150 $\pm$ 956 & 10.35 & 2.34 & 13.90\\
 & $-6$ & 25.97 & 26.87 (eq.4) & 2435 $\pm$ 299 & 9.54 & 3.28 & 15.20\\
 & $-6$ & 25.97 &  26.96 (eq.2) & 2504 $\pm$ 328 & 9.62 & 3.21 & 15.24\\
 & $-6.5$ & 25.47 &  26.44 (eq.4)& 2166 $\pm$ 332 & 9.13 & 2.82 & 11.03\\
 & $-6.5$ & 25.47 &  26.55 (eq.2) & 2229 $\pm$ 313 & 9.23 & 3.04 & 12.46\\
\\
LP791-18c & $-3$ & 27.94 & 27.72 (eq.4)& 2570 $\pm$ 647 & 10.41 & 31.58 & 198.28\\
 & $-3$ & 27.94 & 28.57 (eq.2) & 3390 $\pm$ 1568 & 10.74 & 22.06 & 178.19\\
 & $-4$ & 26.94 & 27.21 (eq.4) & 2223 $\pm$ 305 & 9.65 & 33.3 & 160.91\\
 & $-4$ & 26.94 & 27.75 (eq.2) & 2576 $\pm$ 695 & 10.01 & 23.29 & 126.14\\
 & $-5$ & 25.94 &  26.59 (eq.4)& 2021 $\pm$ 286 & 8.94 & 27.22 & 101.35\\
 & $-5$ & 25.94 &  26.93 (eq.2) & 2180 $\pm$ 296 & 9.23 & 32.34 & 137.66\\
 & $-6$ & 24.94 & 25.84 (eq.4)& 1873 $\pm$ 388 & 8.18 & 9.60 & 29.56\\
 & $-6$ & 24.94 & 26.11 (eq.2)& 1937 $\pm$ 366 & 8.43 & 14.33 & 46.14\\
\hline
\end{tabular}}\label{table:varX}
\end{table*}
%%%%%%%%%%%%%%%%%%%

\bibliographystyle{apj}
\bibliography{reference}

@ARTICLE{Zhangb+2022,
       author = {{Zhang}, Michael and {Cauley}, P. Wilson and {Knutson}, Heather A. and {France}, Kevin and {Kreidberg}, Laura and {Oklop{\v{c}}i{\'c}}, Antonija and {Redfield}, Seth and {Shkolnik}, Evgenya L.},
        title = "{More Evidence for Variable Helium Absorption from HD 189733b}",
      journal = {\aj},
         year = 2022,
        month = dec,
       volume = {164},
       number = {6},
          eid = {237},
        pages = {237},
          doi = {10.3847/1538-3881/ac9675},
archivePrefix = {arXiv},
       eprint = {2204.02985},
 primaryClass = {astro-ph.EP},
       adsurl = {https://ui.adsabs.harvard.edu/abs/2022AJ....164..237Z}
}

@INPROCEEDINGS{Shefov1964,
       author = {{Shefov}, N.~N.},
        title = "{Helium in the upper atmosphere}",
    booktitle = {Theoretical Interpretation of Upper Atmosphere Emission},
         year = 1964,
       editor = {{Bates}, David Robert},
       series = {IAU Symposium},
       volume = {18},
        month = jan,
        pages = {73},
       adsurl = {https://ui.adsabs.harvard.edu/abs/1964IAUS...18...73S}
}

@ARTICLE{Kaifler+2022,
       author = {{Kaifler}, Bernd and {Geach}, Christopher and {B{\"u}denbender}, Hans Christian and {Mezger}, Andreas and {Rapp}, Markus},
        title = "{Measurements of metastable helium in Earth's atmosphere by resonance lidar}",
      journal = {Nature Communications},
         year = 2022,
        month = oct,
       volume = {13},
          eid = {6042},
        pages = {6042},
          doi = {10.1038/s41467-022-33751-6},
       adsurl = {https://ui.adsabs.harvard.edu/abs/2022NatCo..13.6042K}
}

@ARTICLE{Suzuki1983,
       author = {{Suzuki}, K.},
        title = "{Observation of the helium 10830 {\r{A}} airglow emission in midlatitude.}",
      journal = {Journal of Geomagnetism and Geoelectricity},
         year = 1983,
        month = jan,
       volume = {35},
       number = {9},
        pages = {321-330},
          doi = {10.5636/jgg.35.321},
       adsurl = {https://ui.adsabs.harvard.edu/abs/1983JGG....35..321S}
}

@ARTICLE{Palle+2023,
       author = {{Palle}, E. and {Orell-Miquel}, J. and {Brady}, M. and {Bean}, J. and {Hatzes}, A.~P. and {Morello}, G. and {Morales}, J.~C. and {Murgas}, F. and {Molaverdikhani}, K. and {Parviainen}, H. and {Sanz-Forcada}, J. and {B{\'e}jar}, V.~J.~S. and {Caballero}, J.~A. and {Sreenivas}, K.~R. and {Schlecker}, M. and {Ribas}, I. and {Perdelwitz}, V. and {Tal-Or}, L. and {P{\'e}rez-Torres}, M. and {Luque}, R. and {Dreizler}, S. and {Fuhrmeister}, B. and {Aceituno}, F. and {Amado}, P.~J. and {Anglada-Escud{\'e}}, G. and {Caldwell}, D.~A. and {Charbonneau}, D. and {Cifuentes}, C. and {de Leon}, J.~P. and {Collins}, K.~A. and {Dufoer}, S. and {Espinoza}, N. and {Essack}, Z. and {Fukui}, A. and {Chew}, Y. G{\'o}mez Maqueo and {G{\'o}mez-Mu{\~n}oz}, M.~A. and {Henning}, Th. and {Herrero}, E. and {Jeffers}, S.~V. and {Jenkins}, J. and {Kaminski}, A. and {Kasper}, J. and {Kunimoto}, M. and {Latham}, D. and {Lillo-Box}, J. and {L{\'o}pez-Gonz{\'a}lez}, M.~J. and {Montes}, D. and {Mori}, M. and {Narita}, N. and {Quirrenbach}, A. and {Pedraz}, S. and {Reiners}, A. and {Rodr{\'\i}guez}, E. and {Rodr{\'\i}guez-L{\'o}pez}, C. and {Sabin}, L. and {Schanche}, N. and {Schwarz}, R.-P. and {Schweitzer}, A. and {Seifahrt}, A. and {Stefansson}, G. and {Sturmer}, J. and {Trifonov}, T. and {Vanaverbeke}, S. and {Wells}, R.~D. and {Zapatero-Osorio}, M.~R. and {Zechmeister}, M.},
        title = "{GJ 806 (TOI-4481): A bright nearby multi-planetary system with a transiting hot low-density super-Earth}",
      journal = {\aap},
         year = 2023,
        month = oct,
       volume = {678},
          eid = {A80},
        pages = {A80},
          doi = {10.1051/0004-6361/202244261},
archivePrefix = {arXiv},
       eprint = {2301.06873},
 primaryClass = {astro-ph.EP},
       adsurl = {https://ui.adsabs.harvard.edu/abs/2023A&A...678A..80P}
}

@ARTICLE{Krishnamurthy+2023,
       author = {{Krishnamurthy}, Vigneshwaran and {Hirano}, Teruyuki and {Gaidos}, Eric and {Sato}, Bunei and {Kopparapu}, Ravi and {Barclay}, Thomas and {Garcia-Sage}, Katherine and {Harakawa}, Hiroki and {Hodapp}, Klaus and {Jacobson}, Shane and {Konishi}, Mihoko and {Kotani}, Takayuki and {Kudo}, Tomoyuki and {Kurokawa}, Takashi and {Kuzuhara}, Masayuki and {Lopez}, Eric and {Nishikawa}, Jun and {Omiya}, Masashi and {Schlieder}, Joshua E. and {Serizawa}, Takuma and {Tamura}, Motohide and {Ueda}, Akitoshi and {Vievard}, Sebastien},
        title = "{Absence of extended atmospheres in low-mass star radius-gap planets}",
      journal = {\mnras},
         year = 2023,
        month = may,
       volume = {521},
       number = {1},
        pages = {1210-1220},
          doi = {10.1093/mnras/stad404},
archivePrefix = {arXiv},
       eprint = {2302.01479},
 primaryClass = {astro-ph.EP},
       adsurl = {https://ui.adsabs.harvard.edu/abs/2023MNRAS.521.1210K}
}

@ARTICLE{Ahrer+2025,
       author = {{Ahrer}, Eva-Maria and {Radica}, Michael and {Piaulet-Ghorayeb}, Caroline and {Raul}, Eshan and {Wiser}, Lindsey and {Welbanks}, Luis and {Acu{\~n}a}, Lorena and {Allart}, Romain and {Coulombe}, Louis-Philippe and {Louca}, Amy and {MacDonald}, Ryan and {Saidel}, Morgan and {Evans-Soma}, Thomas M. and {Benneke}, Bj{\"o}rn and {Christie}, Duncan and {Beatty}, Thomas G. and {Cadieux}, Charles and {Cloutier}, Ryan and {Doyon}, Ren{\'e} and {Fortney}, Jonathan J. and {Gagnebin}, Anna and {Gapp}, Cyril and {Innes}, Hamish and {Knutson}, Heather A. and {Komacek}, Thaddeus and {Krissansen-Totton}, Joshua and {Miguel}, Yamila and {Pierrehumbert}, Raymond and {Roy}, Pierre-Alexis and {Schlichting}, Hilke E.},
        title = "{Escaping Helium and a Highly Muted Spectrum Suggest a Metal-enriched Atmosphere on Sub-Neptune GJ 3090 b from JWST Transit Spectroscopy}",
      journal = {\apjl},
         year = 2025,
        month = may,
       volume = {985},
       number = {1},
          eid = {L10},
        pages = {L10},
          doi = {10.3847/2041-8213/add010},
archivePrefix = {arXiv},
       eprint = {2504.20428},
 primaryClass = {astro-ph.EP},
       adsurl = {https://ui.adsabs.harvard.edu/abs/2025ApJ...985L..10A}
}

@INPROCEEDINGS{Gully-Santiago+2023,
       author = {{Gully-Santiago}, Michael and {Morley}, Caroline and {HPF Helium Exospheres Program}},
        title = "{A Large and Variable Leading Tail of Helium in HAT-P-67b, a Sub-Saturn Undergoing Runaway Inflation}",
    booktitle = {55th Annual Meeting of the Division for Planetary Sciences},
         year = 2023,
       series = {AAS/Division for Planetary Sciences Meeting Abstracts},
       volume = {55},
        month = oct,
          eid = {200.02},
        pages = {200.02},
       adsurl = {https://ui.adsabs.harvard.edu/abs/2023DPS....5520002G}
}

@ARTICLE{Foreman-Mackey+2013,
       author = {{Foreman-Mackey}, Daniel and {Hogg}, David W. and {Lang}, Dustin and {Goodman}, Jonathan},
        title = "{emcee: The MCMC Hammer}",
      journal = {\pasp},
         year = 2013,
        month = mar,
       volume = {125},
       number = {925},
        pages = {306},
          doi = {10.1086/670067},
archivePrefix = {arXiv},
       eprint = {1202.3665},
 primaryClass = {astro-ph.IM},
       adsurl = {https://ui.adsabs.harvard.edu/abs/2013PASP..125..306F}
}

@ARTICLE{Stassun+2018,
       author = {{Stassun}, Keivan G. and {Corsaro}, Enrico and {Pepper}, Joshua A. and {Gaudi}, B. Scott},
        title = "{Empirical Accurate Masses and Radii of Single Stars with TESS and Gaia}",
      journal = {\aj},
         year = 2018,
        month = jan,
       volume = {155},
       number = {1},
          eid = {22},
        pages = {22},
          doi = {10.3847/1538-3881/aa998a},
archivePrefix = {arXiv},
       eprint = {1710.01460},
 primaryClass = {astro-ph.SR},
       adsurl = {https://ui.adsabs.harvard.edu/abs/2018AJ....155...22S}
}

@ARTICLE{Skrutskie+2006,
       author = {{Skrutskie}, M.~F. and {Cutri}, R.~M. and {Stiening}, R. and {Weinberg}, M.~D. and {Schneider}, S. and {Carpenter}, J.~M. and {Beichman}, C. and {Capps}, R. and {Chester}, T. and {Elias}, J. and {Huchra}, J. and {Liebert}, J. and {Lonsdale}, C. and {Monet}, D.~G. and {Price}, S. and {Seitzer}, P. and {Jarrett}, T. and {Kirkpatrick}, J.~D. and {Gizis}, J.~E. and {Howard}, E. and {Evans}, T. and {Fowler}, J. and {Fullmer}, L. and {Hurt}, R. and {Light}, R. and {Kopan}, E.~L. and {Marsh}, K.~A. and {McCallon}, H.~L. and {Tam}, R. and {Van Dyk}, S. and {Wheelock}, S.},
        title = "{The Two Micron All Sky Survey (2MASS)}",
      journal = {\aj},
         year = 2006,
        month = feb,
       volume = {131},
       number = {2},
        pages = {1163-1183},
          doi = {10.1086/498708},
       adsurl = {https://ui.adsabs.harvard.edu/abs/2006AJ....131.1163S}
}

@ARTICLE{Luque+2022,
       author = {{Luque}, Rafael and {Pall{\'e}}, Enric},
        title = "{Density, not radius, separates rocky and water-rich small planets orbiting M dwarf stars}",
      journal = {Science},
         year = 2022,
        month = sep,
       volume = {377},
       number = {6611},
        pages = {1211-1214},
          doi = {10.1126/science.abl7164},
archivePrefix = {arXiv},
       eprint = {2209.03871},
 primaryClass = {astro-ph.EP},
       adsurl = {https://ui.adsabs.harvard.edu/abs/2022Sci...377.1211L}
}

@ARTICLE{Borucki+2010,
       author = {{Borucki}, William J. and {Koch}, David and {Basri}, Gibor and {Batalha}, Natalie and {Brown}, Timothy and {Caldwell}, Douglas and {Caldwell}, John and {Christensen-Dalsgaard}, J{\o}rgen and {Cochran}, William D. and {DeVore}, Edna and {Dunham}, Edward W. and {Dupree}, Andrea K. and {Gautier}, Thomas N. and {Geary}, John C. and {Gilliland}, Ronald and {Gould}, Alan and {Howell}, Steve B. and {Jenkins}, Jon M. and {Kondo}, Yoji and {Latham}, David W. and {Marcy}, Geoffrey W. and {Meibom}, S{\o}ren and {Kjeldsen}, Hans and {Lissauer}, Jack J. and {Monet}, David G. and {Morrison}, David and {Sasselov}, Dimitar and {Tarter}, Jill and {Boss}, Alan and {Brownlee}, Don and {Owen}, Toby and {Buzasi}, Derek and {Charbonneau}, David and {Doyle}, Laurance and {Fortney}, Jonathan and {Ford}, Eric B. and {Holman}, Matthew J. and {Seager}, Sara and {Steffen}, Jason H. and {Welsh}, William F. and {Rowe}, Jason and {Anderson}, Howard and {Buchhave}, Lars and {Ciardi}, David and {Walkowicz}, Lucianne and {Sherry}, William and {Horch}, Elliott and {Isaacson}, Howard and {Everett}, Mark E. and {Fischer}, Debra and {Torres}, Guillermo and {Johnson}, John Asher and {Endl}, Michael and {MacQueen}, Phillip and {Bryson}, Stephen T. and {Dotson}, Jessie and {Haas}, Michael and {Kolodziejczak}, Jeffrey and {Van Cleve}, Jeffrey and {Chandrasekaran}, Hema and {Twicken}, Joseph D. and {Quintana}, Elisa V. and {Clarke}, Bruce D. and {Allen}, Christopher and {Li}, Jie and {Wu}, Haley and {Tenenbaum}, Peter and {Verner}, Ekaterina and {Bruhweiler}, Frederick and {Barnes}, Jason and {Prsa}, Andrej},
        title = "{Kepler Planet-Detection Mission: Introduction and First Results}",
      journal = {Science},
         year = 2010,
        month = feb,
       volume = {327},
       number = {5968},
        pages = {977},
          doi = {10.1126/science.1185402},
       adsurl = {https://ui.adsabs.harvard.edu/abs/2010Sci...327..977B}
}

@ARTICLE{Yan+2022,
       author = {{Yan}, Dongdong and {Seon}, Kwang-il and {Guo}, Jianheng and {Chen}, Guo and {Li}, Lifang},
        title = "{Modeling the H{\ensuremath{\alpha}} and He 10830 Transmission Spectrum of WASP-52b}",
      journal = {\apj},
         year = 2022,
        month = sep,
       volume = {936},
       number = {2},
          eid = {177},
        pages = {177},
          doi = {10.3847/1538-4357/ac8793},
archivePrefix = {arXiv},
       eprint = {2208.03916},
 primaryClass = {astro-ph.EP},
       adsurl = {https://ui.adsabs.harvard.edu/abs/2022ApJ...936..177Y}
}

@ARTICLE{Roy+2025,
       author = {{Roy}, Pierre-Alexis and {Benneke}, Bj{\"o}rn and {Fournier-Tondreau}, Marylou and {Coulombe}, Louis-Philippe and {Piaulet-Ghorayeb}, Caroline and {Lafreni{\`e}re}, David and {Allart}, Romain and {Cowan}, Nicolas B. and {Dang}, Lisa and {Johnstone}, Doug and {Langeveld}, Adam B. and {Pelletier}, Stefan and {Radica}, Michael and {Taylor}, Jake and {Albert}, Lo{\"\i}c and {Doyon}, Ren{\'e} and {Flagg}, Laura and {Jayawardhana}, Ray and {MacDonald}, Ryan J. and {Turner}, Jake D.},
        title = "{Diversity in the haziness and chemistry of temperate sub-Neptunes}",
      journal = {Nature Astronomy},
         year = 2025,
        month = dec,
         note = {in press (doi: 10.1038/s41550-025-02723-3)},
          doi = {10.1038/s41550-025-02723-3},
archivePrefix = {arXiv},
       eprint = {2512.10876},
 primaryClass = {astro-ph.EP},
       adsurl = {https://ui.adsabs.harvard.edu/abs/2025NatAs.tmp..256R}
}

@ARTICLE{Zhang+2025,
       author = {{Zhang}, Michael and {Bean}, Jacob L. and {Wilson}, David and {Duvvuri}, Girish and {Schneider}, Christian and {Knutson}, Heather A. and {Dai}, Fei and {Collins}, Karen A. and {Watkins}, Cristilyn N. and {Schwarz}, Richard P. and {Barkaoui}, Khalid and {Shporer}, Avi and {Horne}, Keith and {Sefako}, Ramotholo and {Murgas}, Felipe and {Palle}, Enric},
        title = "{Constraining Atmospheric Composition from the Outflow: Helium Observations Reveal the Fundamental Properties of Two Planets Straddling the Radius Gap}",
      journal = {\aj},
         year = 2025,
        month = apr,
       volume = {169},
       number = {4},
          eid = {204},
        pages = {204},
          doi = {10.3847/1538-3881/adb490},
archivePrefix = {arXiv},
       eprint = {2409.08318},
 primaryClass = {astro-ph.EP},
       adsurl = {https://ui.adsabs.harvard.edu/abs/2025AJ....169..204Z}
}

@ARTICLE{Yan+2024,
       author = {{Yan}, Dongdong and {Guo}, Jianheng and {Seon}, Kwang-il and {L{\'o}pez-Puertas}, Manuel and {Czesla}, Stefan and {Lamp{\'o}n}, Manuel},
        title = "{A possibly solar metallicity atmosphere escaping from HAT-P-32b revealed by H{\ensuremath{\alpha}} and He absorption}",
      journal = {\aap},
         year = 2024,
        month = jun,
       volume = {686},
          eid = {A208},
        pages = {A208},
          doi = {10.1051/0004-6361/202348210},
archivePrefix = {arXiv},
       eprint = {2403.17325},
 primaryClass = {astro-ph.EP},
       adsurl = {https://ui.adsabs.harvard.edu/abs/2024A&A...686A.208Y}
}

@ARTICLE{Spake+2022,
       author = {{Spake}, Jessica J. and {Oklop{\v{c}}i{\'c}}, A. and {Hillenbrand}, L.~A. and {Knutson}, Heather A. and {Kasper}, David and {Dai}, Fei and {Orell-Miquel}, Jaume and {Vissapragada}, Shreyas and {Zhang}, Michael and {Bean}, Jacob L.},
        title = "{Non-detection of He I in the Atmosphere of GJ 1214b with Keck/NIRSPEC, at a Time of Minimal Telluric Contamination}",
      journal = {\apjl},
         year = 2022,
        month = nov,
       volume = {939},
       number = {1},
          eid = {L11},
        pages = {L11},
          doi = {10.3847/2041-8213/ac88c9},
archivePrefix = {arXiv},
       eprint = {2209.03502},
 primaryClass = {astro-ph.EP},
       adsurl = {https://ui.adsabs.harvard.edu/abs/2022ApJ...939L..11S}
}

@ARTICLE{Gaidos+2020,
       author = {{Gaidos}, E. and {Hirano}, T. and {Wilson}, D.~J. and {France}, K. and {Rockcliffe}, K. and {Newton}, E. and {Feiden}, G. and {Krishnamurthy}, V. and {Harakawa}, H. and {Hodapp}, K.~W. and {Ishizuka}, M. and {Jacobson}, S. and {Konishi}, M. and {Kotani}, T. and {Kudo}, T. and {Kurokawa}, T. and {Kuzuhara}, M. and {Nishikawa}, J. and {Omiya}, M. and {Serizawa}, T. and {Tamura}, M. and {Ueda}, A. and {Vievard}, S.},
        title = "{Zodiacal exoplanets in time - XI. The orbit and radiation environment of the young M dwarf-hosted planet K2-25b}",
      journal = {\mnras},
         year = 2020,
        month = oct,
       volume = {498},
       number = {1},
        pages = {L119-L124},
          doi = {10.1093/mnrasl/slaa136},
archivePrefix = {arXiv},
       eprint = {2007.12701},
 primaryClass = {astro-ph.EP},
       adsurl = {https://ui.adsabs.harvard.edu/abs/2020MNRAS.498L.119G}
}

@ARTICLE{Aoki+2022,
       author = {{Aoki}, Wako and {Beers}, Timothy C. and {Honda}, Satoshi and {Ishikawa}, Hiroyuki T. and {Matsuno}, Tadafumi and {Placco}, Vinicius M. and {Yoon}, Jinmi and {Harakawa}, Hiroki and {Hirano}, Teruyuki and {Hodapp}, Klaus and {Ishizuka}, Masato and {Jacobson}, Shane and {Kotani}, Takayuki and {Kudo}, Tomoyuki and {Kurokawa}, Takashi and {Kuzuhara}, Masayuki and {Nishikawa}, Jun and {Omiya}, Masashi and {Serizawa}, Takuma and {Tamura}, Motohide and {Ueda}, Akitoshi and {Vievard}, S{\'e}bastien},
        title = "{Silicon and strontium abundances of very metal-poor stars determined from near-infrared spectra}",
      journal = {\pasj},
         year = 2022,
        month = apr,
       volume = {74},
       number = {2},
        pages = {273-282},
          doi = {10.1093/pasj/psab123},
archivePrefix = {arXiv},
       eprint = {2112.07433},
 primaryClass = {astro-ph.SR},
       adsurl = {https://ui.adsabs.harvard.edu/abs/2022PASJ...74..273A}
}

@ARTICLE{Howard+2012,
       author = {{Howard}, Andrew W. and {Marcy}, Geoffrey W. and {Bryson}, Stephen T. and {Jenkins}, Jon M. and {Rowe}, Jason F. and {Batalha}, Natalie M. and {Borucki}, William J. and {Koch}, David G. and {Dunham}, Edward W. and {Gautier}, III, Thomas N. and {Van Cleve}, Jeffrey and {Cochran}, William D. and {Latham}, David W. and {Lissauer}, Jack J. and {Torres}, Guillermo and {Brown}, Timothy M. and {Gilliland}, Ronald L. and {Buchhave}, Lars A. and {Caldwell}, Douglas A. and {Christensen-Dalsgaard}, J{\o}rgen and {Ciardi}, David and {Fressin}, Francois and {Haas}, Michael R. and {Howell}, Steve B. and {Kjeldsen}, Hans and {Seager}, Sara and {Rogers}, Leslie and {Sasselov}, Dimitar D. and {Steffen}, Jason H. and {Basri}, Gibor S. and {Charbonneau}, David and {Christiansen}, Jessie and {Clarke}, Bruce and {Dupree}, Andrea and {Fabrycky}, Daniel C. and {Fischer}, Debra A. and {Ford}, Eric B. and {Fortney}, Jonathan J. and {Tarter}, Jill and {Girouard}, Forrest R. and {Holman}, Matthew J. and {Johnson}, John Asher and {Klaus}, Todd C. and {Machalek}, Pavel and {Moorhead}, Althea V. and {Morehead}, Robert C. and {Ragozzine}, Darin and {Tenenbaum}, Peter and {Twicken}, Joseph D. and {Quinn}, Samuel N. and {Isaacson}, Howard and {Shporer}, Avi and {Lucas}, Philip W. and {Walkowicz}, Lucianne M. and {Welsh}, William F. and {Boss}, Alan and {Devore}, Edna and {Gould}, Alan and {Smith}, Jeffrey C. and {Morris}, Robert L. and {Prsa}, Andrej and {Morton}, Timothy D. and {Still}, Martin and {Thompson}, Susan E. and {Mullally}, Fergal and {Endl}, Michael and {MacQueen}, Phillip J.},
        title = "{Planet Occurrence within 0.25 AU of Solar-type Stars from Kepler}",
      journal = {\apjs},
         year = 2012,
        month = aug,
       volume = {201},
       number = {2},
          eid = {15},
        pages = {15},
          doi = {10.1088/0067-0049/201/2/15},
archivePrefix = {arXiv},
       eprint = {1103.2541},
 primaryClass = {astro-ph.EP},
       adsurl = {https://ui.adsabs.harvard.edu/abs/2012ApJS..201...15H}
}

@ARTICLE{Fossati+2023,
       author = {{Fossati}, L. and {Pillitteri}, I. and {Shaikhislamov}, I.~F. and {Bonfanti}, A. and {Borsa}, F. and {Carleo}, I. and {Guilluy}, G. and {Rumenskikh}, M.~S.},
        title = "{Possible origin of the non-detection of metastable He I in the upper atmosphere of the hot Jupiter WASP-80b}",
      journal = {\aap},
         year = 2023,
        month = may,
       volume = {673},
          eid = {A37},
        pages = {A37},
          doi = {10.1051/0004-6361/202245667},
archivePrefix = {arXiv},
       eprint = {2303.09501},
 primaryClass = {astro-ph.EP},
       adsurl = {https://ui.adsabs.harvard.edu/abs/2023A&A...673A..37F}
}

@ARTICLE{Rousselot+2000,
       author = {{Rousselot}, P. and {Lidman}, C. and {Cuby}, J.-G. and {Moreels}, G. and {Monnet}, G.},
        title = "{Night-sky spectral atlas of OH emission lines in the near-infrared}",
      journal = {\aap},
         year = 2000,
        month = feb,
       volume = {354},
        pages = {1134-1150},
       adsurl = {https://ui.adsabs.harvard.edu/abs/2000A&A...354.1134R}
}

@article{Munoz+2025,
	author = {{Garc{\'i}a Mu{\~n}oz, A.} and {De Fazio, D.} and {Wilson, D. J.} and {France, K.}},
	title = {Vibrationally excited H2 muting the He I triplet line at 1.08 μm on warm exo-Neptunes},
	DOI= "10.1051/0004-6361/202558117",
	url= "https://doi.org/10.1051/0004-6361/202558117",
	journal = {A\&A},
	year = 2025,
	volume = 704,
	pages = "L18",
}

@ARTICLE{Caramazza+2023,
       author = {{Caramazza}, M. and {Stelzer}, B. and {Magaudda}, E. and {Raetz}, St. and {G{\"u}del}, M. and {Orlando}, S. and {Poppenh{\"a}ger}, K.},
        title = "{Complete X-ray census of M dwarfs in the solar neighborhood. I. GJ 745 AB: Coronal-hole stars in the 10 pc sample}",
      journal = {\aap},
         year = 2023,
        month = aug,
       volume = {676},
          eid = {A14},
        pages = {A14},
          doi = {10.1051/0004-6361/202346470},
archivePrefix = {arXiv},
       eprint = {2305.14971},
 primaryClass = {astro-ph.SR},
       adsurl = {https://ui.adsabs.harvard.edu/abs/2023A&A...676A..14C}
}

@ARTICLE{Zhu+2025,
       author = {{Zhu}, E. and {Preibisch}, T.},
        title = "{X-ray activity of nearby G-, K-, and M-type stars and implications for planet habitability around M stars}",
      journal = {\aap},
         year = 2025,
        month = feb,
       volume = {694},
          eid = {A93},
        pages = {A93},
          doi = {10.1051/0004-6361/202452057},
archivePrefix = {arXiv},
       eprint = {2501.07313},
 primaryClass = {astro-ph.SR},
       adsurl = {https://ui.adsabs.harvard.edu/abs/2025A&A...694A..93Z}
}

@ARTICLE{Barkaoui+2025,
       author = {{Barkaoui}, K. and {Korth}, J. and {Gaidos}, E. and {Agol}, E. and {Parviainen}, H. and {Pozuelos}, F.~J. and {Palle}, E. and {Narita}, N. and {Grimm}, S. and {Brady}, M. and {Bean}, J.~L. and {Morello}, G. and {Rackham}, B.~V. and {Burgasser}, A.~J. and {Van Grootel}, V. and {Rojas-Ayala}, B. and {Seifahrt}, A. and {Marfil}, E. and {Passegger}, V.~M. and {Stalport}, M. and {Gillon}, M. and {Collins}, K.~A. and {Shporer}, A. and {Giacalone}, S. and {Yal{\c{c}}{\i}nkaya}, S. and {Ducrot}, E. and {Timmermans}, M. and {Triaud}, A.~H.~M.~J. and {de Wit}, J. and {Soubkiou}, A. and {Watkins}, C.~N. and {Aganze}, C. and {Alonso}, R. and {Amado}, P.~J. and {Basant}, R. and {Ba{\textcommabelow s}t{\"u}rk}, {\"O}. and {Benkhaldoun}, Z. and {Burdanov}, A. and {Calatayud-Borras}, Y. and {Chouqar}, J. and {Conti}, D.~M. and {Collins}, K.~I. and {Davoudi}, F. and {Delrez}, L. and {Dressing}, C.~D. and {de Leon}, J. and {D{\'e}vora-Pajares}, M. and {Demory}, B.~O. and {Dransfield}, G. and {Esparza-Borges}, E. and {Fern{\'a}ndez-Rodriguez}, G. and {Fukuda}, I. and {Fukui}, A. and {Gallardo}, P.~P.~M. and {Garcia}, L. and {Garcia}, N.~A. and {Ghachoui}, M. and {Gerald{\'\i}a-Gonz{\'a}lez}, S. and {G{\'o}mez Maqueo Chew}, Y. and {Gonz{\'a}lez-Rodr{\'\i}guez}, J. and {G{\"u}nther}, M.~N. and {Hayashi}, Y. and {Horne}, K. and {Hooton}, M.~J. and {Hsu}, C.~C. and {Ikuta}, K. and {Isogai}, K. and {Jehin}, E. and {Jenkins}, J.~M. and {Kawauchi}, K. and {Kagetani}, T. and {Kawai}, Y. and {Kasper}, D. and {Kielkopf}, J.~F. and {Klagyivik}, P. and {Lacedelli}, G. and {Latham}, D.~W. and {Libotte}, F. and {Luque}, R. and {Livingston}, J.~H. and {Mancini}, L. and {Massey}, B. and {Mori}, M. and {Mu{\~n}oz Torres}, S. and {Murgas}, F. and {Niraula}, P. and {Orell-Miquel}, J. and {Rapetti}, David and {Rebolo-L{\'o}pez}, R. and {Ricker}, G. and {Papini}, R. and {Pedersen}, P.~P. and {Pel{\'a}ez-Torres}, A. and {P{\'e}rez-Prieto}, J.~A. and {Poultourtzidis}, E. and {Rodriguez}, P.~M. and {Queloz}, D. and {Savel}, A.~B. and {Schanche}, N. and {S{\'a}nchez-Benavente}, M. and {Sibbald}, L. and {Sefako}, R. and {Sohy}, S. and {Sota}, A. and {Schwarz}, R.~P. and {Seager}, S. and {Sebastian}, D. and {Southworth}, J. and {Stangret}, M. and {Stef{\'a}nsson}, G. and {St{\"u}rmer}, J. and {Srdoc}, G. and {Thompson}, S.~J. and {Terada}, Y. and {Vanderspek}, R. and {Wang}, G. and {Watanabe}, N. and {Wilkin}, F.~P. and {Winn}, J. and {Wells}, R.~D. and {Ziegler}, C. and {Z{\'u}{\~n}iga-Fern{\'a}ndez}, S.},
        title = "{TOI-2015 b: A sub-Neptune in strong gravitational interaction with an outer non-transiting planet}",
      journal = {\aap},
         year = 2025,
        month = mar,
       volume = {695},
          eid = {A281},
        pages = {A281},
          doi = {10.1051/0004-6361/202452916},
archivePrefix = {arXiv},
       eprint = {2502.07074},
 primaryClass = {astro-ph.EP},
       adsurl = {https://ui.adsabs.harvard.edu/abs/2025A&A...695A.281B}
}

@ARTICLE{Ilaria+2025,
       author = {{Carleo}, Ilaria and {Nowak}, Grzegorz and {Murgas}, Felipe and {Pall{\'e}}, Enric and {Lacedelli}, Gaia and {Masseron}, Thomas and {Wong}, Emily W. and {Castro-Gonz{\'a}lez}, Amadeo and {Jankowski}, Dawid and {Eggenberger}, Patrick and {Jenkins}, James S. and {Go{\'z}dziewski}, Krzysztof and {Bourrier}, Vincent and {Alves}, Douglas R. and {Vines}, Jos{\'e} I. and {Stassun}, Keivan G. and {Brogi}, Matteo and {Messina}, Sergio and {Clark}, Catherine A. and {Collins}, Karen A. and {Deeg}, Hans J. and {Furlan}, Elise and {Gandolfi}, Davide and {Gonz{\'a}lez}, Samuel Gerald{\'\i}a and {Hellier}, Coel and {Hatzes}, Artie P. and {Howell}, Steve B. and {Korth}, Judith and {Knudstrup}, Emil and {Lillo-Box}, Jorge and {Livingston}, John H. and {Orell-Miquel}, Jaume and {Persson}, Carina and {Redfield}, Seth and {Safonov}, Boris and {Baker}, David and {Delgado}, Rafael Delfin Barrena and {Bieryla}, Allyson and {Boyle}, Andrew and {Bosch-Cabot}, Pau and {Barris}, N{\'u}ria Casasayas and {Chairetas}, Stavros and {Ciardi}, David R. and {Fukui}, Akihiko and {Guerra}, Pere and {Kawauchi}, Kiyoe and {Libotte}, Florence and {Lund}, Michael B. and {Luque}, Rafael and {de Escalante}, Eduardo Lorenzo Mart{\'\i}n Guerrero and {Massey}, Bob and {Michaels}, Edward J. and {Morello}, Giuseppe and {Narita}, Norio and {Parvianien}, Hannu and {Schwarz}, Richard P. and {Shporer}, Avi and {Stangret}, Monika and {Watkins}, Cristilyn N.},
        title = "{Precise mass and radius determination for two new and one known Neptune-sized planets around G Dwarf hosts}",
      journal = {\mnras},
         year = 2025,
        month = nov,
          doi = {10.1093/mnras/staf1958},
archivePrefix = {arXiv},
       eprint = {2511.20119},
 primaryClass = {astro-ph.EP},
       adsurl = {https://ui.adsabs.harvard.edu/abs/2025MNRAS.tmp.1855C}
}

@ARTICLE{Allart+2023,
       author = {{Allart}, R. and {Lem{\'e}e-Joliecoeur}, P.-B. and {Jaziri}, A.~Y. and {Lafreni{\`e}re}, D. and {Artigau}, E. and {Cook}, N. and {Darveau-Bernier}, A. and {Dang}, L. and {Cadieux}, C. and {Boucher}, A. and {Bourrier}, V. and {Deibert}, E.~K. and {Pelletier}, S. and {Radica}, M. and {Benneke}, B. and {Carmona}, A. and {Cloutier}, R. and {Cowan}, N.~B. and {Delfosse}, X. and {Donati}, J.-F. and {Doyon}, R. and {Figueira}, P. and {Forveille}, T. and {Fouqu{\'e}}, P. and {Gaidos}, E. and {Gu}, P.-G. and {H{\'e}brard}, G. and {Kiefer}, F. and {K{\'o}sp{\'a}l}, {\'A}. and {Jayawardhana}, R. and {Martioli}, E. and {Dos Santos}, L.~A. and {Shang}, H. and {Turner}, J.~D. and {Vidotto}, A.~A.},
        title = "{Homogeneous search for helium in the atmosphere of 11 gas giant exoplanets with SPIRou}",
      journal = {\aap},
         year = 2023,
        month = sep,
       volume = {677},
          eid = {A164},
        pages = {A164},
          doi = {10.1051/0004-6361/202245832},
archivePrefix = {arXiv},
       eprint = {2307.05580},
 primaryClass = {astro-ph.EP},
       adsurl = {https://ui.adsabs.harvard.edu/abs/2023A&A...677A.164A}
}

@ARTICLE{Stassun+2019,
       author = {{Stassun}, Keivan G. and {Oelkers}, Ryan J. and {Paegert}, Martin and {Torres}, Guillermo and {Pepper}, Joshua and {De Lee}, Nathan and {Collins}, Kevin and {Latham}, David W. and {Muirhead}, Philip S. and {Chittidi}, Jay and {Rojas-Ayala}, B{\'a}rbara and {Fleming}, Scott W. and {Rose}, Mark E. and {Tenenbaum}, Peter and {Ting}, Eric B. and {Kane}, Stephen R. and {Barclay}, Thomas and {Bean}, Jacob L. and {Brassuer}, C.~E. and {Charbonneau}, David and {Ge}, Jian and {Lissauer}, Jack J. and {Mann}, Andrew W. and {McLean}, Brian and {Mullally}, Susan and {Narita}, Norio and {Plavchan}, Peter and {Ricker}, George R. and {Sasselov}, Dimitar and {Seager}, S. and {Sharma}, Sanjib and {Shiao}, Bernie and {Sozzetti}, Alessandro and {Stello}, Dennis and {Vanderspek}, Roland and {Wallace}, Geoff and {Winn}, Joshua N.},
        title = "{The Revised TESS Input Catalog and Candidate Target List}",
      journal = {\aj},
         year = 2019,
        month = oct,
       volume = {158},
       number = {4},
          eid = {138},
        pages = {138},
          doi = {10.3847/1538-3881/ab3467},
archivePrefix = {arXiv},
       eprint = {1905.10694},
 primaryClass = {astro-ph.SR},
       adsurl = {https://ui.adsabs.harvard.edu/abs/2019AJ....158..138S}
}

@ARTICLE{Orell-Miquel+2023,
       author = {{Orell-Miquel}, J. and {Lamp{\'o}n}, M. and {L{\'o}pez-Puertas}, M. and {Mallorqu{\'\i}n}, M. and {Murgas}, F. and {Pel{\'a}ez-Torres}, A. and {Pall{\'e}}, E. and {Esparza-Borges}, E. and {Sanz-Forcada}, J. and {Tabernero}, H.~M. and {Nortmann}, L. and {Nagel}, E. and {Parviainen}, H. and {Zapatero Osorio}, M.~R. and {Caballero}, J.~A. and {Czesla}, S. and {Cifuentes}, C. and {Morello}, G. and {Quirrenbach}, A. and {Amado}, P.~J. and {Fern{\'a}ndez-Mart{\'\i}n}, A. and {Fukui}, A. and {Henning}, Th. and {Kawauchi}, K. and {de Leon}, J.~P. and {Molaverdikhani}, K. and {Montes}, D. and {Narita}, N. and {Reiners}, A. and {Ribas}, I. and {S{\'a}nchez-L{\'o}pez}, A. and {Schweitzer}, A. and {Stangret}, M. and {Yan}, F.},
        title = "{Confirmation of an He I evaporating atmosphere around the 650-Myr-old sub-Neptune HD 235088 b (TOI-1430 b) with CARMENES}",
      journal = {\aap},
         year = 2023,
        month = sep,
       volume = {677},
          eid = {A56},
        pages = {A56},
          doi = {10.1051/0004-6361/202346445},
archivePrefix = {arXiv},
       eprint = {2307.05191},
 primaryClass = {astro-ph.EP},
       adsurl = {https://ui.adsabs.harvard.edu/abs/2023A&A...677A..56O}
}

@ARTICLE{Linssen+2024,
       author = {{Linssen}, Dion and {Shih}, Jim and {MacLeod}, Morgan and {Oklop{\v{c}}i{\'c}}, Antonija},
        title = "{The open-source sunbather code: Modeling escaping planetary atmospheres and their transit spectra}",
      journal = {\aap},
         year = 2024,
        month = aug,
       volume = {688},
          eid = {A43},
        pages = {A43},
          doi = {10.1051/0004-6361/202450240},
archivePrefix = {arXiv},
       eprint = {2404.12775},
 primaryClass = {astro-ph.EP},
       adsurl = {https://ui.adsabs.harvard.edu/abs/2024A&A...688A..43L}
}

@ARTICLE{Linssen+2022,
       author = {{Linssen}, D.~C. and {Oklop{\v{c}}i{\'c}}, A. and {MacLeod}, M.},
        title = "{Constraining planetary mass-loss rates by simulating Parker wind profiles with Cloudy}",
      journal = {\aap},
         year = 2022,
        month = nov,
       volume = {667},
          eid = {A54},
        pages = {A54},
          doi = {10.1051/0004-6361/202243830},
archivePrefix = {arXiv},
       eprint = {2209.03677},
 primaryClass = {astro-ph.EP},
       adsurl = {https://ui.adsabs.harvard.edu/abs/2022A&A...667A..54L}
}

@ARTICLE{Zhang+2023,
       author = {{Zhang}, Zhoujian and {Morley}, Caroline V. and {Gully-Santiago}, Michael and {MacLeod}, Morgan and {Oklop{\v{c}}i{\'c}}, Antonija and {Luna}, Jessica and {Tran}, Quang H. and {Ninan}, Joe P. and {Mahadevan}, Suvrath and {Krolikowski}, Daniel M. and {Cochran}, William D. and {Bowler}, Brendan P. and {Endl}, Michael and {Stef{\'a}nsson}, Gudmundur and {Tofflemire}, Benjamin M. and {Vanderburg}, Andrew and {Zeimann}, Gregory R.},
        title = "{Giant tidal tails of helium escaping the hot Jupiter HAT-P-32 b}",
      journal = {Science Advances},
         year = 2023,
        month = jun,
       volume = {9},
       number = {23},
          eid = {eadf8736},
        pages = {eadf8736},
          doi = {10.1126/sciadv.adf8736},
archivePrefix = {arXiv},
       eprint = {2306.03913},
 primaryClass = {astro-ph.EP},
       adsurl = {https://ui.adsabs.harvard.edu/abs/2023SciA....9F8736Z}
}

@ARTICLE{Ikuta+2025,
       author = {{Ikuta}, Kai and {Narita}, Norio and {Takarada}, Takuya and {Hirano}, Teruyuki and {Fukui}, Akihiko and {Ishikawa}, Hiroyuki Tako and {Hori}, Yasunori and {Kimura}, Tadahiro and {Kodama}, Takanori and {Ikoma}, Masahiro and {de Leon}, Jerome P. and {Kawauchi}, Kiyoe and {Kuzuhara}, Masayuki and {Lacedelli}, Gaia and {Livingston}, John H. and {Mori}, Mayuko and {Murgas}, Felipe and {Palle}, Enric and {Parviainen}, Hannu and {Watanabe}, Noriharu and {Fukuda}, Izuru and {Harakawa}, Hiroki and {Hayashi}, Yuya and {Hodapp}, Klaus and {Isogai}, Keisuke and {Kagetani}, Taiki and {Kawai}, Yugo and {Krishnamurthy}, Vigneshwaran and {Kudo}, Tomoyuki and {Kurokawa}, Takashi and {Kusakabe}, Nobuhiko and {Nishikawa}, Jun and {Nugroho}, Stevanus K. and {Omiya}, Masashi and {Serizawa}, Takuma and {Takahashi}, Aoi and {Teng}, Huan-Yu and {Terada}, Yuka and {Ueda}, Akitoshi and {Vievard}, S{\'e}bastien and {Zou}, Yujie and {Kotani}, Takayuki and {Tamura}, Motohide},
        title = "{The mass of TOI-654 b: A short-period sub-Neptune transiting a mid-M dwarf}",
      journal = {\pasj},
         year = 2025,
        month = oct,
       volume = {77},
       number = {5},
        pages = {1101-1112},
          doi = {10.1093/pasj/psaf090},
archivePrefix = {arXiv},
       eprint = {2507.16222},
 primaryClass = {astro-ph.EP},
       adsurl = {https://ui.adsabs.harvard.edu/abs/2025PASJ...77.1101I}
}

@ARTICLE{Vissapragada+2022,
       author = {{Vissapragada}, Shreyas and {Knutson}, Heather A. and {Greklek-McKeon}, Michael and {Oklop{\v{c}}i{\'c}}, Antonija and {Dai}, Fei and {dos Santos}, Leonardo A. and {Jovanovic}, Nemanja and {Mawet}, Dimitri and {Millar-Blanchaer}, Maxwell A. and {Paragas}, Kimberly and {Spake}, Jessica J. and {Tinyanont}, Samaporn and {Vasisht}, Gautam},
        title = "{The Upper Edge of the Neptune Desert Is Stable Against Photoevaporation}",
      journal = {\aj},
         year = 2022,
        month = dec,
       volume = {164},
       number = {6},
          eid = {234},
        pages = {234},
          doi = {10.3847/1538-3881/ac92f2},
archivePrefix = {arXiv},
       eprint = {2204.11865},
 primaryClass = {astro-ph.EP},
       adsurl = {https://ui.adsabs.harvard.edu/abs/2022AJ....164..234V}
}

@ARTICLE{Peterson+2023,
       author = {{Peterson}, Merrin S. and {Benneke}, Bj{\"o}rn and {Collins}, Karen and {Piaulet}, Caroline and {Crossfield}, Ian J.~M. and {Ali-Dib}, Mohamad and {Christiansen}, Jessie L. and {Gagn{\'e}}, Jonathan and {Faherty}, Jackie and {Kite}, Edwin and {Dressing}, Courtney and {Charbonneau}, David and {Murgas}, Felipe and {Cointepas}, Marion and {Almenara}, Jose Manuel and {Bonfils}, Xavier and {Kane}, Stephen and {Werner}, Michael W. and {Gorjian}, Varoujan and {Roy}, Pierre-Alexis and {Shporer}, Avi and {Pozuelos}, Francisco J. and {Socia}, Quentin Jay and {Cloutier}, Ryan and {Dietrich}, Jeremy and {Irwin}, Jonathan and {Weiss}, Lauren and {Waalkes}, William and {Berta-Thomson}, Zach and {Evans}, Thomas and {Apai}, Daniel and {Parviainen}, Hannu and {Pall{\'e}}, Enric and {Narita}, Norio and {Howard}, Andrew W. and {Dragomir}, Diana and {Barkaoui}, Khalid and {Gillon}, Micha{\"e}l and {Jehin}, Emmanuel and {Ducrot}, Elsa and {Benkhaldoun}, Zouhair and {Fukui}, Akihiko and {Mori}, Mayuko and {Nishiumi}, Taku and {Kawauchi}, Kiyoe and {Ricker}, George and {Latham}, David W. and {Winn}, Joshua N. and {Seager}, Sara and {Isaacson}, Howard and {Bixel}, Alex and {Gibbs}, Aidan and {Jenkins}, Jon M. and {Smith}, Jeffrey C. and {Chavez}, Jose Perez and {Rackham}, Benjamin V. and {Henning}, Thomas and {Gabor}, Paul and {Chen}, Wen-Ping and {Espinoza}, Nestor and {Jensen}, Eric L.~N. and {Collins}, Kevin I. and {Schwarz}, Richard P. and {Conti}, Dennis M. and {Wang}, Gavin and {Kielkopf}, John F. and {Mao}, Shude and {Horne}, Keith and {Sefako}, Ramotholo and {Quinn}, Samuel N. and {Moldovan}, Dan and {Fausnaugh}, Michael and {F{\.z}{\.z}r{\'e}sz}, G{\'a}bor and {Barclay}, Thomas},
        title = "{A temperate Earth-sized planet with tidal heating transiting an M6 star}",
      journal = {\nat},
         year = 2023,
        month = may,
       volume = {617},
       number = {7962},
        pages = {701-705},
          doi = {10.1038/s41586-023-05934-8},
       adsurl = {https://ui.adsabs.harvard.edu/abs/2023Natur.617..701P}
}

@ARTICLE{Newton+2017,
       author = {{Newton}, Elisabeth R. and {Irwin}, Jonathan and {Charbonneau}, David and {Berlind}, Perry and {Calkins}, Michael L. and {Mink}, Jessica},
        title = "{The H{\ensuremath{\alpha}} Emission of Nearby M Dwarfs and its Relation to Stellar Rotation}",
      journal = {\apj},
         year = 2017,
        month = jan,
       volume = {834},
       number = {1},
          eid = {85},
        pages = {85},
          doi = {10.3847/1538-4357/834/1/85},
archivePrefix = {arXiv},
       eprint = {1611.03509},
 primaryClass = {astro-ph.SR},
       adsurl = {https://ui.adsabs.harvard.edu/abs/2017ApJ...834...85N}
}

@ARTICLE{Cloutier+2021,
       author = {{Cloutier}, Ryan and {Charbonneau}, David and {Stassun}, Keivan G. and {Murgas}, Felipe and {Mortier}, Annelies and {Massey}, Robert and {Lissauer}, Jack J. and {Latham}, David W. and {Irwin}, Jonathan and {Haywood}, Rapha{\"e}lle D. and {Guerra}, Pere and {Girardin}, Eric and {Giacalone}, Steven A. and {Bosch-Cabot}, Pau and {Bieryla}, Allyson and {Winn}, Joshua and {Watson}, Christopher A. and {Vanderspek}, Roland and {Udry}, St{\'e}phane and {Tamura}, Motohide and {Sozzetti}, Alessandro and {Shporer}, Avi and {S{\'e}gransan}, Damien and {Seager}, Sara and {Savel}, Arjun B. and {Sasselov}, Dimitar and {Rose}, Mark and {Ricker}, George and {Rice}, Ken and {Quintana}, Elisa V. and {Quinn}, Samuel N. and {Piotto}, Giampaolo and {Phillips}, David and {Pepe}, Francesco and {Pedani}, Marco and {Parviainen}, Hannu and {Palle}, Enric and {Narita}, Norio and {Molinari}, Emilio and {Micela}, Giuseppina and {McDermott}, Scott and {Mayor}, Michel and {Matson}, Rachel A. and {Martinez Fiorenzano}, Aldo F. and {Lovis}, Christophe and {L{\'o}pez-Morales}, Mercedes and {Kusakabe}, Nobuhiko and {Jensen}, Eric L.~N. and {Jenkins}, Jon M. and {Huang}, Chelsea X. and {Howell}, Steve B. and {Harutyunyan}, Avet and {F{\H{u}}r{\'e}sz}, G{\'a}bor and {Fukui}, Akihiko and {Esquerdo}, Gilbert A. and {Esparza-Borges}, Emma and {Dumusque}, Xavier and {Dressing}, Courtney D. and {Fabrizio}, Luca Di and {Collins}, Karen A. and {Cameron}, Andrew Collier and {Christiansen}, Jessie L. and {Cecconi}, Massimo and {Buchhave}, Lars A. and {Boschin}, Walter and {Andreuzzi}, Gloria},
        title = "{TOI-1634 b: An Ultra-short-period Keystone Planet Sitting inside the M-dwarf Radius Valley}",
      journal = {\aj},
         year = 2021,
        month = aug,
       volume = {162},
       number = {2},
          eid = {79},
        pages = {79},
          doi = {10.3847/1538-3881/ac0157},
archivePrefix = {arXiv},
       eprint = {2103.12790},
 primaryClass = {astro-ph.EP},
       adsurl = {https://ui.adsabs.harvard.edu/abs/2021AJ....162...79C}
}

@ARTICLE{Crossfield+2019,
       author = {{Crossfield}, Ian J.~M. and {Waalkes}, William and {Newton}, Elisabeth R. and {Narita}, Norio and {Muirhead}, Philip and {Ment}, Kristo and {Matthews}, Elisabeth and {Kraus}, Adam and {Kostov}, Veselin and {Kosiarek}, Molly R. and {Kane}, Stephen R. and {Isaacson}, Howard and {Halverson}, Sam and {Gonzales}, Erica and {Everett}, Mark and {Dragomir}, Diana and {Collins}, Karen A. and {Chontos}, Ashley and {Berardo}, David and {Winters}, Jennifer G. and {Winn}, Joshua N. and {Scott}, Nicholas J. and {Rojas-Ayala}, Barbara and {Rizzuto}, Aaron C. and {Petigura}, Erik A. and {Peterson}, Merrin and {Mocnik}, Teo and {Mikal-Evans}, Thomas and {Mehrle}, Nicholas and {Matson}, Rachel and {Kuzuhara}, Masayuki and {Irwin}, Jonathan and {Huber}, Daniel and {Huang}, Chelsea and {Howell}, Steve and {Howard}, Andrew W. and {Hirano}, Teruyuki and {Fulton}, Benjamin J. and {Dupuy}, Trent and {Dressing}, Courtney D. and {Dalba}, Paul A. and {Charbonneau}, David and {Burt}, Jennifer and {Berta-Thompson}, Zachory and {Benneke}, Bj{\"o}rn and {Watanabe}, Noriharu and {Twicken}, Joseph D. and {Tamura}, Motohide and {Schlieder}, Joshua and {Seager}, S. and {Rose}, Mark E. and {Ricker}, George and {Quintana}, Elisa and {L{\'e}pine}, S{\'e}bastien and {Latham}, David W. and {Kotani}, Takayuki and {Jenkins}, Jon M. and {Hori}, Yasunori and {Colon}, Knicole and {Caldwell}, Douglas A.},
        title = "{A Super-Earth and Sub-Neptune Transiting the Late-type M Dwarf LP 791-18}",
      journal = {\apjl},
         year = 2019,
        month = sep,
       volume = {883},
       number = {1},
          eid = {L16},
        pages = {L16},
          doi = {10.3847/2041-8213/ab3d30},
archivePrefix = {arXiv},
       eprint = {1906.09267},
 primaryClass = {astro-ph.EP},
       adsurl = {https://ui.adsabs.harvard.edu/abs/2019ApJ...883L..16C}
}

@ARTICLE{Hirano+2021,
       author = {{Hirano}, Teruyuki and {Livingston}, John H. and {Fukui}, Akihiko and {Narita}, Norio and {Harakawa}, Hiroki and {Ishikawa}, Hiroyuki Tako and {Miyakawa}, Kohei and {Kimura}, Tadahiro and {Nakayama}, Akifumi and {Fujita}, Naho and {Hori}, Yasunori and {Stassun}, Keivan G. and {Bieryla}, Allyson and {Cadieux}, Charles and {Ciardi}, David R. and {Collins}, Karen A. and {Ikoma}, Masahiro and {Vanderburg}, Andrew and {Barclay}, Thomas and {Brasseur}, C.~E. and {de Leon}, Jerome P. and {Doty}, John P. and {Doyon}, Ren{\'e} and {Esparza-Borges}, Emma and {Esquerdo}, Gilbert A. and {Furlan}, Elise and {Gaidos}, Eric and {Gonzales}, Erica J. and {Hodapp}, Klaus and {Howell}, Steve B. and {Isogai}, Keisuke and {Jacobson}, Shane and {Jenkins}, Jon M. and {Jensen}, Eric L.~N. and {Kawauchi}, Kiyoe and {Kotani}, Takayuki and {Kudo}, Tomoyuki and {Kurita}, Seiya and {Kurokawa}, Takashi and {Kusakabe}, Nobuhiko and {Kuzuhara}, Masayuki and {Lafreni{\`e}re}, David and {Latham}, David W. and {Massey}, Bob and {Mori}, Mayuko and {Murgas}, Felipe and {Nishikawa}, Jun and {Nishiumi}, Taku and {Omiya}, Masashi and {Paegert}, Martin and {Palle}, Enric and {Parviainen}, Hannu and {Quinn}, Samuel N. and {Ricker}, George R. and {Schwarz}, Richard P. and {Seager}, Sara and {Tamura}, Motohide and {Tenenbaum}, Peter and {Terada}, Yuka and {Vanderspek}, Roland K. and {Vievard}, S{\'e}bastien and {Watanabe}, Noriharu and {Winn}, Joshua N.},
        title = "{Two Bright M Dwarfs Hosting Ultra-Short-Period Super-Earths with Earth-like Compositions}",
      journal = {\aj},
         year = 2021,
        month = oct,
       volume = {162},
       number = {4},
          eid = {161},
        pages = {161},
          doi = {10.3847/1538-3881/ac0fdc},
archivePrefix = {arXiv},
       eprint = {2103.12760},
 primaryClass = {astro-ph.EP},
       adsurl = {https://ui.adsabs.harvard.edu/abs/2021AJ....162..161H}
}

@ARTICLE{Ballabio+2025,
       author = {{Ballabio}, Giulia and {Owen}, James E.},
        title = "{Understanding what helium absorption tells us about atmospheric escape from exoplanets}",
      journal = {\mnras},
         year = 2025,
        month = feb,
       volume = {537},
       number = {2},
        pages = {1305-1319},
          doi = {10.1093/mnras/staf073},
archivePrefix = {arXiv},
       eprint = {2501.06149},
 primaryClass = {astro-ph.EP},
       adsurl = {https://ui.adsabs.harvard.edu/abs/2025MNRAS.537.1305B}
}

@ARTICLE{Biassoni+2024,
       author = {{Biassoni}, Federico and {Caldiroli}, Andrea and {Gallo}, Elena and {Haardt}, Francesco and {Spinelli}, Riccardo and {Borsa}, Francesco},
        title = "{Self-consistent modeling of metastable helium exoplanet transits}",
      journal = {\aap},
         year = 2024,
        month = feb,
       volume = {682},
          eid = {A115},
        pages = {A115},
          doi = {10.1051/0004-6361/202347517},
archivePrefix = {arXiv},
       eprint = {2310.13052},
 primaryClass = {astro-ph.EP},
       adsurl = {https://ui.adsabs.harvard.edu/abs/2024A&A...682A.115B}
}

@ARTICLE{Krishnamurthy+2024,
       author = {{Krishnamurthy}, Vigneshwaran and {Cowan}, Nicolas B.},
        title = "{Helium in Exoplanet Exospheres: Orbital and Stellar Influences}",
      journal = {\aj},
         year = 2024,
        month = jul,
       volume = {168},
       number = {1},
          eid = {30},
        pages = {30},
          doi = {10.3847/1538-3881/ad5441},
archivePrefix = {arXiv},
       eprint = {2403.15575},
 primaryClass = {astro-ph.EP},
       adsurl = {https://ui.adsabs.harvard.edu/abs/2024AJ....168...30K}
}

@ARTICLE{Orell-Miquel+2024,
       author = {{Orell-Miquel}, J. and {Murgas}, F. and {Pall{\'e}}, E. and {Mallorqu{\'\i}n}, M. and {L{\'o}pez-Puertas}, M. and {Lamp{\'o}n}, M. and {Sanz-Forcada}, J. and {Nortmann}, L. and {Czesla}, S. and {Nagel}, E. and {Ribas}, I. and {Stangret}, M. and {Livingston}, J. and {Knudstrup}, E. and {Albrecht}, S.~H. and {Carleo}, I. and {Caballero}, J.~A. and {Dai}, F. and {Esparza-Borges}, E. and {Fukui}, A. and {Heng}, K. and {Henning}, Th. and {Kagetani}, T. and {Lesjak}, F. and {de Leon}, J.~P. and {Montes}, D. and {Morello}, G. and {Narita}, N. and {Quirrenbach}, A. and {Amado}, P.~J. and {Reiners}, A. and {Schweitzer}, A. and {Vico Linares}, J.~I.},
        title = "{The MOPYS project: A survey of 70 planets in search of extended He I and H atmospheres: No evidence of enhanced evaporation in young planets}",
      journal = {\aap},
         year = 2024,
        month = sep,
       volume = {689},
          eid = {A179},
        pages = {A179},
          doi = {10.1051/0004-6361/202449411},
archivePrefix = {arXiv},
       eprint = {2404.16732},
 primaryClass = {astro-ph.EP},
       adsurl = {https://ui.adsabs.harvard.edu/abs/2024A&A...689A.179O}
}

@ARTICLE{Oklopcic2019,
       author = {{Oklop{\v{c}}i{\'c}}, Antonija},
        title = "{Helium Absorption at 1083 nm from Extended Exoplanet Atmospheres: Dependence on Stellar Radiation}",
      journal = {\apj},
         year = 2019,
        month = aug,
       volume = {881},
       number = {2},
          eid = {133},
        pages = {133},
          doi = {10.3847/1538-4357/ab2f7f},
archivePrefix = {arXiv},
       eprint = {1903.02576},
 primaryClass = {astro-ph.EP},
       adsurl = {https://ui.adsabs.harvard.edu/abs/2019ApJ...881..133O}
}

@ARTICLE{Oklopcic+2018,
       author = {{Oklop{\v{c}}i{\'c}}, Antonija and {Hirata}, Christopher M.},
        title = "{A New Window into Escaping Exoplanet Atmospheres: 10830 {\r{A}} Line of Helium}",
      journal = {\apjl},
         year = 2018,
        month = mar,
       volume = {855},
       number = {1},
          eid = {L11},
        pages = {L11},
          doi = {10.3847/2041-8213/aaada9},
archivePrefix = {arXiv},
       eprint = {1711.05269},
 primaryClass = {astro-ph.EP},
       adsurl = {https://ui.adsabs.harvard.edu/abs/2018ApJ...855L..11O}
}

@ARTICLE{Eggleton1983,
       author = {{Eggleton}, P.~P.},
        title = "{Aproximations to the radii of Roche lobes.}",
      journal = {\apj},
         year = 1983,
        month = may,
       volume = {268},
        pages = {368-369},
          doi = {10.1086/160960},
       adsurl = {https://ui.adsabs.harvard.edu/abs/1983ApJ...268..368E}
}

@ARTICLE{Erkaev+2007,
       author = {{Erkaev}, N.~V. and {Kulikov}, Yu. N. and {Lammer}, H. and {Selsis}, F. and {Langmayr}, D. and {Jaritz}, G.~F. and {Biernat}, H.~K.},
        title = "{Roche lobe effects on the atmospheric loss from ``Hot Jupiters''}",
      journal = {\aap},
         year = 2007,
        month = sep,
       volume = {472},
       number = {1},
        pages = {329-334},
          doi = {10.1051/0004-6361:20066929},
archivePrefix = {arXiv},
       eprint = {astro-ph/0612729},
 primaryClass = {astro-ph},
       adsurl = {https://ui.adsabs.harvard.edu/abs/2007A&A...472..329E}
}

@ARTICLE{Loyd+2016,
       author = {{Loyd}, R.~O.~P. and {France}, Kevin and {Youngblood}, Allison and {Schneider}, Christian and {Brown}, Alexander and {Hu}, Renyu and {Linsky}, Jeffrey and {Froning}, Cynthia S. and {Redfield}, Seth and {Rugheimer}, Sarah and {Tian}, Feng},
        title = "{The MUSCLES Treasury Survey. III. X-Ray to Infrared Spectra of 11 M and K Stars Hosting Planets}",
      journal = {\apj},
         year = 2016,
        month = jun,
       volume = {824},
       number = {2},
          eid = {102},
        pages = {102},
          doi = {10.3847/0004-637X/824/2/102},
archivePrefix = {arXiv},
       eprint = {1604.04776},
 primaryClass = {astro-ph.SR},
       adsurl = {https://ui.adsabs.harvard.edu/abs/2016ApJ...824..102L}
}

@ARTICLE{Youngblood+2016,
       author = {{Youngblood}, Allison and {France}, Kevin and {Loyd}, R.~O. Parke and {Linsky}, Jeffrey L. and {Redfield}, Seth and {Schneider}, P. Christian and {Wood}, Brian E. and {Brown}, Alexander and {Froning}, Cynthia and {Miguel}, Yamila and {Rugheimer}, Sarah and {Walkowicz}, Lucianne},
        title = "{The MUSCLES Treasury Survey. II. Intrinsic LY{\ensuremath{\alpha}} and Extreme Ultraviolet Spectra of K and M Dwarfs with Exoplanets*}",
      journal = {\apj},
         year = 2016,
        month = jun,
       volume = {824},
       number = {2},
          eid = {101},
        pages = {101},
          doi = {10.3847/0004-637X/824/2/101},
archivePrefix = {arXiv},
       eprint = {1604.01032},
 primaryClass = {astro-ph.SR},
       adsurl = {https://ui.adsabs.harvard.edu/abs/2016ApJ...824..101Y}
}

@ARTICLE{France+2016,
       author = {{France}, Kevin and {Loyd}, R.~O. Parke and {Youngblood}, Allison and {Brown}, Alexander and {Schneider}, P. Christian and {Hawley}, Suzanne L. and {Froning}, Cynthia S. and {Linsky}, Jeffrey L. and {Roberge}, Aki and {Buccino}, Andrea P. and {Davenport}, James R.~A. and {Fontenla}, Juan M. and {Kaltenegger}, Lisa and {Kowalski}, Adam F. and {Mauas}, Pablo J.~D. and {Miguel}, Yamila and {Redfield}, Seth and {Rugheimer}, Sarah and {Tian}, Feng and {Vieytes}, Mariela C. and {Walkowicz}, Lucianne M. and {Weisenburger}, Kolby L.},
        title = "{The MUSCLES Treasury Survey. I. Motivation and Overview}",
      journal = {\apj},
         year = 2016,
        month = apr,
       volume = {820},
       number = {2},
          eid = {89},
        pages = {89},
          doi = {10.3847/0004-637X/820/2/89},
archivePrefix = {arXiv},
       eprint = {1602.09142},
 primaryClass = {astro-ph.SR},
       adsurl = {https://ui.adsabs.harvard.edu/abs/2016ApJ...820...89F}
}

@ARTICLE{Caldiroli+2022,
       author = {{Caldiroli}, Andrea and {Haardt}, Francesco and {Gallo}, Elena and {Spinelli}, Riccardo and {Malsky}, Isaac and {Rauscher}, Emily},
        title = "{Irradiation-driven escape of primordial planetary atmospheres. II. Evaporation efficiency of sub-Neptunes through hot Jupiters}",
      journal = {\aap},
         year = 2022,
        month = jul,
       volume = {663},
          eid = {A122},
        pages = {A122},
          doi = {10.1051/0004-6361/202142763},
archivePrefix = {arXiv},
       eprint = {2112.00744},
 primaryClass = {astro-ph.EP},
       adsurl = {https://ui.adsabs.harvard.edu/abs/2022A&A...663A.122C}
}

@ARTICLE{Caldiroli+2021,
       author = {{Caldiroli}, Andrea and {Haardt}, Francesco and {Gallo}, Elena and {Spinelli}, Riccardo and {Malsky}, Isaac and {Rauscher}, Emily},
        title = "{Irradiation-driven escape of primordial planetary atmospheres. I. The ATES photoionization hydrodynamics code}",
      journal = {\aap},
         year = 2021,
        month = nov,
       volume = {655},
          eid = {A30},
        pages = {A30},
          doi = {10.1051/0004-6361/202141497},
archivePrefix = {arXiv},
       eprint = {2106.10294},
 primaryClass = {astro-ph.EP},
       adsurl = {https://ui.adsabs.harvard.edu/abs/2021A&A...655A..30C}
}

@ARTICLE{DosSantos+2022,
       author = {{Dos Santos}, Leonardo A. and {Vidotto}, Aline A. and {Vissapragada}, Shreyas and {Alam}, Munazza K. and {Allart}, Romain and {Bourrier}, Vincent and {Kirk}, James and {Seidel}, Julia V. and {Ehrenreich}, David},
        title = "{p-winds: An open-source Python code to model planetary outflows and upper atmospheres}",
      journal = {\aap},
         year = 2022,
        month = mar,
       volume = {659},
          eid = {A62},
        pages = {A62},
          doi = {10.1051/0004-6361/202142038},
archivePrefix = {arXiv},
       eprint = {2111.11370},
 primaryClass = {astro-ph.EP},
       adsurl = {https://ui.adsabs.harvard.edu/abs/2022A&A...659A..62D}
}

@ARTICLE{S2025,
       author = {{Sanz-Forcada}, J. and {L{\'o}pez-Puertas}, M. and {Lamp{\'o}n}, M. and {Czesla}, S. and {Nortmann}, L. and {Caballero}, J.~A. and {Zapatero Osorio}, M.~R. and {Amado}, P.~J. and {Murgas}, F. and {Orell-Miquel}, J. and {Pall{\'e}}, E. and {Quirrenbach}, A. and {Reiners}, A. and {Ribas}, I. and {S{\'a}nchez-L{\'o}pez}, A. and {Solano}, E.},
        title = "{Connection between planetary He I {\ensuremath{\lambda}}10 830 {\r{A}} absorption and extreme-ultraviolet emission of planet-host stars}",
      journal = {\aap},
         year = 2025,
        month = jan,
       volume = {693},
          eid = {A285},
        pages = {A285},
          doi = {10.1051/0004-6361/202451680},
archivePrefix = {arXiv},
       eprint = {2501.03716},
 primaryClass = {astro-ph.EP},
       adsurl = {https://ui.adsabs.harvard.edu/abs/2025A&A...693A.285S}
}

@ARTICLE{Donati+1997,
       author = {{Donati}, J. -F. and {Semel}, M. and {Carter}, B.~D. and {Rees}, D.~E. and {Collier Cameron}, A.},
        title = "{Spectropolarimetric observations of active stars}",
      journal = {\mnras},
         year = 1997,
        month = nov,
       volume = {291},
       number = {4},
        pages = {658-682},
          doi = {10.1093/mnras/291.4.658},
       adsurl = {https://ui.adsabs.harvard.edu/abs/1997MNRAS.291..658D}
}

@ARTICLE{Kawauchi+2018,
       author = {{Kawauchi}, Kiyoe and {Narita}, Norio and {Sato}, Bun'ei and
         {Hirano}, Teruyuki and {Kawashima}, Yui and {Nakamoto}, Taishi and
         {Yamashita}, Takuya and {Tamura}, Motohide},
        title = "{Earth's atmosphere's lowest layers probed during a lunar eclipse}",
      journal = {\pasj},
         year = "2018",
        month = "Oct",
       volume = {70},
       number = {5},
          eid = {84},
        pages = {84},
          doi = {10.1093/pasj/psy079},
archivePrefix = {arXiv},
       eprint = {1806.09085},
 primaryClass = {astro-ph.EP},
       adsurl = {https://ui.adsabs.harvard.edu/abs/2018PASJ...70...84K}
}

@INPROCEEDINGS{Tamura+2012,
       author = {{Tamura}, M. and {Suto}, H. and {Nishikawa}, J. and {Kotani}, T. and {Sato}, B. and {Aoki}, W. and {Usuda}, T. and {Kurokawa}, T. and {Kashiwagi}, K. and {Nishiyama}, S. and {Ikeda}, Y. and {Hall}, D. and {Hodapp}, K. and {Hashimoto}, J. and {Morino}, J. and {Inoue}, S. and {Mizuno}, Y. and {Washizaki}, Y. and {Tanaka}, Y. and {Suzuki}, S. and {Kwon}, J. and {Suenaga}, T. and {Oh}, D. and {Narita}, N. and {Kokubo}, E. and {Hayano}, Y. and {Izumiura}, H. and {Kambe}, E. and {Kudo}, T. and {Kusakabe}, N. and {Ikoma}, M. and {Hori}, Ya. and {Omiya}, M. and {Genda}, H. and {Fukui}, A. and {Fujii}, Y. and {Guyon}, O. and {Harakawa}, H. and {Hayashi}, M. and {Hidai}, M. and {Hirano}, T. and {Kuzuhara}, M. and {Machida}, M. and {Matsuo}, T. and {Nagata}, T. and {Ohnuki}, H. and {Ogihara}, M. and {Oshino}, S. and {Suzuki}, R. and {Takami}, H. and {Takato}, N. and {Takahashi}, Y. and {Tachinami}, C. and {Terada}, H.},
        title = "{Infrared Doppler instrument for the Subaru Telescope (IRD)}",
    booktitle = {Ground-based and Airborne Instrumentation for Astronomy IV},
         year = 2012,
       editor = {{McLean}, Ian S. and {Ramsay}, Suzanne K. and {Takami}, Hideki},
       series = {Society of Photo-Optical Instrumentation Engineers (SPIE) Conference Series},
       volume = {8446},
        month = sep,
          eid = {84461T},
        pages = {84461T},
          doi = {10.1117/12.925885},
       adsurl = {https://ui.adsabs.harvard.edu/abs/2012SPIE.8446E..1TT}
}

@INPROCEEDINGS{Kotani+2018,
       author = {{Kotani}, Takayuki and {Tamura}, Motohide and {Nishikawa}, Jun and {Ueda}, Akitoshi and {Kuzuhara}, Masayuki and {Omiya}, Masashi and {Hashimoto}, Jun and {Ishizuka}, Masato and {Hirano}, Teruyuki and {Suto}, Hiroshi and {Kurokawa}, Takashi and {Kokubo}, Tsukasa and {Mori}, Takahiro and {Tanaka}, Yosuke and {Kashiwagi}, Ken and {Konishi}, Mihoko and {Kudo}, Tomoyuki and {Sato}, Bun'ei and {Jacobson}, Shane and {Hodapp}, Klaus W. and {Hall}, Donald B. and {Aoki}, Wako and {Usuda}, Tomonori and {Nishiyama}, Shogo and {Nakajima}, Tadashi and {Ikeda}, Yuji and {Yamamuro}, Tomoyasu and {Morino}, Jun-Ichi and {Baba}, Haruka and {Hosokawa}, Ko and {Ishikawa}, Hiroyuki and {Narita}, Norio and {Kokubo}, Eiichiro and {Hayano}, Yutaka and {Izumiura}, Hideyuki and {Kambe}, Eiji and {Kusakabe}, Nobuhiko and {Kwon}, Jungmi and {Ikoma}, Masahiro and {Hori}, Yasunori and {Genda}, Hidenori and {Fukui}, Akihiko and {Fujii}, Yuka and {Kawahara}, Hajime and {Olivier}, Guyon and {Jovanovic}, Nemanja and {Harakawa}, Hiroki and {Hayashi}, Masahiko and {Hidai}, Masahide and {Machida}, Masahiro and {Matsuo}, Taro and {Nagata}, Tetsuya and {Ogihara}, Masahiro and {Takami}, Hideki and {Takato}, Naruhisa and {Terada}, Hiroshi and {Oh}, Daehyeon},
        title = "{The infrared Doppler (IRD) instrument for the Subaru telescope: instrument description and commissioning results}",
    booktitle = {Ground-based and Airborne Instrumentation for Astronomy VII},
         year = 2018,
       editor = {{Evans}, Christopher J. and {Simard}, Luc and {Takami}, Hideki},
       series = {Society of Photo-Optical Instrumentation Engineers (SPIE) Conference Series},
       volume = {10702},
        month = jul,
          eid = {1070211},
        pages = {1070211},
          doi = {10.1117/12.2311836},
       adsurl = {https://ui.adsabs.harvard.edu/abs/2018SPIE10702E..11K}
}

@INPROCEEDINGS{Tody1993,
       author = {{Tody}, Doug},
        title = "{IRAF in the Nineties}",
    booktitle = {Astronomical Data Analysis Software and Systems II},
         year = 1993,
       editor = {{Hanisch}, R.~J. and {Brissenden}, R.~J.~V. and {Barnes}, J.},
       series = {Astronomical Society of the Pacific Conference Series},
       volume = {52},
        month = jan,
        pages = {173},
       adsurl = {https://ui.adsabs.harvard.edu/abs/1993ASPC...52..173T}
}

@ARTICLE{Hirano+2020,
       author = {{Hirano}, Teruyuki and {Kuzuhara}, Masayuki and {Kotani}, Takayuki and {Omiya}, Masashi and {Kudo}, Tomoyuki and {Harakawa}, Hiroki and {Vievard}, S{\'e}bastien and {Kurokawa}, Takashi and {Nishikawa}, Jun and {Tamura}, Motohide and {Hodapp}, Klaus and {Ishizuka}, Masato and {Jacobson}, Shane and {Konishi}, Mihoko and {Serizawa}, Takuma and {Ueda}, Akitoshi and {Gaidos}, Eric and {Sato}, Bun'ei},
        title = "{Precision radial velocity measurements by the forward-modeling technique in the near-infrared}",
      journal = {\pasj},
         year = 2020,
        month = dec,
       volume = {72},
       number = {6},
          eid = {93},
        pages = {93},
          doi = {10.1093/pasj/psaa085},
archivePrefix = {arXiv},
       eprint = {2007.11013},
 primaryClass = {astro-ph.EP},
       adsurl = {https://ui.adsabs.harvard.edu/abs/2020PASJ...72...93H}
}

@INPROCEEDINGS{Kuzuhara+2018,
       author = {{Kuzuhara}, Masayuki and {Hirano}, Teruyuki and {Kotani}, Takayuki and {Ishizuka}, Masato and {Omiya}, Masashi and {Konishi}, Mihoko and {Kudo}, Tomoyuki and {Nishikawa}, Jun and {Ueda}, Akitoshi and {Hosokawa}, Ko and {Kusakabe}, Nobuhiko and {Kurokawa}, Takashi and {Kokubo}, Tsukasa and {Mori}, Takahiro and {Tanaka}, Yosuke and {Jacobson}, Shane and {Hodapp}, Klaus and {Tamura}, Motohide},
        title = "{Performance tests of Subaru/IRD for very precise and stable infrared radial velocity observations}",
    booktitle = {Ground-based and Airborne Instrumentation for Astronomy VII},
         year = 2018,
       editor = {{Evans}, Christopher J. and {Simard}, Luc and {Takami}, Hideki},
       series = {Society of Photo-Optical Instrumentation Engineers (SPIE) Conference Series},
       volume = {10702},
        month = jul,
          eid = {1070260},
        pages = {1070260},
          doi = {10.1117/12.2311832},
       adsurl = {https://ui.adsabs.harvard.edu/abs/2018SPIE10702E..60K}
}

@ARTICLE{Krishnamurthy+2021,
       author = {{Krishnamurthy}, Vigneshwaran and {Hirano}, Teruyuki and {Stef{\'a}nsson}, Gumundur and {Ninan}, Joe P. and {Mahadevan}, Suvrath and {Gaidos}, Eric and {Kopparapu}, Ravi and {Sato}, Bunei and {Hori}, Yasunori and {Bender}, Chad F. and {Ca{\~n}as}, Caleb I. and {Diddams}, Scott A. and {Halverson}, Samuel and {Harakawa}, Hiroki and {Hawley}, Suzanne and {Hearty}, Fred and {Hebb}, Leslie and {Hodapp}, Klaus and {Jacobson}, Shane and {Kanodia}, Shubham and {Konishi}, Mihoko and {Kotani}, Takayuki and {Kowalski}, Adam and {Kudo}, Tomoyuki and {Kurokawa}, Takashi and {Kuzuhara}, Masayuki and {Lin}, Andrea and {Maney}, Marissa and {Metcalf}, Andrew J. and {Morris}, Brett and {Nishikawa}, Jun and {Omiya}, Masashi and {Robertson}, Paul and {Roy}, Arpita and {Schwab}, Christian and {Serizawa}, Takuma and {Tamura}, Motohide and {Ueda}, Akitoshi and {Vievard}, S{\'e}bastien and {Wisniewski}, John},
        title = "{Nondetection of Helium in the Upper Atmospheres of TRAPPIST-1b, e, and f}",
      journal = {\aj},
         year = 2021,
        month = sep,
       volume = {162},
       number = {3},
          eid = {82},
        pages = {82},
          doi = {10.3847/1538-3881/ac0d57},
archivePrefix = {arXiv},
       eprint = {2106.11444},
 primaryClass = {astro-ph.EP},
       adsurl = {https://ui.adsabs.harvard.edu/abs/2021AJ....162...82K}
}

@ARTICLE{Cutri+2003,
       author = {{Cutri}, R.~M. and {Skrutskie}, M.~F. and {van Dyk}, S. and {Beichman}, C.~A. and {Carpenter}, J.~M. and {Chester}, T. and {Cambresy}, L. and {Evans}, T. and {Fowler}, J. and {Gizis}, J. and {Howard}, E. and {Huchra}, J. and {Jarrett}, T. and {Kopan}, E.~L. and {Kirkpatrick}, J.~D. and {Light}, R.~M. and {Marsh}, K.~A. and {McCallon}, H. and {Schneider}, S. and {Stiening}, R. and {Sykes}, M. and {Weinberg}, M. and {Wheaton}, W.~A. and {Wheelock}, S. and {Zacarias}, N.},
        title = "{VizieR Online Data Catalog: 2MASS All-Sky Catalog of Point Sources (Cutri+ 2003)}",
      journal = {VizieR Online Data Catalog},
         year = 2003,
        month = jun,
          eid = {II/246},
        pages = {II/246},
       adsurl = {https://ui.adsabs.harvard.edu/abs/2003yCat.2246....0C}
}

@ARTICLE{Zeng2019,
       author = {{Zeng}, Li and {Jacobsen}, Stein B. and {Sasselov}, Dimitar D. and
         {Petaev}, Michail I. and {Vanderburg}, Andrew and
         {Lopez-Morales}, Mercedes and {Perez-Mercader}, Juan and
         {Mattsson}, Thomas R. and {Li}, Gongjie and {Heising}, Matthew Z. and
         {Bonomo}, Aldo S. and {Damasso}, Mario and {Berger}, Travis A. and
         {Cao}, Hao and {Levi}, Amit and {Wordsworth}, Robin D.},
        title = "{Growth model interpretation of planet size distribution}",
      journal = {Proceedings of the National Academy of Science},
         year = 2019,
        month = may,
       volume = {116},
       number = {20},
        pages = {9723-9728},
          doi = {10.1073/pnas.1812905116},
archivePrefix = {arXiv},
       eprint = {1906.04253},
 primaryClass = {astro-ph.EP},
       adsurl = {https://ui.adsabs.harvard.edu/abs/2019PNAS..116.9723Z}
}

@INPROCEEDINGS{Allard+2014,
       author = {{Allard}, F.},
        title = "{The BT-Settl Model Atmospheres for Stars, Brown Dwarfs and Planets}",
    booktitle = {Exploring the Formation and Evolution of Planetary Systems},
         year = 2014,
       editor = {{Booth}, Mark and {Matthews}, Brenda C. and {Graham}, James R.},
       volume = {299},
        month = jan,
        pages = {271-272},
          doi = {10.1017/S1743921313008545},
       adsurl = {https://ui.adsabs.harvard.edu/abs/2014IAUS..299..271A}
}

@ARTICLE{Ricker+2015,
       author = {{Ricker}, George R. and {Winn}, Joshua N. and {Vanderspek}, Roland and {Latham}, David W. and {Bakos}, G{\'a}sp{\'a}r {\'A}. and {Bean}, Jacob L. and {Berta-Thompson}, Zachory K. and {Brown}, Timothy M. and {Buchhave}, Lars and {Butler}, Nathaniel R. and {Butler}, R. Paul and {Chaplin}, William J. and {Charbonneau}, David and {Christensen-Dalsgaard}, J{\o}rgen and {Clampin}, Mark and {Deming}, Drake and {Doty}, John and {De Lee}, Nathan and {Dressing}, Courtney and {Dunham}, Edward W. and {Endl}, Michael and {Fressin}, Francois and {Ge}, Jian and {Henning}, Thomas and {Holman}, Matthew J. and {Howard}, Andrew W. and {Ida}, Shigeru and {Jenkins}, Jon M. and {Jernigan}, Garrett and {Johnson}, John Asher and {Kaltenegger}, Lisa and {Kawai}, Nobuyuki and {Kjeldsen}, Hans and {Laughlin}, Gregory and {Levine}, Alan M. and {Lin}, Douglas and {Lissauer}, Jack J. and {MacQueen}, Phillip and {Marcy}, Geoffrey and {McCullough}, Peter R. and {Morton}, Timothy D. and {Narita}, Norio and {Paegert}, Martin and {Palle}, Enric and {Pepe}, Francesco and {Pepper}, Joshua and {Quirrenbach}, Andreas and {Rinehart}, Stephen A. and {Sasselov}, Dimitar and {Sato}, Bun'ei and {Seager}, Sara and {Sozzetti}, Alessandro and {Stassun}, Keivan G. and {Sullivan}, Peter and {Szentgyorgyi}, Andrew and {Torres}, Guillermo and {Udry}, Stephane and {Villasenor}, Joel},
        title = "{Transiting Exoplanet Survey Satellite (TESS)}",
      journal = {Journal of Astronomical Telescopes, Instruments, and Systems},
         year = 2015,
        month = jan,
       volume = {1},
          eid = {014003},
        pages = {014003},
          doi = {10.1117/1.JATIS.1.1.014003},
       adsurl = {https://ui.adsabs.harvard.edu/abs/2015JATIS...1a4003R}
}

@ARTICLE{Fulton+2017,
       author = {{Fulton}, Benjamin J. and {Petigura}, Erik A. and {Howard}, Andrew W. and {Isaacson}, Howard and {Marcy}, Geoffrey W. and {Cargile}, Phillip A. and {Hebb}, Leslie and {Weiss}, Lauren M. and {Johnson}, John Asher and {Morton}, Timothy D. and {Sinukoff}, Evan and {Crossfield}, Ian J.~M. and {Hirsch}, Lea A.},
        title = "{The California-Kepler Survey. III. A Gap in the Radius Distribution of Small Planets}",
      journal = {\aj},
         year = 2017,
        month = sep,
       volume = {154},
       number = {3},
          eid = {109},
        pages = {109},
          doi = {10.3847/1538-3881/aa80eb},
archivePrefix = {arXiv},
       eprint = {1703.10375},
 primaryClass = {astro-ph.EP},
       adsurl = {https://ui.adsabs.harvard.edu/abs/2017AJ....154..109F}
}

@ARTICLE{Lopez&Rice2018,
       author = {{Lopez}, Eric D. and {Rice}, Ken},
        title = "{How formation time-scales affect the period dependence of the transition between rocky super-Earths and gaseous sub-Neptunesand implications for {\ensuremath{\eta}}$_{{\ensuremath{\oplus}}}$}",
      journal = {\mnras},
         year = 2018,
        month = oct,
       volume = {479},
       number = {4},
        pages = {5303-5311},
          doi = {10.1093/mnras/sty1707},
archivePrefix = {arXiv},
       eprint = {1610.09390},
 primaryClass = {astro-ph.EP},
       adsurl = {https://ui.adsabs.harvard.edu/abs/2018MNRAS.479.5303L}
}

@ARTICLE{Gupta&Schlichting2019,
       author = {{Gupta}, Akash and {Schlichting}, Hilke E.},
        title = "{Sculpting the valley in the radius distribution of small exoplanets as a by-product of planet formation: the core-powered mass-loss mechanism}",
      journal = {\mnras},
         year = 2019,
        month = jul,
       volume = {487},
       number = {1},
        pages = {24-33},
          doi = {10.1093/mnras/stz1230},
archivePrefix = {arXiv},
       eprint = {1811.03202},
 primaryClass = {astro-ph.EP},
       adsurl = {https://ui.adsabs.harvard.edu/abs/2019MNRAS.487...24G}
}

@ARTICLE{Madhusudhan+2021,
       author = {{Madhusudhan}, Nikku and {Piette}, Anjali A.~A. and {Constantinou}, Savvas},
        title = "{Habitability and Biosignatures of Hycean Worlds}",
      journal = {\apj},
         year = 2021,
        month = sep,
       volume = {918},
       number = {1},
          eid = {1},
        pages = {1},
          doi = {10.3847/1538-4357/abfd9c},
archivePrefix = {arXiv},
       eprint = {2108.10888},
 primaryClass = {astro-ph.EP},
       adsurl = {https://ui.adsabs.harvard.edu/abs/2021ApJ...918....1M}
}

@ARTICLE{Lepine2005,
       author = {{L{\'e}pine}, S{\'e}bastien and {Shara}, Michael M.},
        title = "{A Catalog of Northern Stars with Annual Proper Motions Larger than 0.15'' (LSPM-NORTH Catalog)}",
      journal = {\aj},
         year = 2005,
        month = mar,
       volume = {129},
       number = {3},
        pages = {1483-1522},
          doi = {10.1086/427854},
archivePrefix = {arXiv},
       eprint = {astro-ph/0412070},
 primaryClass = {astro-ph},
       adsurl = {https://ui.adsabs.harvard.edu/abs/2005AJ....129.1483L}
}

@ARTICLE{Ginzburg+2016,
       author = {{Ginzburg}, Sivan and {Schlichting}, Hilke E. and {Sari}, Re'em},
        title = "{Super-Earth Atmospheres: Self-consistent Gas Accretion and Retention}",
      journal = {\apj},
         year = 2016,
        month = jul,
       volume = {825},
       number = {1},
          eid = {29},
        pages = {29},
          doi = {10.3847/0004-637X/825/1/29},
archivePrefix = {arXiv},
       eprint = {1512.07925},
 primaryClass = {astro-ph.EP},
       adsurl = {https://ui.adsabs.harvard.edu/abs/2016ApJ...825...29G}
}

@ARTICLE{Ginzburg+2018,
       author = {{Ginzburg}, Sivan and {Schlichting}, Hilke E. and {Sari}, Re'em},
        title = "{Core-powered mass-loss and the radius distribution of small exoplanets}",
      journal = {\mnras},
         year = 2018,
        month = may,
       volume = {476},
       number = {1},
        pages = {759-765},
          doi = {10.1093/mnras/sty290},
archivePrefix = {arXiv},
       eprint = {1708.01621},
 primaryClass = {astro-ph.EP},
       adsurl = {https://ui.adsabs.harvard.edu/abs/2018MNRAS.476..759G}
}

@ARTICLE{Gupta&Schlichting2021,
       author = {{Gupta}, Akash and {Schlichting}, Hilke E.},
        title = "{Caught in the act: core-powered mass-loss predictions for observing atmospheric escape}",
      journal = {\mnras},
         year = 2021,
        month = jul,
       volume = {504},
       number = {3},
        pages = {4634-4648},
          doi = {10.1093/mnras/stab1128},
archivePrefix = {arXiv},
       eprint = {2103.08785},
 primaryClass = {astro-ph.EP},
       adsurl = {https://ui.adsabs.harvard.edu/abs/2021MNRAS.504.4634G}
}

@ARTICLE{Mordasini2020,
       author = {{Mordasini}, C.},
        title = "{Planetary evolution with atmospheric photoevaporation. I. Analytical derivation and numerical study of the evaporation valley and transition from super-Earths to sub-Neptunes}",
      journal = {\aap},
         year = 2020,
        month = jun,
       volume = {638},
          eid = {A52},
        pages = {A52},
          doi = {10.1051/0004-6361/201935541},
archivePrefix = {arXiv},
       eprint = {2002.02455},
 primaryClass = {astro-ph.EP},
       adsurl = {https://ui.adsabs.harvard.edu/abs/2020A&A...638A..52M}
}

@ARTICLE{Owen&Wu2017,
       author = {{Owen}, James E. and {Wu}, Yanqin},
        title = "{The Evaporation Valley in the Kepler Planets}",
      journal = {\apj},
         year = 2017,
        month = sep,
       volume = {847},
       number = {1},
          eid = {29},
        pages = {29},
          doi = {10.3847/1538-4357/aa890a},
archivePrefix = {arXiv},
       eprint = {1705.10810},
 primaryClass = {astro-ph.EP},
       adsurl = {https://ui.adsabs.harvard.edu/abs/2017ApJ...847...29O}
}

@ARTICLE{Owen&Wu2013,
       author = {{Owen}, James E. and {Wu}, Yanqin},
        title = "{Kepler Planets: A Tale of Evaporation}",
      journal = {\apj},
         year = 2013,
        month = oct,
       volume = {775},
       number = {2},
          eid = {105},
        pages = {105},
          doi = {10.1088/0004-637X/775/2/105},
archivePrefix = {arXiv},
       eprint = {1303.3899},
 primaryClass = {astro-ph.EP},
       adsurl = {https://ui.adsabs.harvard.edu/abs/2013ApJ...775..105O}
}

@ARTICLE{Jin+2014,
       author = {{Jin}, Sheng and {Mordasini}, Christoph and {Parmentier}, Vivien and {van Boekel}, Roy and {Henning}, Thomas and {Ji}, Jianghui},
        title = "{Planetary Population Synthesis Coupled with Atmospheric Escape: A Statistical View of Evaporation}",
      journal = {\apj},
         year = 2014,
        month = nov,
       volume = {795},
       number = {1},
          eid = {65},
        pages = {65},
          doi = {10.1088/0004-637X/795/1/65},
archivePrefix = {arXiv},
       eprint = {1409.2879},
 primaryClass = {astro-ph.EP},
       adsurl = {https://ui.adsabs.harvard.edu/abs/2014ApJ...795...65J}
}

@ARTICLE{Elkins-Tanton2008,
       author = {{Elkins-Tanton}, Linda T. and {Seager}, Sara},
        title = "{Ranges of Atmospheric Mass and Composition of Super-Earth Exoplanets}",
      journal = {\apj},
         year = 2008,
        month = oct,
       volume = {685},
       number = {2},
        pages = {1237-1246},
          doi = {10.1086/591433},
archivePrefix = {arXiv},
       eprint = {0808.1909},
 primaryClass = {astro-ph},
       adsurl = {https://ui.adsabs.harvard.edu/abs/2008ApJ...685.1237E}
}

@ARTICLE{Orell-Miquel+2022,
       author = {{Orell-Miquel}, J. and {Murgas}, F. and {Pall{\'e}}, E. and {Lamp{\'o}n}, M. and {L{\'o}pez-Puertas}, M. and {Sanz-Forcada}, J. and {Nagel}, E. and {Kaminski}, A. and {Casasayas-Barris}, N. and {Nortmann}, L. and {Luque}, R. and {Molaverdikhani}, K. and {Sedaghati}, E. and {Caballero}, J.~A. and {Amado}, P.~J. and {Bergond}, G. and {Czesla}, S. and {Hatzes}, A.~P. and {Henning}, Th. and {Khalafinejad}, S. and {Montes}, D. and {Morello}, G. and {Quirrenbach}, A. and {Reiners}, A. and {Ribas}, I. and {S{\'a}nchez-L{\'o}pez}, A. and {Schweitzer}, A. and {Stangret}, M. and {Yan}, F. and {Zapatero Osorio}, M.~R.},
        title = "{A tentative detection of He I in the atmosphere of GJ 1214 b}",
      journal = {\aap},
         year = 2022,
        month = mar,
       volume = {659},
          eid = {A55},
        pages = {A55},
          doi = {10.1051/0004-6361/202142455},
archivePrefix = {arXiv},
       eprint = {2201.11120},
 primaryClass = {astro-ph.EP},
       adsurl = {https://ui.adsabs.harvard.edu/abs/2022A&A...659A..55O}
}

@ARTICLE{Czesla+2022,
       author = {{Czesla}, S. and {Lamp{\'o}n}, M. and {Sanz-Forcada}, J. and {Garc{\'\i}a Mu{\~n}oz}, A. and {L{\'o}pez-Puertas}, M. and {Nortmann}, L. and {Yan}, D. and {Nagel}, E. and {Yan}, F. and {Schmitt}, J.~H.~M.~M. and {Aceituno}, J. and {Amado}, P.~J. and {Caballero}, J.~A. and {Casasayas-Barris}, N. and {Henning}, Th. and {Khalafinejad}, S. and {Molaverdikhani}, K. and {Montes}, D. and {Pall{\'e}}, E. and {Reiners}, A. and {Schneider}, P.~C. and {Ribas}, I. and {Quirrenbach}, A. and {Zapatero Osorio}, M.~R. and {Zechmeister}, M.},
        title = "{H{\ensuremath{\alpha}} and He I absorption in HAT-P-32 b observed with CARMENES. Detection of Roche lobe overflow and mass loss}",
      journal = {\aap},
         year = 2022,
        month = jan,
       volume = {657},
          eid = {A6},
        pages = {A6},
          doi = {10.1051/0004-6361/202039919},
archivePrefix = {arXiv},
       eprint = {2110.13582},
 primaryClass = {astro-ph.EP},
       adsurl = {https://ui.adsabs.harvard.edu/abs/2022A&A...657A...6C}
}

@ARTICLE{Kawauchi+2022b,
       author = {{Kawauchi}, K. and {Murgas}, F. and {Palle}, E. and {Narita}, N. and {Fukui}, A. and {Hirano}, T. and {Parviainen}, H. and {Ishikawa}, H.~T. and {Watanabe}, N. and {Esparaza-Borges}, E. and {Kuzuhara}, M. and {Orell-Miquel}, J. and {Krishnamurthy}, V. and {Mori}, M. and {Kagetani}, T. and {Zou}, Y. and {Isogai}, K. and {Livingston}, J.~H. and {Howell}, S.~B. and {Crouzet}, N. and {de Leon}, J.~P. and {Kimura}, T. and {Kodama}, T. and {Korth}, J. and {Kurita}, S. and {Laza-Ramos}, A. and {Luque}, R. and {Madrigal-Aguado}, A. and {Miyakawa}, K. and {Morello}, G. and {Nishiumi}, T. and {Rodr{\'\i}guez}, G.~E.~F. and {S{\'a}nchez-Benavente}, M. and {Stangret}, M. and {Teng}, H. and {Terada}, Y. and {Gnilka}, C.~L. and {Guerrero}, N. and {Harakawa}, H. and {Hodapp}, K. and {Hori}, Y. and {Ikoma}, M. and {Jacobson}, S. and {Konishi}, M. and {Kotani}, T. and {Kudo}, T. and {Kurokowa}, T. and {Kusakabe}, N. and {Nishikawa}, J. and {Omiya}, M. and {Serizawa}, T. and {Tamura}, M. and {Ueda}, A. and {Vievard}, S.},
        title = "{Validation and atmospheric exploration of the sub-Neptune TOI-2136b around a nearby M3 dwarf}",
      journal = {\aap},
         year = 2022,
        month = oct,
       volume = {666},
          eid = {A4},
        pages = {A4},
          doi = {10.1051/0004-6361/202243381},
archivePrefix = {arXiv},
       eprint = {2202.10182},
 primaryClass = {astro-ph.EP},
       adsurl = {https://ui.adsabs.harvard.edu/abs/2022A&A...666A...4K}
}

\end{document}